\documentclass[11pt]{article}
\pdfoutput=1
\usepackage{color}
\usepackage{mhchem}
\usepackage{xcolor}
\usepackage{cite}
\usepackage{hyperref}
\hypersetup{colorlinks=true,linkcolor=red,anchorcolor=black,citecolor=green}
\usepackage[toc,page]{appendix}
\usepackage{mathtools}
\usepackage{amsfonts}
\usepackage{bbold}
\usepackage{textcomp}
\usepackage[DIV13]{typearea}
\usepackage{amsmath, amsthm, amssymb, mathtools,empheq,latexsym,dsfont}
\usepackage{bbm}
\usepackage{amsmath,amssymb,latexsym,dsfont}
\usepackage{slashed, simplewick}
\usepackage[utf8]{inputenc}
\usepackage{graphicx}
\usepackage{graphicx,placeins}
\usepackage{makeidx}
\usepackage[font=small,labelfont=bf]{caption}
\usepackage{nicefrac}
\usepackage{subfigure}
\usepackage{array, bigdelim,multirow,multicol}
\usepackage{adjustbox}
\usepackage{booktabs}
\usepackage{indentfirst}
\usepackage{booktabs}
\usepackage{tabulary}
\usepackage{longtable}
\usepackage{fancybox}
\usepackage{graphicx}
\usepackage{bm}
\usepackage{bbm}
\usepackage{tikz}
\usepackage{diagbox}
\usepackage{enumitem}

\graphicspath{{immagini/}}

\renewcommand{\theequation}{\thesection.\arabic{equation}}

\definecolor{Gray}{gray}{0.92}

\newcommand{\ignore}[1]{}

\newcommand{\be}{\begin{equation}}
\newcommand{\ee}{\end{equation}}
\newcommand{\bea}{\begin{eqnarray}}
\newcommand{\eea}{\end{eqnarray}}

\newcounter{Thm}[section]
\renewcommand{\theThm}{\arabic{section}.\arabic{Thm}}

\newcounter{nodecount}
\newcommand\tabnode[1]{\addtocounter{nodecount}{1} \tikz \node (\arabic{nodecount}) {#1};}

\tikzstyle{every picture}+=[remember picture,baseline]
\tikzstyle{every node}+=[inner sep=0pt,anchor=base,
text depth=.25ex,outer sep=1.5pt]
\tikzstyle{every path}+=[thick, rounded corners]
\tikzset{
        plabel/.style={inner sep=2pt}
}

\makeatletter
\@addtoreset{equation}{section}
\makeatother
\begin{document}

\title{\begin{center}
{\Large\bf Minimal eclectic flavor group $Q_{8}\rtimes S_3$ and neutrino mixing}
\end{center}}
\date{}
\author{Cai-Chang Li$^{a,b,c}$\footnote{E-mail: {\tt
ccli@nwu.edu.cn}},  \
Jun-Nan Lu$^{d}$\footnote{E-mail: {\tt hitman@mail.ustc.edu.cn}},  \
Gui-Jun Ding$^{d}$\footnote{E-mail: {\tt dinggj@ustc.edu.cn}} \ \\*[20pt]
\centerline{\begin{minipage}{\linewidth}
\begin{center}
$^a${\it\small School of Physics, Northwest University, Xi'an 710127, China}\\[2mm]
$^b${\it\small Shaanxi Key Laboratory for Theoretical Physics Frontiers, Xi'an 710127, China}\\[2mm]
$^c${\it\small NSFC-SPTP Peng Huanwu Center for Fundamental Theory, Xi'an 710127, China}\\[2mm]
$^d${\it \small Department of Modern Physics, University of Science and Technology of China,\\
Hefei, Anhui 230026, China}\\[2mm]
\end{center}
\end{minipage}} \\[10mm]}
\maketitle

\centerline{\large\bf Abstract}
\begin{quote}
\indent
We perform a comprehensive analysis of the minimal eclectic flavor group $Q_{8}\rtimes S_3$ which is isomorphic to $GL(2,3)$, and all its irreducible representations are induced from the irreducible representations of $Q_{8}$ and $S_{3}$. The consistency conditions between EFG and generalized CP (gCP)  symmetry are revisited, and we find the gCP symmetry compatible with the minimal EFG $Q_{8}\rtimes S_3$. The most general forms of K\"ahler potential and superpotential based on $Q_{8}\rtimes S_3$ are discussed, and the corresponding fermion mass matrices are presented. A concrete lepton model invariant under $Q_{8}\rtimes S_3$ and gCP is constructed, in which the experimental data of all six lepton masses and six mixing parameters can be successfully described through seven real input parameters. The model predicts  a vanishing effective mass $m_{\beta\beta}$ in neutrinoless double beta decay.
\end{quote}

\newpage

\section{Introduction}

The origin of fermion mass hierarchy, flavor mixing and CP violation is one of the fundamental problems in particle physics.  So far, there is no an organizing principle or model that can successfully solve the puzzle.  The discrete flavor symmetry has been widely exploited to understand the flavor structure of the standard model (SM)~\cite{Altarelli:2010gt,Ishimori:2010au,King:2013eh,King:2014nza,King:2017guk,Petcov:2017ggy,Xing:2020ijf,Feruglio:2019ybq,Almumin:2022rml,Ding:2024ozt}, and the modular invariance approach~\cite{Feruglio:2017spp} provides an origin of the flavor symmetry, see Refs.~\cite{Kobayashi:2023zzc,Ding:2023htn} for recent review.  In the approach of breaking traditional flavor symmetry and gCP, lepton mixing  angles and CP violation phases  can be predicted in terms one to two real parameters for certain residual symmetries of the charged lepton and neutrino sectors~\cite{Feruglio:2012cw,Lu:2016jit,Li:2017abz,Lu:2019gqp}, while lepton masses are unconstrained. Usually a large number of scalar fields called flavons are required to be break the traditional flavor symmetry, and their vacuum expectation values (VEVs) are along specific directions in flavor space. The modular flavor symmetry can avoid the complication of vacuum alignment and allows to construct quite predictive fermion mass models. For example all the six lepton masses as well as six lepton mixing parameters can be determined only in terms of six real input parameters in the known  minimal lepton model~\cite{Ding:2022nzn}, and the masses and mixing parameters of both quarks and leptons can be described  by 14 real free parameters in the minimal model~\cite{Ding:2023ydy}. However, there are still some unsolved problems such as modulus stabilization~\cite{Feruglio:2019ybq,Ding:2023htn} and the issue of less constrained K\"ahler potential~\cite{Chen:2019ewa,Lu:2019vgm}  in modular flavor symmetry.

In order to solve the unconstrained K\"ahler potential problem, recently the modular symmetry  has been extended to include traditional flavor symmetry. This approach takes the advantage that the K\"ahler potential is suppressed by powers of $\langle\Phi\rangle/\Lambda$ in the traditional flavor symmetry models~\cite{King:2003xq,King:2004tx,Antusch:2007vw,Chen:2012ha,Chen:2013aya}, where $\Lambda$ denotes the cutoff scale and $\langle\Phi\rangle$ represents the VEV of flavon. The simplest idea is to impose the so-called quasi-eclectic flavor symmetry which is the direct product of a finite modular symmetry and a traditional flavor symmetry~\cite{Chen:2021prl}. In this approach, traditional flavor group and  finite modular group can be freely chosen. On the other hand, the top-down constructions motivated by string theory generally give rise to both traditional flavor symmetry such as $D_4$ and $\Delta(54)$~\cite{Kobayashi:2004ya,Kobayashi:2006wq,Abe:2009vi} and modular symmetry. Thus the full symmetry group is a nontrivial product of traditional flavor symmetry and
modular symmetry, and it is called eclectic flavor group (EFG)~\cite{Baur:2019kwi,Baur:2019iai,Nilles:2020tdp,Nilles:2020gvu,Nilles:2020nnc,Nilles:2020kgo}.
The scheme of EFG has also been discussed in a bottom-up way~\cite{Nilles:2020nnc,Nilles:2020kgo,Ding:2023ynd,Li:2023dvm}. The modular transformations are found to be the automorphisms of the traditional flavor group $G_{f}$, and hence the finite modular group $\Gamma_{N}$ (or $\Gamma^{\prime}_{N}$) is a subgroup of the full automorphism group of $G_{f}$. The mathematical structure of EFG is found to be a semi-direct product of the traditional flavor group $G_{f}$ and the finite modular group $\Gamma_{N}$ (or $\Gamma'_{N}$)~\cite{Ding:2023ynd,Li:2023dvm}. The scheme of EFG allows one to combine the advantages of both the traditional flavor and modular invariance approaches while largely avoiding their limitations. The superpotential and  K\"ahler potential in eclectic models would be severely restricted at the same time. Hence the EFG scheme is more predictive than any one of other two schemes alone.

In eclectic models, both flavon fields and complex modulus $\tau$  are necessary, in which $\tau$ is invariant under the action of traditional flavor transformation and flavon fields transform nontrivially under both traditional flavor and modular symmetries. The EFG has to be broken to obtain realistic fermion masses and mixing angles, the breaking of modular symmetry can arise from VEV of flavon or modulus $\tau$, while the breaking of traditional symmetry can only  arise from the VEV of flavon. The EFG scheme has already provided us with completely realistic predictive models in both top-down approach and bottom-up approach. The first string-derived model based on the EFG $\Omega(2)\cong [1944 , 3448]$ has been constructed~\cite{Baur:2022hma}, here we adopt the naming scheme of \texttt{GAP}~\cite{GAP,SmallGroups}. In the bottom-up approach, the phenomenological example models for the lepton sector based on the EFG $\Omega(1)\cong\Delta(27)\rtimes T^{\prime}$~\cite{Ding:2023ynd} and $\Delta(27)\rtimes S_{3}$~\cite{Li:2023dvm} have been constructed. These eclectic models are highly predictive and can present a fit to the data with much fewer parameters than observables.

In the present work, we shall study the EFG $Q_{8}\rtimes S_{3}\cong GL(2,3)$ with group ID $[48,29]$ in a bottom-up way. It is the minimal EFG and is an extension of the traditional flavor group $Q_{8}$ by the finite modular group $S_{3}$ which is a subgroup of the full automorphism group of $Q_{8}$. All modular transformations correspond to the outer automorphisms  of $Q_{8}$, and certain consistency conditions have to be fulfilled in order to consistently combine $S_{3}$ with $Q_{8}$.  The modular transformation matrices in the $Q_{8}$ multiplets  can be derived  by solving the consistency condition between traditional flavor group and finite modular group. Then all the eight irreducible representations of the EFG $Q_{8}\rtimes S_{3}$ are induced from the irreducible representations of $Q_{8}$ and $S_{3}$. Furthermore, we discuss the consistency conditions between EFG and generalized CP (gCP) symmetry. The gCP symmetry compatible with the EFG $Q_{8}\rtimes S_{3}$ is given, then the group ID of resulting group with gCP is $[96,193]$ in \texttt{GAP}. Following the EFG scheme, we perform a comprehensive analysis for the superpotential and K\"ahler potential which are invariant under the EFG $Q_{8}\rtimes S_{3}$ for all possible assignments of matter fields and flavons. We find that the corrections from K\"ahler potential to the fermion masses and flavor mixing parameters are under control by $Q_{8}\rtimes S_{3}$ when three generations of matter fields are assigned to a singlet $\bm{1_{a}}$ plus a doublet $\bm{2_{b}}$. The mass matrices of matter fields for each possible representation assignments of the matter fields and flavon are derived. Then we propose construct a concrete lepton model based on $Q_{8}\rtimes S_{3}$ and gCP. In the model,  the K\"ahler potential is strongly restricted by EFG and it is constrained to be the minimal form plus higher order corrections suppressed by powers of $\langle\Phi\rangle^2/\Lambda^2$. The six observable fermion masses and six mixing parameters can be described very well by only seven real input parameters, and the effective mass $m_{\beta\beta}$ in neutrinoless double beta decay is always predicted to be zero.

The layout of this paper is as follows. In section~\ref{sec:EFG}, we present the basic theory of EFG. The EFG $Q_{8}\rtimes S_{3}$ is illustrated, and all its irreducible representations are obtained. We study the implications of combining the gCP symmetry with EFG in the construction of models. In section~\ref{sec:mod_general}, we perform a comprehensive analysis for the K\"ahler potential and superpotential  based on the EFG $Q_{8}\rtimes S_{3}$. In section~\ref{sec:example-model-EFG}, we give a viable lepton model invariant under both $Q_{8}\rtimes S_{3}$ and gCP symmetries. Section~\ref{sec:conclusion} is devoted to conclusion. The group theory of $Q_{8}\rtimes S_{3}\cong GL(2,3)$ are presented in Appendix~\ref{sec:EFGQ8S3_group}.

\section{\label{sec:EFG}Eclectic flavor group $Q_{8}\rtimes S_{3}$ and generalized CP}

The top-down constructions motivated by string theory generally gives rise to the EFG which combines the traditional flavor symmetry with modular symmetry~\cite{Baur:2019kwi,Baur:2019iai,Nilles:2020nnc,Nilles:2020kgo,Nilles:2020tdp,Nilles:2020gvu}. The full modular group $\Gamma\cong SL(2, \mathbb{Z})$ can be generated in terms of two generators $S$ and $T$ fulfilling the following relations
\begin{equation}
S^4=(ST)^3=\mathbb{1}_2,~~~~S^2T=TS^2\,,
\end{equation}
where $\mathbb{1}_2$ denotes the two-dimensional identity matrix. The matrix representations of the two generators can be written to be
\begin{equation}
S=\begin{pmatrix}
 0  & 1\\-1  & 0
\end{pmatrix},\qquad T=\begin{pmatrix}
1 & 1 \\
0  & 1
\end{pmatrix}\,.
\end{equation}
In the approach of EFG, the modulus $\tau$ is invariant under the action of the traditional flavor symmetry $G_{f}$ and the modular group $ SL(2, \mathbb{Z})$ acts on it  with linear fractional transformation:
\begin{equation}
\tau\stackrel{g}{\longrightarrow}g\tau=\tau, \quad
\tau \stackrel{\gamma}{\longrightarrow}\gamma \tau\equiv\frac{a\tau+b}{c\tau+d}, \qquad \forall g\in G_{f}, \quad  \gamma=\begin{pmatrix}
a  &  b\\
c  & d
\end{pmatrix}\in SL(2, \mathbb{Z})\,,
\end{equation}
where the imaginary part of $\tau$ is positive. The traditional flavor symmetry $G_{f}$ and the modular group $ SL(2, \mathbb{Z})$  act on a  generic matter field multiplet $\psi$ as follow
\begin{equation}\label{eq:psi_trans}
\psi\stackrel{g}{\longrightarrow}\rho(g)\psi, \qquad ~\psi\stackrel{\gamma}{\longrightarrow}(c\tau+d)^{-k_{\psi}}\rho(\gamma)\psi \,,
\end{equation}
where $k_{\psi}$ is the modular weight of $\psi$, and $\rho(g)$ and $\rho(\gamma)$ refer to unitary representations of  $G_f$ and the inhomogeneous finite modular group $\Gamma_{N}$ (or the homogeneous finite modular group $\Gamma^\prime_{N}$), respectively. The finite modular group $\Gamma_{N}$  and its double covering $\Gamma^\prime_{N}$ are defined as the quotient groups  $\Gamma_N\equiv\Gamma/(\pm\Gamma(N))$ and $\Gamma^{\prime}_N\equiv\Gamma/\Gamma(N)$ respectively, where $\Gamma(N)$ denotes as the principal congruence subgroup of level $N$ which includes $T^{N}$, consequently they can be defined by the presentation rules~\cite{Feruglio:2017spp,Liu:2019khw}
\begin{equation}\label{eq:GammaN_defing}
S^{N_{s}}=(ST)^3=T^N=1\,, \qquad S^2T=TS^2\,,
\end{equation}
with $N_s=2$ for $\Gamma_N$ and $N_s=4$ for $\Gamma^{\prime}_N$, and additional relations are necessary to render the group finite for level $N>5$~\cite{deAdelhartToorop:2011re}. The weight $k$ and level $N$ modular forms which are a holomorphic function of the complex modulus $\tau$ can be arranged into some modular multiplets of $\Gamma_{N}$ for even $k$ and  of $\Gamma^{\prime}_N$ for integer $k$, while they are invariant under the traditional flavor group $G_f$. Under the action of $\Gamma_N$ ($\Gamma^\prime_N$), the corresponding modular multiplets $Y^{(k)}(\tau)$ transforms as~\cite{Feruglio:2017spp,Liu:2019khw}:
\begin{equation}
Y^{(k)}(\tau) \stackrel{\gamma}{\longrightarrow}Y^{(k)}\left(\gamma\tau\right)=\left(c\tau+d\right)^{k}\rho_{\bm{r}}(\gamma)\,Y^{(k)}(\tau)\,,
\end{equation}
where $\rho_{\bm{r}}(\gamma)$ refers to unitary irreducible  representation of $\Gamma_N$ ($\Gamma^\prime_N$).

In the approach of EFG, the traditional flavor symmetry and finite modular symmetry must be compatible with each other. The consistency
condition between them must be satisfied~\cite{Nilles:2020nnc,Ding:2023ynd}
\begin{equation}\label{eq:cons_con}
\rho(\gamma) \rho(g) \rho^{-1}(\gamma)=\rho(u_{\gamma}(g)) \,, \qquad \forall g\in G_{f}\,, \quad \gamma\in SL(2, \mathbb{Z})\,,
\end{equation}
where $\rho(g)$ and $\rho(\gamma)$ are defined in Eq.~\eqref{eq:psi_trans}. Obviously the modular transformation $\rho(\gamma)$ must be related to an automorphism $u_{\gamma}:G_{f}\rightarrow G_{f}$, where $u_{\gamma}$ can be  inner or outer automorphism of $G_f$.
It implies that the finite modular group $\Gamma_{N}$ (or $\Gamma^{\prime}_{N}$) generated by $u_{\gamma}$ must be a subgroup of the automorphism group of $G_{f}$, and the mathematical structure of EFG is a semi-direct product $G_{f}\rtimes \Gamma_{N}$ (or $G_{f}\rtimes \Gamma^{\prime}_{N}$).  Note that if $u_{\gamma}$ is the trivial identity automorphism with $u_{\gamma}(g)=g$ for every $\gamma\in \Gamma$ and for all $g\in G_{f}$, the mathematical structure of $G_{f}$ and $\Gamma_{N}$ (or $\Gamma^{\prime}_{N}$) is generally a direct product instead of a semi-direct product and the result group is the so-called quasi-eclectic flavor symmetry~\cite{Chen:2021prl}. In the present work, we are not interested in the approach of quasi-eclectic flavor symmetry. Therefore, we shall be concerned with the case that $u_{\gamma}$ is nontrivial at least for some modular transformation $\gamma$.

As $S$ and $T$ generate the entire modular groups $\Gamma_N$ and $\Gamma^\prime_{N}$, it is sufficient to consider the consistency condition of Eq.~\eqref{eq:cons_con}  for $\gamma=S, T$:
\begin{equation}\label{eq:ST_Cons}
\rho(S)\,\rho(g)\,\rho^{-1}(S)= \rho(u_{S}(g)), \qquad \rho(T)\,\rho(g)\,\rho^{-1}(T)~=~ \rho(u_{T}(g))\,.
\end{equation}
Here $\rho(S)$ and $\rho(T)$ are the representation matrices of $S$ and $T$ respectively, and they have to satisfy the presentation rules in Eq.~\eqref{eq:GammaN_defing}, so do the automorphisms $u_{S}$ and $u_{T}$, i.e.
\begin{equation}\label{eq:uS_uT_rules}
\left(u_{S}\right)^{N_s} =\left(u_{T}\right)^N =\left(u_{S} \circ u_{T}\right)^3=1,  \qquad \left(u_{S}\right)^2  \circ u_{T} = u_{T} \circ \left(u_{S}\right)^2\,.
\end{equation}

\subsection{\label{eq:EFG_Q8S3}The minimal EFG $Q_{8}\rtimes S_{3}$}

\begin{table}[tb]\centering
\begin{tabular}{c|ccccc}
Classes & \tabnode{$1C_1$} & \tabnode{$1C_2$} & \tabnode{$2C_4$} & \tabnode{$2C^{\prime}_4$}  & \tabnode{$2C^{\prime\prime}_4$}\\ \hline
Representatives & $1$ & $A^2$ & $A$ & $B$ & $AB$\\ \hline
~\tabnode{$\bm{1_{0,0}}$} & 1 &1 &1 & 1 & 1\\
~\tabnode{$\bm{1_{0,1}}$} & 1 & $1$ & $1$ & $-1$ & $-1$\\
~\tabnode{$\bm{1_{1,0}}$} & 1 & $1$ & $-1$ & $1$ & $-1$\\
~\tabnode{$\bm{1_{1,1}}$} & 1 & $1$ & $-1$ & $-1$ & $1$\\
~\tabnode{$\bm{2}$}  & 2 & $-2$ & 0 & 0 & $0$ \\
\end{tabular}
\begin{tikzpicture}[overlay]
\node [above=.3cm,minimum width=0pt] at (1) (c1){};
\node [above=.3cm,minimum width=0pt] at (2) (c2){};
\node [above=.3cm,minimum width=0pt] at (3) (c3){};
\node [above=.3cm,minimum width=0pt] at (4) (c4){};
\node [above=.3cm,minimum width=0pt] at (5) (c5){};
\node [left=.3cm,minimum width=0pt] at (6) (r00){};
\node [left=.3cm,minimum width=0pt] at (7) (r01){};
\node [left=.3cm,minimum width=0pt] at (8) (r10){};
\node [left=.3cm,minimum width=0pt] at (9) (r11){};
\node [left=.3cm,minimum width=0pt] at (10) (r2){};
\node [right=.2cm,minimum width=0pt] at (6) (r00ri){};
\node [right=.2cm,minimum width=0pt] at (7) (r01ri){};
\node [right=.2cm,minimum width=0pt] at (8) (r10ri){};
\node [right=.2cm,minimum width=0pt] at (9) (r11ri){};
\node [right=.3cm,minimum width=0pt] at (10) (r2ri){};
\node [above=.25cm,minimum width=0pt] at (10) (r2ab){};
\node [below=.25cm,minimum width=0pt] at (10) (r2be){};
\draw [<->,out=45,in=135,blue!50, very thick,below=1cm] (c4) to (c5);
\draw [<->,out=45,in=135,green!50, very thick,below=1cm] (c3) to (c4);
\draw [<->,out=215,in=145,green!50, very thick,below=1cm] (r01) to (r10);
\draw [<->,out=215,in=145,blue!50, very thick,below=1cm] (r10) to (r11);
\end{tikzpicture}
\caption{\label{tab:Q8character}Character table of $Q_{8}$. The arrows illustrate the generators of the outer automorphism group $u_S$(blue) and $u_T$(green) which act on the character table by interchanging the conjugacy classes and irreducible representations. All other conjugacy classes and irreducible representations remain unchanged under the actions of $u_{S}$ or $u_{T}$.}
\end{table}

For  non-Abelian traditional flavor symmetry, the minimal EFG turns out to be $Q_8\rtimes S_3$ in which the traditional flavor symmetry group
$Q_8$~\cite{Ishimori:2010au,Frigerio:2004jg} is extended by the modular symmetry $S_3$~\cite{Nilles:2020nnc}. Notice that the minimal EFG is $(Z_2\times Z_2)\rtimes S_3\cong S_4 $ if the traditional flavor symmetry is an Abelian group. $Q_8$ is  a non-Abelian group of order 8 with \texttt{GAP} ID [8,4] and can be generated by two generators $A$ and $B$ satisfying the multiplication rules
\begin{equation}\label{eq:Q8_mul_ruls}
A^4=1,\qquad B^2=A^2,\qquad B^{-1}AB=A^{-1}\,.
\end{equation}
The eight group elements of $Q_{8}$ can be divided into 5 conjugacy classes
\begin{equation}\label{eq:Q8_CC}
1C_1=\{1\}, \quad 1C_2=\{A^2\},\quad
2C_4= \{A,A^3\}, \quad  2C^{\prime}_4= \{B,B^3\},  \quad  2C^{\prime\prime}_4= \{AB,AB^3\}\,,
\end{equation}
where $kC_{n}$ denotes a conjugacy class which contains $k$ elements of order $n$. It is easy to see that $Q_{8}$ has five irreducible representations: four singlet representations $\bm{1_{a,b}}$ with $a,b=0,1$, and one doublet representation $\bm{2}$. In the present work, we shall work in the basis where the representation matrix of the generator $A$ is diagonal. The explicit forms of the representation matrices of generators $A$ and $B$ in the five irreducible representations are taken to be~\cite{Ishimori:2010au}
\begin{eqnarray}
\nonumber  \bm{1_{a,b}}&:&~~ \rho_{\bm{1_{a,b}}}(A)=(-1)^{a}, \qquad \rho_{\bm{1_{a,b}}}(B)=(-1)^{b} \,,  \\
\label{eq:Q8_irre} \bm{2}&:&~~ \rho_{\bm{2}}(A)=
\left(\begin{array}{cc}
i &~ 0 \\
0 &~ -i \\
\end{array}\right), \qquad \rho_{\bm{2}}(B)=\left(\begin{array}{cc}
0 &~ i \\
i &~ 0 \\
\end{array}\right) \,.
\end{eqnarray}
The character table of $Q_{8}$ follows immediately, as shown in table~\ref{tab:Q8character}.

Following the discussion above, we have indicated that the finite modular group which is compatible the traditional flavor group $Q_{8}$ must be a subgroup of the full automorphism group of  $Q_{8}$. The automorphism group of $ Q_{8}$ is $\mathrm{Aut}\left(Q_{8}\right) \cong S_4$. It only contains three subgroups $S_3$, $A_4$ and $S_4$ which can be regarded as finite modular groups. Hence the three EFGs $Q_{8}\rtimes S_{3}\cong GL(2,3)\cong [48,29]$, $Q_{8}\rtimes A_{4}\cong [96,204]$ and $Q_{8}\rtimes S_{4}\cong[192,1494]$ can be obtained from the extensions of the traditional flavor group $Q_{8}$ by the finite modular group $\Gamma_2\cong S_3$, $\Gamma_3\cong A_4$ and $\Gamma_4\cong S_4$, respectively. In the present work, the minimal EFG will be considered, i.e. the traditional flavor group $Q_{8}$ is extended by $\Gamma_{2}\cong S_{3}$.

The group $\Gamma_{2}\cong S_{3}$ is a symmetric group of degree three. The multiplication rules of it can be obtained from Eq.~\eqref{eq:GammaN_defing} with $N_{s}=N=2$, i.e.
\begin{equation}\label{eq:Fin_Modu_S3}
S^2=T^2=(ST)^3=1\,.
\end{equation}
The $S_{3}$ group has two singlet irreducible representations $\bm{1^{a}}$ with $a=0, 1$ and one two-dimensional representation $\bm{2^{\prime}}$.  Our choice for the representation matrices of the generators $S$ and $T$ are
\begin{eqnarray}
\nonumber  \bm{1^{a}}&:&~~ \rho_{\bm{1^{a}}}(S)=(-1)^{a}, \qquad \rho_{\bm{1^{a}}}(T)=(-1)^{a} \,,  \\
\label{eq:S3_irre}  \bm{2^{\prime}}&:&~~ \rho_{\bm{2^{\prime}}}(S)=-\frac{1}{2}
\left(\begin{array}{cc} 1 &~  \sqrt{3} \\
 \sqrt{3} &~ -1 \\
\end{array}\right), \qquad \rho_{\bm{2^{\prime}}}(T)=\left(\begin{array}{cc}
1 &~ 0 \\
0 &~ -1 \\
\end{array}\right) \,.
\end{eqnarray}

In our working basis, we can choose the following two automorphisms $u_{S}$ and $ u_{T}$ of $Q_{8}$ to generate the finite modular group $\Gamma_{2}\cong S_{3}$~\cite{Nilles:2020nnc}
\begin{equation}\label{eq:S3_uSuT}
u_{S}(A)  = A^3,  \qquad  u_{S}(B)  = AB,  \qquad   u_{T}(A)  = B, \qquad  u_{T}(B)  = A\,.
\end{equation}
Note both automorphisms $u_{S}$ and $u_{T}$ are outer automorphisms of $Q_{8}$.  One can verify that $u_{S}$ and $ u_{T}$ satisfy the following relations
\begin{equation}\label{eq:auto_S3_rules}
\left(u_{S}\right)^{2} = \left(u_{T}\right)^2=\left(u_{S} \circ u_{T}\right)^3=1\,,
\end{equation}
which are the multiplication rules of the finite modular group $\Gamma_{2}\cong S_{3}$. If the traditional flavor symmetry $Q_{8}$ and modular symmetry whose generators $S$ and $T$ satisfy the conditions in Eq.~\eqref{eq:S3_uSuT} are considered in a model at the same time, the traditional flavor group $Q_{8}$ is extended to the EFG $Q_{8}\rtimes S_{3}$. Furthermore, we find that all nontrivial modular transformations of $S_{3}$ correspond to the outer automorphisms of $Q_{8}$.

As we know, the outer automorphisms of a group map elements from one conjugacy class to another as well as one irreducible representation to another, while keep the character table invariant. Under the action of the outer automorphism $u_S$, the $Q_8$ conjugacy classes $2C^{\prime}_{4}$ and $2C^{\prime\prime}_{4}$ are exchanged while others are left invariant. Analogously $u_T$ interchanges the conjugacy classes $2C_{4}$ and $2C^{\prime}_{4}$. The action of $u_S$ and $u_T$ is shown in  table~\ref{tab:Q8character}. The consistency condition Eq.~\eqref{eq:cons_con}  may be regard as a similarity transformation between the representations $\rho$ and $\rho^\prime=\rho\circ u_{\gamma}$ of $Q_{8}$ under the action of the automorphism $u_{\gamma}$, i.e.
\begin{equation}\label{eq:rep_trans}
u_{\gamma}:\rho\rightarrow\rho^\prime=\rho\circ u_{\gamma}\,,
\end{equation}
which implies that different irreducible  representations of $Q_{8}$ are interchanged by the automorphism $u_{\gamma}$ and the corresponding  modular transformation $\rho(\gamma)$ can only be implemented if all representations connected by the corresponding automorphism are present in the theory. Hence one need to define modular transformation $\rho(\gamma)$ on the reducible representation of $Q_{8}$. In order to determine the explicit expressions of the representation matrices $\rho(S)$ and $\rho(T)$ corresponding to the outer automorphisms $u_{S}$ and $u_{T}$, one has to find  all irreducible representations of $Q_{8}$ which are  connected by the corresponding automorphism $u_{S}$ and $u_{T}$.

It is straightforward to verify that the trivial singlet representation $\bm{1_{0,0}}$ and the two-dimensional representations $\bm{2}$ of $Q_{8}$ are both invariant under the actions of $u_{S}$ and $u_{T}$, i.e.
\begin{equation}
u_{S},~u_{T}~:~ \bm{1_{0,0}}\rightarrow \bm{1_{0,0}}\,,\qquad  \bm{2}\rightarrow \bm{2}\,,
\end{equation}
which implies that the two irreducible representations already contain all representations that are connected via the outer automorphisms $u_{S}$ and $u_{T}$, and they need not be extended to include other irreducible representations of $Q_{8}$. The two outer automorphisms $u_S$ and $u_T$ interchange  the three nontrivial singlet irreducible representations of $Q_{8}$ as follows
\begin{equation}\label{eq:uS_uT_singlets}
u_{S}~:~ \bm{1_{0,1}}\leftrightarrow \bm{1_{0,1}}\,,\quad  \bm{1_{1,0}}\leftrightarrow \bm{1_{1,1}}\,, \qquad \qquad
u_{T}~:~ \bm{1_{0,1}}\leftrightarrow \bm{1_{1,0}}\,,\quad  \bm{1_{1,1}}\leftrightarrow \bm{1_{1,1}}\,,
\end{equation}
which is displayed in table~\ref{tab:Q8character}. As a consequence, the three nontrivial one-dimensional representations of $Q_{8}$ are required to form a reducible three-dimensional representation to fulfill the consistency condition Eq.~\eqref{eq:cons_con} with $u_{S}$ and $u_{T}$ in Eq.~\eqref{eq:S3_uSuT}. The reducible three-dimensional representation of $Q_{8}$  is taken to be
\begin{equation}\label{eq:Q8_3D_def}
\bm{3}\equiv(\bm{1_{0,1}}, \bm{1_{1,0}}, \bm{1_{1,1}})^T\,.
\end{equation}
For the triplet representation $\bm{3}$, the representation matrices of the $Q_{8}$ generators $A$ and $B$ are given by
\begin{equation}\label{eq:Q8_3D_Reps}
\bm{3}:~~\rho_{\bm{3}}(A)=\text{diag}(1,-1,-1), \qquad \rho_{\bm{3}}(B)=\text{diag}(-1,1,-1)\,.
\end{equation}
For the representative automorphisms $u_S:(A,\,B)\rightarrow (A^3,\,AB)$ and $u_T:(A,\,B)\rightarrow (B,\,A)$, we find that the corresponding modular transformations $\rho(S)$ and $\rho(T)$ are determined by the consistency equations
\begin{eqnarray}
 \nonumber &&\rho_{\bm{r}}(S)\,\rho_{\bm{r}}(A)\,\rho^{-1}_{\bm{r}}(S) = \rho_{\bm{r}}(A^3),\qquad \rho_{\bm{r}}(S)\,\rho_{\bm{r}}(B)\,\rho^{-1}_{\bm{r}}(S)  = \rho_{\bm{r}}(AB)\,,\\
 \label{eq:cons_Q8_S3}&& \rho_{\bm{r}}(T)\,\rho_{\bm{r}}(A)\,\rho^{-1}_{\bm{r}}(T) = \rho_{\bm{r}}(B), \qquad \rho_{\bm{r}}(T)\,\rho_{\bm{r}}(B)\,\rho^{-1}_{\bm{r}}(T) = \rho_{\bm{r}}(A)\,,
\end{eqnarray}
where $\bm{r}$ can be the trivial singlet representation $\bm{1_{0,0}}$, irreducible doublet representation $\bm{2}$ and the reducible triplet representation $\bm{3}$ of $Q_{8}$. As elements $S$ and $T$ are the generators of $S_{3}$, then the representation matrices $\rho_{\bm{r}}(S)$ and $\rho_{\bm{r}}(T)$ have to satisfy the defining relations in Eq.~\eqref{eq:Fin_Modu_S3}, i.e.
\begin{equation}\label{eq:S3_mul_rules_rhoST}
\rho^2_{\bm{r}}(S)=\rho^2_{\bm{r}}(T)=\rho^3_{\bm{r}}(ST)=\mathbb{1}_{\bm{r}}\,.
\end{equation}
Furthermore, if $\bm{r}$ is irreducible representation of $S_{3}$ with $\rho_{\bm{r}}(A)=\rho_{\bm{r}}(B)=\mathbb{1}_{\bm{r}}$, the consistency equations in Eq.~\eqref{eq:cons_Q8_S3} and the multiplication rules in Eq.~\eqref{eq:S3_mul_rules_rhoST} are satisfied automatically. Then we obtain two one-dimensional irreducible representations and one two-dimensional representation of the EFG $Q_{8}\rtimes S_{3}$. They are induced from the trivial singlet representation of $Q_{8}$ and are defined as $\bm{1_{a}}$ ($a=0,1$) and $\bm{2^{0}}$, in which the representation matrices of generators $A$, $B$, $S$ and $T$ are
\begin{eqnarray}
\nonumber \bm{1_{a}} &:& \rho_{\bm{1_{a}}}(A)=1, \qquad \rho_{\bm{1_{a}}}(B)=1,\qquad   \rho_{\bm{1_{a}}}(S)=(-1)^a, \qquad \rho_{\bm{1_{a}}}(T)=(-1)^a \,, \\
\bm{2^{0}}  &:&\rho_{\bm{2^{0}}}(A)=\mathbb{1}_{2}, \qquad \rho_{\bm{2^{0}}}(B)=\mathbb{1}_{2}, \qquad \rho_{\bm{2^{0}}}(S)=\rho_{\bm{2^{\prime}}}(S), \qquad \rho_{\bm{2^{0}}}(T)=\rho_{\bm{2^{\prime}}}(T)\,.~~
\end{eqnarray}
These can also be derived from the  corresponding consistency conditions for the traditional flavor symmetry and modular  transformations acting on the modular multiplets. For the irreducible two-dimensional representation $\bm{2}$ of $Q_{8}$, the modular transformations $\rho_{\bm{r}}(S)$ and $\rho_{\bm{r}}(T)$ which fulfil the consistency conditions in Eq.~\eqref{eq:cons_Q8_S3} and the multiplication rules in Eq.~\eqref{eq:S3_mul_rules_rhoST}  are determined to be
\begin{equation}\label{eq:EFG_Q8S3_2D}
\rho_{{\bm{2_{a}}}}({S}) = (-1)^a\left(
\begin{array}{cc}
0 & -\xi \\
\xi^3 & 0
\end{array}\right)\;, \qquad
\rho_{{\bm{2_{a}}}}({T}) = \frac{(-1)^a}{\sqrt{2}}\left(\begin{array}{cc}       1 & 1 \\
1 & -1 \end{array}\right) \;,
\end{equation}
with $a=0,1$ and $\xi=e^{\pi i/4}$. Then we obtain two irreducible representations $\bm{2_{a}}$ of the EFG $Q_{8}\rtimes S_{3}$ with $\rho_{\bm{2_{a}}}(A)=\rho_{\bm{2}}(A)$ and $\rho_{\bm{2_{a}}}(B)=\rho_{\bm{2}}(B)$. They are induced from the  two-dimensional representation $\bm{2}$ of $Q_{8}$. Furthermore, for the reducible three-dimensional representation $\bm{3}$ which is defined in Eq.~\eqref{eq:Q8_3D_Reps}, we find two independent solutions for $\rho_{\bm{r}}(S)$ and $\rho_{\bm{r}}(T)$, and they can be taken to be
\begin{equation}\label{eq:EFG_Q8S3_3D}
\rho_{{\bm{ 3_{a}}}}(S) = (-1)^{a}\left(\begin{array}{ccc}1&0&0\\0&0&i\\0&-i&0\end{array}\right) \;,\qquad
\rho_{{\bm{ 3_{a}}}}(T) =(-1)^{a}\left(
\begin{array}{ccc}
0 &1  & 0 \\
1 & 0   & 0 \\
0 & 0  & 1 \\
\end{array}
\right) \;,
\end{equation}
with $a = 0,1$. Then two irreducible three-dimensional representations $\bm{ 3_{a}}$ of $Q_{8}\rtimes S_{3}$ are achieved, in which the representation matrices of generators $A$ and $B$ are $\rho_{\bm{3_{a}}}(A)=\rho_{\bm{3}}(A)$ and $\rho_{\bm{3_{a}}}(B)=\rho_{\bm{3}}(B)$, respectively.

In short, from the representations of $Q_{8}$ flavor symmetry we have achieved partial irreducible representations of the EFG $Q_{8}\rtimes S_{3}$ denoted as $\bm{1_{a}}$, $\bm{2^{0}}$, $\bm {2_{a}}$, $\bm{3_a}$ with $a=0,1$. The remain four-dimensional  irreducible representation of the EFG $Q_{8}\rtimes S_{3}$ can be constructed by~\cite{semidireci_Reps}
\begin{equation}
\bm{4}=\bm{2_0}\otimes\bm{2^{0}},
\end{equation}
where the operation $\otimes$ denotes as Kronecker product. Then the representation matrices of the EFG generators $A$, $B$, $S$ and $T$ take the following form
\begin{eqnarray}
\nonumber &&\hskip-0.3in \rho_{\bm{4}}(A)=\left(\begin{array}{cc}
i\mathbb{1}_{2} &~ \mathbb{0}_{2} \\
\mathbb{0}_{2} &~ -i\mathbb{1}_{2} \\
\end{array}\right), \qquad \rho_{\bm{4}}(B)=\left(\begin{array}{cc}
\mathbb{0}_{2} &~ i\mathbb{1}_{2} \\
i\mathbb{1}_{2} &~ \mathbb{0}_{2} \\
\end{array}\right)\,,\\
\label{eq:EFG_Q8S3_4D}&&\hskip-0.3in \rho_{\bm{4}}(S)=\frac{1}{2}     \left(\begin{array}{cccc}
0& 0 & \xi & \sqrt{3}\xi \\
0& 0 & \sqrt{3}\xi & -\xi \\
- \xi^3 & -\sqrt{3}\xi^3 & 0 & 0 \\
-\sqrt{3}\xi^3 & \xi^3 & 0 & 0
\end{array}\right), \quad \rho_{\bm{4}}(T)=\frac{1}{\sqrt{2}}\left(
\begin{array}{cccc}
1 & 0 & 1 & 0 \\
0 & -1 & 0 & -1 \\
1 & 0 & -1 & 0 \\
0 & -1 & 0 & 1 \\
\end{array}
\right)\,,~~~~
\end{eqnarray}
which can be easily obtained from the representation matrices of the representations $\bm{2_0}$ and $\bm{2^{0}}$ of $Q_{8}\rtimes S_{3}$. Now all eight irreducible representations of $Q_{8}\rtimes S_{3}$ have been obtained and they are summarized in table~\ref{tab:EFG_Q8S3_Reps}. In our working basis, all Clebsch-Gordan (CG)  coefficients of the EFG $Q_{8}\rtimes S_{3}$ are real and are shown in  table~\ref{tab:2O_CG-1st}.

\begin{table}[t!]
\begin{center}
\renewcommand{\tabcolsep}{0.8mm}
\renewcommand{\arraystretch}{1.1}
\begin{tabular}{|c|c|c|c|c|c|}\hline\hline
~~  &  $A$  &   $B$  &  $S$  &   $T$ \\ \hline

$\bm{1_{a}}$ & $1$   &  $1$ & $(-1)^{a}$   &  $(-1)^{a}$    \\ \hline

$\bm{2^{0}}$ &
$\mathbb{1}_{2}$ & $\mathbb{1}_{2}$
&  $\rho_{\bm{2^{\prime}}}(S)$ in Eq.~\eqref{eq:S3_irre}
&  $\rho_{\bm{2^{\prime}}}(T)$ in Eq.~\eqref{eq:S3_irre} \\ \hline

$\bm{2_{a}}$ &  $\rho_{\bm{2}}(A) $ in Eq.~\eqref{eq:Q8_irre} & $\rho_{\bm{2}}(B) $ in Eq.~\eqref{eq:Q8_irre}
&  $\rho_{\bm{2_{a}}}(S)$ in Eq.~\eqref{eq:EFG_Q8S3_2D}
&  $\rho_{\bm{2_{a}}}(T)$ in Eq.~\eqref{eq:EFG_Q8S3_2D} \\ \hline

$\bm{3_{a}}$ & $\rho_{\bm{3}}(A)$ in Eq.~\eqref{eq:Q8_3D_Reps}      & $\rho_{\bm{3}}(B)$ in Eq.~\eqref{eq:Q8_3D_Reps}
& $\rho_{\bm{3_{a}}}(S)$ in Eq.~\eqref{eq:EFG_Q8S3_3D} & $\rho_{\bm{3_{a}}}(T)$ in Eq.~\eqref{eq:EFG_Q8S3_3D} \\ \hline

$\bm{4}=\bm{2_{0}}\otimes\bm{2^{0}}$ & $\rho_{\bm{4}}(A)$ in Eq.~\eqref{eq:EFG_Q8S3_4D}
& $\rho_{\bm{4}}(B)$ in Eq.~\eqref{eq:EFG_Q8S3_4D}
&  $\rho_{\bm{4}}(S)$ in Eq.~\eqref{eq:EFG_Q8S3_4D}
&  $\rho_{\bm{4}}(T)$ in Eq.~\eqref{eq:EFG_Q8S3_4D} \\ \hline\hline
\end{tabular}
\caption{\label{tab:EFG_Q8S3_Reps}The representation matrices of the generators $A$, $B$, $S$ and $T$ for all  irreducible representations of the EFG group  $Q_{8}\rtimes S_{3}\cong\text{GL}(2,3)\cong[48,29]$ in the chosen basis, where $a=0,1$.  }
\end{center}
\end{table}

In the eclectic approach, modular forms are holomorphic functions of the complex modulus $\tau$ which is invariant under the actions of traditional flavor transformations. It implies that the transformation properties of modular forms of level 2 and weight $k$ under the action of $g\gamma$ in the eclectic approach are the same as $\gamma$ in modular invariant approach, where $g\gamma$  represents any element of EFG  with $g\in Q_{8}$ and $\gamma \in S_{3}$. In the modular invariant approach, two linearly independent modular forms of level 2 and weight 2 can be assigned into a doublet $\bm{2^{\prime}}$ of $S_{3}$~\cite{Kobayashi:2018vbk}. Hence the level 2 and weight 2 modular forms can be assigned into a doublet $\bm{2^{0}}$ of the EFG $Q_{8}\rtimes S_{3}$, i.e.
\begin{equation}
Y^{(2)}_{\bm{2^{0}}}(\tau)\equiv\left(\begin{array}{c}
Y_1(\tau) \\ Y_2(\tau)
\end{array}\right)\,,
\end{equation}
where the modular forms $Y_{1}(\tau)$ and $Y_{2}(\tau)$ take the following form~\cite{Kobayashi:2018vbk}
\begin{eqnarray}
Y_1(\tau) &=&\frac{i}{4\pi}\left[ \frac{\eta'(\tau/2)}{\eta(\tau/2)}  +\frac{\eta'((\tau +1)/2)}{\eta((\tau+1)/2)}
- 8\frac{\eta'(2\tau)}{\eta(2\tau)}  \right] ,\nonumber \\
\label{eq:MF_level2_weight2}    Y_2(\tau) &=& \frac{\sqrt{3}i}{4\pi}\left[ \frac{\eta'(\tau/2)}{\eta(\tau/2)}  -\frac{\eta'((\tau +1)/2)}{\eta((\tau+1)/2)}   \right]\,,
\end{eqnarray}
with the Dedekind function $\eta(\tau)$ being
\begin{equation}
\eta(\tau)=q^{1/24}\prod_{n=1}^\infty \left(1-q^n \right),\qquad
q\equiv e^{2 \pi i\tau}\,.
\end{equation}
Then the modular forms $Y_{1}(\tau)$ and $Y_{2}(\tau)$ have the following q-expressions
\begin{eqnarray}
\nonumber \hskip-0.3in Y_1(\tau) &=& 1/8+3 q+3 q^2+12 q^3+3 q^4+18 q^5+12 q^6+24 q^7+3 q^8+39 q^9+18 q^{10} \cdots , \\
\hskip-0.3in Y_2(\tau) &=& \sqrt 3 q^{1/2} (1+4 q+6 q^2+8 q^3+13 q^4+12 q^5+14 q^6+24 q^7+18 q^8+20 q^9 \cdots  ).
\end{eqnarray}
The explicit expression of the modular doublet $Y^{(2)}_{\bm{2^{0}}}(\tau)$ can also obtained from Ref.~\cite{Li:2021buv}, and it corresponds to the level 6 and weight 2 modular multiplet $Y^{(2)}_{\bm{2_{0}}}$ which transforms as $\bm{2_{0}}$ under $\Gamma^{\prime}_{6}$. The expressions of the linearly independent higher weight modular multiplets can be obtained from the tensor products of lower weight modular multiplets. The explicit expressions of the modular multiplets of level $N=2$ up to weight 8 are
\begin{eqnarray}\label{eq:Yw2to8}
\nonumber &&\hskip-0.3in
Y^{(4)}_{\bm{1_{0}}}
=\left(Y^{(2)}_{\bm{2^{0}}}Y^{(2)}_{\bm{2^{0}}}\right)_{\bm{1_{0}}}=Y_{1}^2+Y_{2}^2\,, \quad
Y^{(4)}_{\bm{2^{0}}}
=\left(Y^{(2)}_{\bm{2^{0}}}Y^{(2)}_{\bm{2^{0}}}\right)_{\bm{2^{0}}}
=\left(\begin{array}{c}Y_{2}^2-Y_{1}^2\\ 2 Y_{1} Y_{2}\end{array}\right)\,, \\
\nonumber &&\hskip-0.3in Y^{(6)}_{\bm{1_{0}}}
=\left(Y^{(2)}_{\bm{2^{0}}}Y^{(4)}_{\bm{2^{0}}}\right)_{\bm{1_{0}}} =3 Y_{1} Y_{2}^2-Y_{1}^3, \quad
Y^{(6)}_{\bm{1_{1}}}
=\left(Y^{(2)}_{\bm{2^{0}}}Y^{(4)}_{\bm{2^{0}}}\right)_{\bm{1_{1}}}=3 Y_{1}^2 Y_{2}-Y_{2}^3 \,,\\
\nonumber &&\hskip-0.3in Y^{(6)}_{\bm{2^{0}}}
=\left(Y^{(2)}_{\bm{2^{0}}}Y^{(4)}_{\bm{1_{0}}}\right)_{\bm{2^{0}}}=\left(\begin{array}{c}Y_{1} \left(Y_{1}^2+Y_{2}^2\right) \\ Y_{2} \left(Y_{1}^2+Y_{2}^2\right)\end{array}\right) \,, \quad
Y^{(8)}_{\bm{1_{0}}}=\left(Y^{(4)}_{\bm{1_{0}}}Y^{(4)}_{\bm{1_{0}}}\right)_{\bm{1_{0}}}=\left(Y_{1}^2+Y_{2}^2\right)^2\,, \\
&&\hskip-0.3in Y^{(8)}_{\bm{2a}}=\left(Y^{(4)}_{\bm{1_{0}}}Y^{(4)}_{\bm{2^{0}}}\right)_{\bm{2^{0}}}=\left(\begin{array}{c}Y_{2}^4-Y_{1}^4 \\ 2 Y_{1} Y_{2} \left(Y_{1}^2+Y_{2}^2\right)\end{array}\right),~~
Y^{(8)}_{\bm{2b}}=\left(Y^{(4)}_{\bm{2^{0}}}Y^{(4)}_{\bm{2^{0}}}\right)_{\bm{2^{0}}}=\left(\begin{array}{c}6 Y_{1}^2 Y_{2}^2-Y_{1}^4-Y_{2}^4 \\ 4 Y_{1} Y_{2} \left(Y_{2}^2-Y_{1}^2\right)\end{array}\right) \,,~~~~~~~~
\end{eqnarray}
which are useful in model building.

\subsection{\label{sec:EFG_gCP}Generalized CP consistent with EFG }

In this section, we shall derive the most general form of the gCP transformation  consistent with EFG.  Correspondingly, a new generator $\mathcal{CP}$ corresponding to gCP transformation should be introduced. Under the action of $\mathcal{CP}$, the modulus $\tau$ and chiral superfield $\psi(x)$ with modular weight $k_{\psi}$ transform as~\cite{Baur:2019kwi,Baur:2019iai,Dent:2001cc,Giedt:2002ns,Novichkov:2019sqv}
\begin{equation}
 \tau \stackrel{\mathcal{CP}}{\longrightarrow}-\bar{\tau} ,\qquad \psi(x) \stackrel{\mathcal{CP}}{\longrightarrow}X_{\bm{r}}\bar{\psi}(x_{\mathcal{P}})\,,
\end{equation}
where $X_{\bm{r}}$ is the gCP transformation matrix and $x_{\mathcal{P}}=(t,-\bm{x})$. As the mathematical structure of EFG is a semi-direct product  of the traditional flavor group $G_{f}$ and finite modular group $\Gamma_{N}$ ($\Gamma^{\prime}_{N}$), then each element of EFG can be written as $g\gamma$ with $g\in G_{f}$ and $\gamma \in SL(2,\mathbb{Z})$. The consistency of the gCP symmetry and the EFG group requires the following consistency condition has to be satisfied for both modulus and matter fields
\begin{equation}\label{eq:CC_chain}
\mathcal{CP}\rightarrow g\gamma \rightarrow \mathcal{CP}^{-1}=g^{\prime} \gamma^{\prime}\equiv u_{\mathcal{CP}}(g\gamma), \qquad g^{\prime}\in G_{f}, \quad \gamma^{\prime}\in SL(2,\mathbb{Z})\,,
\end{equation}
or equivalently
\begin{equation}
\label{eq:CC-chain-fur} \mathcal{CP}\rightarrow g \rightarrow \mathcal{CP}^{-1}\equiv u_{\mathcal{CP}}(g),  \qquad \mathcal{CP}\rightarrow \gamma \rightarrow \mathcal{CP}^{-1}\equiv u_{\mathcal{CP}}(\gamma)\,,
\end{equation}
The gCP symmetry is physically well-defined if and only $u_{\mathcal{CP}}$ is a class-inverting automorphism of the EFG~\cite{Chen:2014tpa}. As a consequence, $u_{\mathcal{CP}}(g)$ and $u_{\mathcal{CP}}(\gamma)$ should be in the same conjugacy class as $g^{-1}$ and $\gamma^{-1}$ respectively. It is sufficient to focus the traditional flavor symmetry $G_{f}$ and the modular generators $S$ and $T$. Under the action of consistency condition chain of Eq.~\eqref{eq:CC-chain-fur}, the modulus $\tau$ transforms as
\begin{eqnarray}
\nonumber &&    \tau  \stackrel{\mathcal{CP}}{\longrightarrow}-\bar{\tau}\stackrel{g}{\longrightarrow} -\bar{\tau}\stackrel{\mathcal{CP}^{-1}}{\longrightarrow} \tau=g^{\prime}\tau\equiv u_{\mathcal{CP}}(g)\tau\,, \\
\nonumber &&    \tau  \stackrel{\mathcal{CP}}{\longrightarrow}-\bar{\tau}\stackrel{S}{\longrightarrow} \frac{1}{\bar{\tau}}\stackrel{\mathcal{CP}^{-1}}{\longrightarrow} -\frac{1}{\tau}=g_{1}S^{-1}\tau\equiv u_{\mathcal{CP}}(S)\tau\,, \\
&&\tau\stackrel{\mathcal{CP}}{\longrightarrow}-\bar{\tau}\stackrel{T}{\longrightarrow} -\bar{\tau}-1\stackrel{\mathcal{CP}^{-1}}{\longrightarrow} \tau-1=g_{2}T^{-1}\tau\equiv u_{\mathcal{CP}}(T)\tau\,, \qquad g^{\prime},g_{1},g_{2}\in G_{f}\,,
\end{eqnarray}
where  we have used the property that the modulus $\tau$ is invariant under the action of any traditional flavor transformation. This implies $u_{\mathcal{CP}}(S)=g_{1}S^{-1}$ and $u_{\mathcal{CP}}(T)=g_{2}T^{-1}$. Notice that both $g_1$ and $g_2$ are identity element in the theory of modular symmetry alone. Then let us apply the consistency condition chain in Eq.~\eqref{eq:CC-chain-fur} to a chiral superfield $\psi(x)$ transforming in an irreducible representation $\bm{r}$ of the EFG $G_{f}\rtimes \Gamma_{N}$ (or $G_{f}\rtimes \Gamma^{\prime}_{N}$),
\begin{eqnarray}
\nonumber  &&\hskip-0.3in  \psi(x)  \xrightarrow{\mathcal{CP}\circ g\circ \mathcal{CP}^{-1}} X_{\bm{r}}\rho^{*}_{\bm{r}}(g)X_{\bm{r}}^{-1}\psi(x)=\rho_{\bm{r}}(g^{\prime})\psi(x)\,, \qquad \\
\nonumber  && \hskip-0.3in  \psi(x) \xrightarrow{\mathcal{CP}\circ S\circ \mathcal{CP}^{-1}}(\tau)^{-k_{\psi}} X_{\bm{r}}\rho^{*}_{\bm{r}}(S)X_{\bm{r}}^{-1}\psi(x)=(\tau)^{-k_{\psi}} \rho_{\bm{r}}(g_{1}S^{-1})\psi(x)\,, \\
&&\hskip-0.3in \psi(x)  \xrightarrow{\mathcal{CP}\circ T\circ \mathcal{CP}^{-1}} X_{\bm{r}}\rho^{*}_{\bm{r}}(T)X_{\bm{r}}^{-1}\psi(x)= \rho_{\bm{r}}(g_{2}T^{-1})\psi(x)\,.
\end{eqnarray}
Consequently, the gCP transformation $X_{\bm{r}}$ has to satisfy the following equations
\begin{eqnarray}
\nonumber &&X_{\bm{r}}\rho^*_{\bm{r}}(g)X^{-1}_{\bm{r}}
=\rho_{\bm{r}}(g^{\prime})\equiv\rho_{\bm{r}}(u_{\mathcal{CP}}(g)), \\
\nonumber &&X_{\bm{r}}\rho^{*}_{\bm{r}}(S)X_{\bm{r}}^{-1}
=\rho_{\bm{r}}(g_{1}S^{-1})\equiv\rho_{\bm{r}}(u_{\mathcal{CP}}(S)), \\
&&X_{\bm{r}}\rho^{*}_{\bm{r}}(T)X_{\bm{r}}^{-1}
=\rho_{\bm{r}}(g_{2}T^{-1})\equiv\rho_{\bm{r}}(u_{\mathcal{CP}}(T))\,.
\end{eqnarray}
The consistency conditions of  $u_{\mathcal{CP}}(S)=S^{-1}$ and $u_{\mathcal{CP}}(T)=T^{-1}$ are imposed in Ref~\cite{Nilles:2020nnc},  while this is true if there is only modular symmetry. However the presence of traditional flavor symmetry in EFG makes the consistency conditions to become more general  $u_{\mathcal{CP}}(S)=g_1S^{-1}$ and $u_{\mathcal{CP}}(T)=g_2T^{-1}$.
It was shown that the physically well-defined CP transformations require that $u_{\mathcal{CP}}$ must be class-inverting automorphisms of EFG $G_{f}\rtimes \Gamma_{N}$ (or $G_{f}\rtimes \Gamma^{\prime}_{N}$)~\cite{Chen:2014tpa}. Therefore, the elements $g^{\prime}$, $g_{1}S^{-1}$ and $g_{2}T^{-1}$ must belong to the classes of the inverse elements of $g$, $S$ and $T$, respectively.
In the present work, the gCP transformation $\mathcal{CP}$ is required to be order 2. It implies that
\begin{equation}
 \mathcal{CP}^2=1, \qquad  X_{\bm{r}}X^{*}_{\bm{r}}=\mathbb{1}_{\bm{r}}, \qquad u^2_{\mathcal{CP}}=\mathbb{1}\,.
\end{equation}
As a consequence, the automorphisms $u_{\mathcal{CP}}$, $u_S$ and $u_T$ should satisfy the following relations
\begin{equation}
u_{\mathcal{CP}}\circ u_S\circ u_{\mathcal{CP}}=\mu_{g_{1}}\circ u^{-1}_S, \qquad u_{\mathcal{CP}}\circ u_T\circ u_{\mathcal{CP}}=\mu_{g_{2}}\circ u^{-1}_T\,,
\end{equation}
where $\mu_{h}$ represents an inner automorphism $\mu_{h}~:~g\rightarrow hgh^{-1}$  with $g,h\in G_{f}$.

In the terminology of Ref.~\cite{Chen:2014tpa}, we find that the EFG $Q_{8}\rtimes S_{3}\cong GL(2,3)$ is a group of type II A. For this kind of groups, there exists a basis in which all CG coefficients are real, and there also exists an automorphism $u_{\mathcal{CP}}$ which can be used to define a proper CP transformation. From  table~\ref{tab:2O_CG-1st}, we see that all the CG coefficients of $Q_{8}\rtimes S_{3}$ are already real in our working basis.  As the physical CP transformation should be connected to a class-inverting automorphisms of $Q_{8}\rtimes S_{3}$, the class-inverting automorphisms of $Q_{8}\rtimes S_{3}$ and its actions on the generators $A$, $B$, $S$ and $T$ can be taken to be\footnote{In  Ref~\cite{Nilles:2020nnc}, the authors claimed that gCP can not be defined because the consistency conditions $u_{\mathcal{CP}}(S)=S^{-1}$ and $u_{\mathcal{CP}}(T)=T^{-1}$ are required.}
\begin{equation}
u_{\mathcal{CP}}(A)=A^{3}, \qquad u_{\mathcal{CP}}(B)=B^{3}, \qquad u_{\mathcal{CP}}(S)=A^{3}S, \qquad u_{\mathcal{CP}}(T)=T\,.
\end{equation}
One can check that $u_{\mathcal{CP}}$ is a class-inverting automorphisms of $Q_{8}\rtimes S_{3}$ from the conjugacy classes given in Eq.~\eqref{eq:Q8S3_CC}. The group ID of the EFG group including gCP is $[96,193]$. The explicit form of gCP transformation matrix $X_{\bm{r}}$ is determined by the following consistency equations~\cite{Novichkov:2019sqv,Ding:2021iqp}
\begin{eqnarray}
\nonumber &&X_{\bm{r}}\rho^*_{\bm{r}}(A)X^{-1}_{\bm{r}}
=\rho_{\bm{r}}(A^{3}), \qquad X_{\bm{r}}\rho^*_{\bm{r}}(B)X^{-1}_{\bm{r}}
=\rho_{\bm{r}}(B^{3})\,,\\
\label{eq:con_uK} &&X_{\bm{r}}\rho^{*}_{\bm{r}}(S)X_{\bm{r}}^{-1}
=\rho_{\bm{r}}(A^{3}S),\qquad X_{\bm{r}}\rho^{*}_{\bm{r}}(T)X_{\bm{r}}^{-1}
=\rho_{\bm{r}}(T)\,,
\end{eqnarray}
where $\bm{r}$ denotes any irreducible representation of the EFG $Q_{8}\rtimes S_{3}\cong GL(2,3)$.  As shown in table~\ref{tab:EFG_Q8S3_Reps}, we find the following relations for all irreducible representations of $Q_{8}\rtimes S_{3}$ are fulfilled in our working basis,
\begin{equation}
\rho^*_{\bm{r}}(A)=\rho_{\bm{r}}(A^{3}), \quad \rho^*_{\bm{r}}(B)=\rho_{\bm{r}}(B^{3}), \quad
\rho^*_{\bm{r}}(S)=\rho_{\bm{r}}(A^{3}S), \quad
\rho^*_{\bm{r}}(T)=\rho_{\bm{r}}(T)\,,
\end{equation}
which implies that the gCP transformation matrix $X_{\bm{r}}$ is fixed
to be unity matrix up to an overall phase, i.e.
\begin{equation}
X_{\bm{r}}=\mathbb{1}_{\bm{r}}\,.
\end{equation}
As a consequence, the gCP constrains the couplings constants in the EFG $Q_{8}\rtimes S_{3}$ models to be real.

\section{\label{sec:mod_general}K\"ahler potential and superpotential under $Q_{8}\rtimes S_{3}$  }

From the discussion of section~\ref{sec:EFG}, we have obtained all the nonequivalent irreducible representations of the EFG $Q_{8}\rtimes S_{3}$, and the results are summarized in table~\ref{tab:EFG_Q8S3_Reps}. Furthermore, the physical CP transformations compatible with the EFG $Q_{8}\rtimes S_{3}$ are found. Now we shall perform a comprehensive analysis on the K\"ahler potential $\mathcal{K}$ and the superpotential $\mathcal{W}$ invariant under the action of $Q_{8}\rtimes S_{3}$  in the framework of $\mathcal{N}=1$ global supersymmetry. In the present work, the Higgs fields $H_u$ and $H_d$ are assumed to be trivial singlet $\bm{1_{0}}$. The three generations matter fields (quarks and leptons) can be assigned to transform as triplet $\bm{3_{a}}$, a singlet $\bm{1_{a}}$ plus a doublet $\bm{2_{b}}$, a singlet $\bm{1_{a}}$ plus a doublet $\bm{2^{0}}$ or three singlets $\bm{1_{a}}$ with $a,b=0,1$, as shown in table~\ref{tab:EFG_Q8S3_Reps}. In the case that the three generations of matter fields are all assigned to singlets, the EFG would play less role.  Besides, if the matter fields are assigned to $\bm{1_{a}}\oplus \bm{2^{0}}$, the EFG would be reduced to the $S_3$ modular symmetry, so that the corresponding K\"ahler potential would be less constrained.  As a result, we only consider the remaining two kinds of assignments of matter fields.
Nevertheless, flavons can be assigned to all the irreducible representations of the EFG $Q_{8}\rtimes S_{3}$. In section~\ref{eq:EFG_Q8S3}, we have indicated that the modular forms of level 2 can be organized into modular multiplets which transform as  $\bm{\mathcal{R}}\in\{\bm{1_{0}}, \bm{1_{1}}, \bm{2^{0}}\}$, and this convention will be used throughout the paper.  To obtain the general forms of K\"ahler potential and superpotential for each kind of assignments of matter fields, we assume that the modular multiplets of level 2 and weight $k$ comprise all these three irreducible multiplets, i.e. we shall impose all the three modular multiplets $Y^{(k)}_{\bm{1_{0}}}(\tau)$, $Y^{(k)}_{\bm{1_{1}}}(\tau)$ and $Y^{(k)}_{\bm{2^{0}}}(\tau)$ for each modular weight $k$. If some modular multiplets are absent for a given weight, the corresponding terms must be set to zero. Notice that the contributions of the linearly independent modular form multiplets in the same representation take a similar form. In the following, we will give the general forms of K\"ahler potential and superpotential for all possible assignments of fields one by one.

\subsection{\label{sec:Kahlerpotential} K\"ahler potential}

As we know, the EFG approach provides a way to control the K\"ahler potential~\cite{Nilles:2020nnc,Nilles:2020kgo}, in which the higher order corrections of the K\"ahler potential are suppressed by powers of $\langle\Phi\rangle/\Lambda$, where $\langle\Phi\rangle$ refers to the VEV of flavon $\Phi$ which is necessary to broken traditional flavor symmetry in the EFG approach. The flavon $\Phi$ is assumed to be singlet of standard model in the following. In this section, we shall focus on the K\"ahler potential $\mathcal{K}$ for matter fields $\psi$ and construct the most general $\mathcal{K}$ consistent with the EFG $Q_{8}\rtimes S_{3}$. As explained above, the three generations of matter fields $\psi$ can only be assigned to transform as reducible triplet $\bm{1_{a}}\oplus \bm{2_{b}}$  or irreducible triplet $\bm{3_{a}}$.

\subsubsection{\label{subsubsec:kahler1} K\"ahler potential for $\psi\sim \mathbf{1_{a}}\oplus\mathbf{2_{b}}$}

We first consider the case that the three generations of matter fields $\psi$ are assigned to transform as reducible triplet $\bm{1_{a}}\oplus \bm{2_{b}}$.
Without loss of generality, we assume that the first generation is a singlet $\psi_{1}\sim\bm{1_{a}}$ and the other two generations form a doublet $\psi_{d}=(\psi_{2},\psi_{3})^T\sim\bm{2_{b}}$. The modular weights of $\psi_{1}$ and $\psi_{d}$ are defined as $k_{\psi_{1}}$ and $k_{\psi_{d}}$, respectively.  At leading order, the most general K\"ahler potential for the matter fields $\psi$ takes the following form
\begin{equation}\label{eq:kahler_LO1}
 \mathcal{K}_{\rm LO} = \sum_{k,\bm{r_1},\bm{r_2},s} (-i \tau+ i \bar \tau)^{k-k_{\psi_1}}
\left( Y^{(k)\dagger}_{\bm{r_1}} Y^{(k)}_{\bm{r_2}}  \psi_{1}^{\dagger} \psi_{1}\right)_{\bm{1_{0}},s}+(-i \tau+ i \bar \tau)^{k-k_{\psi_d}}
\left( Y^{(k)\dagger}_{\bm{r_1}} Y^{(k)}_{\bm{r_2}}  \psi_{d}^{\dagger} \psi_{d}\right)_{\bm{1_{0}},s}\,,
\end{equation}
where we have dropped the coupling constants of each independent term, and  we have to sum over the even modular weights $k\in2\mathbbm{N}$, all linearly independent modular multiplets $Y^{(k)}_{\bm{r}}$ of weight $k$ and all $Q_{8}\rtimes S_{3}$ singlet contractions labeled by the index $s$\footnote{The modular form of weight 0 is taken to be $Y^{(0)}_{\bm{r}}=1$ and the minimal K\"ahler potential can be obtained from the terms of $k=0$.}.  From the Kronecker products of two irreducible representations of $Q_{8}\rtimes S_{3}$  given in Eq.~\eqref{eq:Q8S3_KP} and the transformation properties of modular form multiplets, we find that the contractions of any two modular form multiplets can be written as  $\left(Y^{(k_{1})\dagger}_{\bm{r_1}} Y^{(k_{2})}_{\bm{r_2}}\right)_{\bm{\mathcal{R}}}$. In order to obtain $\mathcal{K}_{\rm LO}$, the contractions $\psi^{\dagger}_{1}\psi_{1}$ and $\psi^{\dagger}_{d}\psi_{d}$ must transform as $\bm{\mathcal{R}}$. The CG coefficients of the  tensor product $\bm{2_{b}}\otimes\bm{\bar{2}_{b}} = \bm{1_{0}}\oplus \bm{3_{1}}$ can be obtained from those of $\bm{2_{b}}\otimes \bm{2_{[b+1]}}$  and Eq.~\eqref{eq:Reps_conju}, where the notation $[b+1]$ is defined as $b+1$ modulo 2, and we adopt this convention in the whole article. Then we find the contractions $\psi^{\dagger}_{1}\psi_{1}$ and $\psi^{\dagger}_{d}\psi_{d}$ as follows
\begin{equation}\label{eq:psi1psid_contra}
\left(\psi^{\dagger}_{1}\psi_{1}\right)_{\bm{1_{0}}}=\psi^{\dagger}_1\psi_1\,, \quad \left(\psi^{\dagger}_{d}\psi_{d}\right)_{\bm{1_{0}}}=\psi^{\dagger}_2\psi_2+\psi^{\dagger}_3\psi_3\,,
\quad  \left(\psi^{\dagger}_{d}\psi_{d}\right)_{\bm{3_{1}}}=\left(\begin{array}{c}
\psi^{\dagger}_2\psi_2 -\psi^{\dagger}_3\psi _3 \\
-\psi^{\dagger}_2\psi_3-\psi^{\dagger}_3\psi_2 \\
\psi^{\dagger}_2\psi_3-\psi^{\dagger}_3\psi_2
\end{array}\right)\,,
\end{equation}
where the last contraction is not a term which transforms as $\bm{\mathcal{R}}$, thus it does not lead to a term of LO K\"ahler potential invariant under the EFG $Q_{8}\rtimes S_{3}$. Then one can easily obtain the LO K\"ahler potential as follows
\begin{equation}\label{eq:kahler_LO1fin}
\mathcal{K}_{\rm LO}=g_{1}(\tau,\bar{\tau})\psi_{1}^{\dagger} \psi_{1}+g_{2}(\tau,\bar{\tau})\left(\psi_{2}^{\dagger} \psi_{2}+\psi_{3}^{\dagger} \psi_{3}\right)\,,
\end{equation}
where the functions $g_{1}(\tau,\bar{\tau})$ and $g_{2}(\tau,\bar{\tau})$ are defined as
\begin{eqnarray}
\nonumber && g_{1}(\tau,\bar{\tau})\equiv \sum_{k,\bm{r_1},\bm{r_2}}(-i \tau+ i \bar \tau)^{k-k_{\psi_1}}\left( Y^{(k)\dagger}_{\bm{r_1}} Y^{(k)}_{\bm{r_2}} \right)_{\bm{1_{0}}}\,, \\
&& g_{2}(\tau,\bar{\tau})\equiv \sum_{k,\bm{r_1},\bm{r_2}}(-i \tau+ i \bar \tau)^{k-k_{\psi_d}}\left( Y^{(k)\dagger}_{\bm{r_1}} Y^{(k)}_{\bm{r_2}} \right)_{\bm{1_{0}}} \,,
\end{eqnarray}
The resulting K\"ahler metric in Eq.~\eqref{eq:kahler_LO1fin} is diagonal $\text{diag}\left(g_{1}(\tau,\bar{\tau}),g_{2}(\tau,\bar{\tau}),g_{2}(\tau,\bar{\tau})\right)$. In order to get canonical kinetic terms, we have to rescale the matter fields $\psi_{1}$ and $\psi_{d}$. One can check that the effect of rescaling on fermion masses can be absorbed into the parameters of the superpotential. As a consequence, no  corrections from LO K\"ahler potential to fermion masses and flavor mixing parameters will be introduced.

Now let us consider the next-to-leading-order (NLO) corrections to the K\"ahler potential. In comparison with LO K\"ahler potential, NLO terms are suppressed by $\langle\Phi\rangle/\Lambda$ and the general form of them can be written as
\begin{eqnarray}
\nonumber \hskip-0.3in \mathcal{K}_{\rm NLO} &=&\frac{1}{\Lambda} \sum_{k,\bm{r_1},\bm{r_2},\bm{\mathcal{R}}} (-i \tau+ i \bar \tau)^{k-k_{\psi_{1}}}
\left[  \left( Y^{(k)\dagger}_{\bm{r_1}} Y^{(k+k_{\Phi_{1}})}_{\bm{r_2}} \right)_{\bm{\mathcal{R}}}\left( \psi^{\dagger}_{1} \psi_{1}\Phi_{1}\right)_{\bm{\mathcal{R}}}\right]_{\bm{1_{0}}} \\
\nonumber &&\hskip-0.1in+\frac{1}{\Lambda} \sum_{k,\bm{r_1},\bm{r_2},\bm{\mathcal{R}}} (-i \tau+ i \bar \tau)^{k-k_{\psi_{d}}}
\left[  \left( Y^{(k)\dagger}_{\bm{r_1}} Y^{(k+k_{\Phi_{2}})}_{\bm{r_2}} \right)_{\bm{\mathcal{R}}}\left( \psi^{\dagger}_{d} \psi_{d}\Phi_{2}\right)_{\bm{\mathcal{R}}}\right]_{\bm{1_{0}}}\\
\label{eq:kahler_NLO1}  &&\hskip-0.1in+\frac{1}{\Lambda} \sum_{k,\bm{r_1},\bm{r_2},\bm{\mathcal{R}}} (-i \tau+ i \bar \tau)^{k-k_{\psi_{1}}}
\left[  \left( Y^{(k)\dagger}_{\bm{r_1}} Y^{(k-k_{\psi_{1}}+k_{\psi_{d}}+k_{\Phi_{3}})}_{\bm{r_2}} \right)_{\bm{\mathcal{R}}}\left( \psi^{\dagger}_{1} \psi_{d}\Phi_{3}\right)_{\bm{\mathcal{R}}}\right]_{\bm{1_{0}}}+\text{h.c.}\,,
\end{eqnarray}
where we have imposed three flavons $\Phi_{i}$ ($i=1,2,3$) with modular weights $k_{\Phi_{i}}$, and some of them may be the same flavon in a model. If some of them are absent in a certain model, the corresponding terms are vanishing. Comparing the expressions of $\mathcal{K}_{\rm LO}$ in Eq.~\eqref{eq:kahler_LO1} and $\mathcal{K}_{\rm NLO}$ in Eq.~\eqref{eq:kahler_NLO1}, we find that the flavons $\Phi_{1}$ and $\Phi_{2}$ can contribute to $\mathcal{K}_{\rm NLO}$ if and only if they are  invariant under all the auxiliary symmetry groups in a model, and the flavon $\Phi_{3}$ can contribute to $\mathcal{K}_{\rm NLO}$ if and only if the charge of $\psi_{1}$ equal to the sum of the charges of $\psi_{d}$ and $\Phi_{3}$ under the auxiliary symmetry groups. In Eq.~\eqref{eq:kahler_NLO1} we have used the fact that $Y^{(k_{1})\dagger}_{\bm{r_1}} Y^{(k_{2})}_{\bm{r_2}}$ transform as $\bm{\mathcal{R}}$, so that the contractions of $\psi^{\dagger}_{1}\psi_{1}\Phi_{1}$, $\psi^{\dagger}_{d}\psi_{d}\Phi_{2}$ and $\psi^{\dagger}_{1}\psi_{d}\Phi_{3}$ should also transform as $\bm{\mathcal{R}}$ to obtain invariant contractions of $Q_{8}\rtimes S_{3}$. The transformation properties of the contractions $\psi^{\dagger}_{1}\psi_{1}$ and $\psi^{\dagger}_{d}\psi_{d}$ are given in Eq.~\eqref{eq:psi1psid_contra} and the contraction  $\psi^{\dagger}_{1}\psi_{d}$ is given by
\begin{equation}
\left(\psi^{\dagger}_{1}\psi_{d}\right)_{\bm{2_{[a+b]}}}=\psi^{\dagger}_1\left(\begin{array}{c}
\psi_2 \\
\psi_3
\end{array}\right)\,.
\end{equation}
 As a consequence, the flavon $\Phi_{1}$ must transform as $\bm{1_{c}}$, the flavon $\Phi_{2}$ must transform as $\bm{1_{c}}$ or $\bm{3_{c}}$, and the flavon $\Phi_{3}$ must transform as $\bm{2_{c}}$ or $\bm{4}$ under $Q_{8}\rtimes S_{3}$, where $c=0,1$. We find that the contraction result of the first line in Eq.~\eqref{eq:kahler_NLO1} is proportional to $\psi^{\dagger}_1\psi_1$, and it can be absorbed into the first term of LO K\"ahler potential in Eq.~\eqref{eq:kahler_LO1fin}. It implies that this term provides no correction to mixing parameters. The contraction result of the second line in Eq.~\eqref{eq:kahler_NLO1} can be written as
{\small\begin{equation}\label{eq:NLO_Phi2}
\sum_{\bm{\mathcal{R}}}\left[   \left( Y^{(k)\dagger}_{\bm{r_1}} Y^{(k+k_{\Phi_{2}})}_{\bm{r_2}} \right)_{\bm{\mathcal{R}}}\left( \psi^{\dagger}_{d} \psi_{d}\Phi_{2}\right)_{\bm{\mathcal{R}}}\right]_{\bm{1_{0}}}=\left\{\begin{array}{ll}\left( Y^{(k)\dagger}_{\bm{r_1}} Y^{(k+k_{\Phi_{2}})}_{\bm{r_2}} \right)_{\bm{1_{c}}}\left[\left(\psi^{\dagger}_{d} \psi_{d}\right)_{\bm{1_{0}}}\Phi_{2}\right]_{\bm{1_{c}}}, & \Phi_{2}\sim\bm{1_{c}}, \\ \\
\begin{array}{l}\left( Y^{(k)\dagger}_{\bm{r_1}} Y^{(k+k_{\Phi_{2}})}_{\bm{r_2}} \right)_{\bm{1_{[c+1]}}}\left[\left(\psi^{\dagger}_{d} \psi_{d}\right)_{\bm{3_{1}}}\Phi_{2}\right]_{\bm{1_{[c+1]}}} \\
+\left\{\left( Y^{(k)\dagger}_{\bm{r_1}} Y^{(k+k_{\Phi_{2}})}_{\bm{r_2}} \right)_{\bm{2^{0}}}\left[\left(\psi^{\dagger}_{d} \psi_{d}\right)_{\bm{3_{1}}}\Phi_{2}\right]_{\bm{2^{0}}}\right\}_{\bm{1_{0}}}\end{array}, &  \Phi_{2}\sim\bm{3_{c}},
\end{array}\right.
\end{equation}}
For singlet $\Phi_{2}$, the corresponding contraction is proportional to $\psi_{2}^{\dagger} \psi_{2}+\psi_{3}^{\dagger} \psi_{3}$, and it can be absorbed into the second term of LO K\"ahler potential in Eq.~\eqref{eq:kahler_LO1fin} and provides no contribution to fermion masses and flavor observables. If a triplet flavon $\Phi_{2}\sim \bm{3_{c}}$ which is chargeless under all auxiliary groups is introduced in a model, the explicit expression of Eq.~\eqref{eq:NLO_Phi2} can be obtained from the result of $\left(\psi^{\dagger}_{d} \psi_{d}\right)_{\bm{3_{1}}}$ in Eq.~\eqref{eq:psi1psid_contra} and the CG coefficients of $Q_{8}\rtimes S_{3}$ in table~\ref{tab:2O_CG-1st}. Then it is easy to check that the corresponding K\"ahler metric is not diagonal and the non-canonical corrections are suppressed by $\langle\Phi\rangle/\Lambda$ in comparison with $\mathcal{K}_{\rm LO}$. These terms  generally induce corrections to fermion masses and mixing parameters, but the corrections are suppressed by $\langle\Phi\rangle/\Lambda$. If a model involves a flavon $\Phi_{3}\sim\bm{2_{c}}$ or $\Phi_{3}\sim\bm{4}$ with auxiliary symmetry charge equal to the charge of $\psi_{1}$ minus the charge of $\psi_{d}$, the third line of Eq.~\eqref{eq:kahler_NLO1} is not vanishing. It always gives rise to off-diagonal elements of the K\"ahler metric and these terms yield corrections to the observable fermion masses and mixing parameters. The corresponding corrections are also suppressed by $\langle\Phi\rangle/\Lambda$.

If no doublet, triplet or quartet flavon meeting the above criteria is introduced in a model,  NLO K\"ahler potential yields no corrections to the observable fermion masses and mixing. Then the corrections to fermion masses and mixing parameters could arise from the next-to-next-to-leading order (NNLO) terms of the K\"ahler potential which takes the following form

\begin{eqnarray}
\nonumber \hskip-0.3in \mathcal{K}_{\rm NNLO} &=& \frac{1}{\Lambda^2}\sum_{k, \bm{r_1},\bm{r_2}, s} (-i \tau+ i \bar \tau)^{k-k_{\psi_{1}}-k_{\Theta}}\left( Y^{(k)\dagger}_{\bm{r_1}} Y^{(k+k_{\Phi}-k_{\Theta})}_{\bm{r_2}} \psi^{\dagger}_{1}\psi_{1}\Theta^{\dagger}\Phi\right)_{\bm{1_{0}},s}\\
\nonumber &&\hskip-0.2in+\frac{1}{\Lambda^2}\sum_{k, \bm{r_1},\bm{r_2}, s} (-i \tau+ i \bar \tau)^{k-k_{\psi_{d}}-k_{\Theta}}\left( Y^{(k)\dagger}_{\bm{r_1}} Y^{(k+k_{\Phi}-k_{\Theta})}_{\bm{r_2}} \psi^{\dagger}_{d}\psi_{d}\Theta^{\dagger}\Phi\right)_{\bm{1_{0}},s}\\
\label{eq:kahler_NNLO1}&&\hskip-0.2in+\frac{1}{\Lambda^2}\sum_{k, \bm{r_1},\bm{r_2}, s} (-i \tau+ i \bar \tau)^{k-k_{\psi_{1}}-k_{\Theta}}\left( Y^{(k)\dagger}_{\bm{r_1}} Y^{(k-k_{\psi_{1}}+k_{\psi_{d}}+k_{\Phi}-k_{\Theta})}_{\bm{r_2}} \psi^{\dagger}_{1}\psi_{d}\Theta^{\dagger}\Phi\right)_{\bm{1_{0}},s}+\text{h.c.}\,,
\end{eqnarray}
where the two generic flavons $\Phi$ and $\Theta$  could be identical, and the contribution from the first line can be absorbed by the LO K\"ahler potential. In general, the second line in Eq.~\eqref{eq:kahler_NNLO1} always contains the following contractions
\begin{eqnarray}
\nonumber \hskip-0.3in \sum_{s}\left( Y^{(k)\dagger}_{\bm{r_1}} Y^{(k)}_{\bm{r_2}} \psi^{\dagger}_{d}\psi_{d}\Phi^{\dagger}\Phi\right)_{\bm{1_{0}},s}&=&\sum_{a}\left( Y^{(k)\dagger}_{\bm{r_1}} Y^{(k)}_{\bm{r_2}} \right)_{\bm{1_{a}}}\left(\psi^{\dagger}_{d}\psi_{d}\right)_{\bm{1_{0}}}\left(\Phi^{\dagger}\Phi\right)_{\bm{1_{a}}} \\
\nonumber &&+\sum_{a}\left( Y^{(k)\dagger}_{\bm{r_1}} Y^{(k)}_{\bm{r_2}} \right)_{\bm{1_{[1+a]}}}\left[\left(\psi^{\dagger}_{d}\psi_{d}\right)_{\bm{3_{1}}}\left(\Phi^{\dagger}\Phi\right)_{\bm{3_{a}}}\right]_{\bm{1_{[a+1]}}} \\
&&+\sum_{a}\left\{\left( Y^{(k)\dagger}_{\bm{r_1}} Y^{(k)}_{\bm{r_2}} \right)_{\bm{2^{0}}}\left[\left(\psi^{\dagger}_{d}\psi_{d}\right)_{\bm{3_{1}}}\left(\Phi^{\dagger}\Phi\right)_{\bm{3_{a}}}\right]_{\bm{2^{0}}}\right\}_{\bm{1_{0}}}\,,
\end{eqnarray}
whose contributions is of the same form as the LO minimal K\"ahler when the flavon $\Phi$ transforms as $\bm{1_{0}}$, $\bm{1_{1}}$ or $\bm{2^{0}}$. On the other hand, if a flavon transforming as $\bm{2_{a}}$, $\bm{3_{a}}$ or $\bm{4}$ is introduced, the K\"ahler potential $\mathcal{K}_{\rm NNLO}$ generally yields deviations from canonical kinetic terms and leads to corrections to fermion masses and mixing parameters, while the corresponding corrections are suppressed by $\langle\Phi\rangle^2/\Lambda^2$ and consequently they should be negligible.

\subsubsection{\label{subsubsec:kahler2} K\"ahler potential for $\psi\sim \mathbf{3_{a}}$}

In this case, the LO K\"ahler potential can be written as
\begin{equation}\label{eq:kahler_LO2}
\mathcal{K}_{\rm LO} = \sum_{k,\bm{r_1},\bm{r_2},\bm{\mathcal{R}}} (-i \tau+ i \bar \tau)^{k-k_{\psi}} \left[\left( Y^{(k)\dagger}_{\bm{r_1}} Y^{(k)}_{\bm{r_2}}\right)_{\bm{\mathcal{R}}} \left( \psi^{\dagger} \psi\right)_{\bm{\mathcal{R}}} \right]_{\bm{1_{0}}}\,,
\end{equation}
where we have to sum over all possible representations $\bm{\mathcal{R}}$. From the tensor product $\bm{3_{a}}\otimes\bm{\bar{3}_{a}} = \bm{1_{0}}\oplus \bm{2^{0}}\oplus \bm{3_{0}}\oplus \bm{3_{1}}$, we find that the representations $\bm{\mathcal{R}}$ in Eq.~\eqref{eq:kahler_LO2} can be $ \bm{1_{0}}$ and $\bm{2^{0}}$. Then the contraction  $\left( \psi^{\dagger} \psi\right)_{\bm{\mathcal{R}}}$ gives rise to
\begin{equation}
\left(\psi^{\dagger}\psi\right)_{\bm{\mathcal{R}}}=\left\{\begin{array}{cc}
\psi^{\dagger}_1\psi_1+\psi^{\dagger}_2\psi_2+\psi^{\dagger}_3 \psi_{3}& \text{for} ~~ \bm{\mathcal{R}}=\bm{1_{0}}\,, \\
\left(\begin{array}{cc}
\psi^{\dagger}_1\psi_1+\psi^{\dagger}_2\psi_2-2\psi^{\dagger}_3\psi_3\\
\sqrt{3}\left(-\psi^{\dagger}_1\psi_1+\psi^{\dagger}_2\psi_2\right) \\
\end{array}\right) & \text{for}~~ \bm{\mathcal{R}}=\bm{2^{0}}\,.
\end{array}
\right.
\end{equation}
The LO  K\"ahler potential can be expanded as
\begin{eqnarray}
\nonumber \mathcal{K}_{\rm LO}&=&\psi^{\dagger}_1\psi_1\sum_{k}  (-i \tau+ i \bar \tau)^{k-k_{\psi}}\sum_{\bm{r_{1}},\bm{r_{2}}}\left\{
\left(Y^{(k)\dagger}_{\bm{r_{1}}} Y^{(k)}_{\bm{r_{2}}} \right)_{\bm{1_{0}}}+
y_{1}-\sqrt{3}y_{2}\right\} \\
\nonumber&&+ \psi^{\dagger}_2\psi_2\sum_{k}  (-i \tau+ i \bar \tau)^{k-k_{\psi}}\sum_{\bm{r_{1}},\bm{r_{2}}}\left\{
\left(Y^{(k)\dagger}_{\bm{r_{1}}} Y^{(k)}_{\bm{r_{2}}} \right)_{\bm{1_{0}}}+y_{1}+\sqrt{3}y_{2}\right\} \\
\label{eq:Kaehler_LO2fin}&&+\psi^{\dagger}_3\psi_3\sum_{k}  (-i \tau+ i \bar \tau)^{k-k_{\psi}}\sum_{\bm{r_{1}},\bm{r_{2}}}\left\{
\left(Y^{(k)\dagger}_{\bm{r_{1}}} Y^{(k)}_{\bm{r_{2}}} \right)_{\bm{1_{0}}}-2y_{1}\right\} \,,
\end{eqnarray}
where $y_{1}$ and $y_{2}$ are defined as
\begin{equation}
\left(\begin{array}{c}
y_{1}(\tau) \\ y_{2}(\tau)
\end{array}\right)\equiv \left( Y^{(k)\dagger}_{\bm{r_{1}}}(\tau) Y^{(k)}_{\bm{r_{2}}} (\tau)\right)_{\bm{2^{0}}}\,.
\end{equation}
In general, the coefficients of $\psi^{\dagger}_1\psi_1$, $\psi^{\dagger}_2\psi_2$ and $\psi^{\dagger}_3\psi_3$ are not equal to each other. Hence the resulting LO K\"ahler metric is diagonal while it is not proportional to unit matrix.   In order to get canonical kinetic terms, we have to rescale the matter fields $\psi_{1,2,3}$ separately. This rescaling leads to nontrivial effects on the fermion masses and flavor mixing parameters, and it might significantly modify the predictions of a model.

In short, if the three generations of  matter fields $\psi$ are assigned to  be a singlet $\bm{1_{a}}$ plus a doublet $\bm{2_{b}}$ of $Q_{8}\rtimes S_{3}$, the EFG  can efficiently restrict the K\"ahler potential, otherwise the predictive power of the  EFG $Q_{8}\rtimes S_{3}$ would be reduced. Hence we shall focus on the reducible assignments $\psi\sim \bm{1_{a}}\oplus\bm{2_{b}}$  for matter fields in the following EFG model construction.

\subsection{\label{sec:general_superpotential}Superpotential for fermion masses }

Now let us perform a systematical analysis of the superpotential and mass matrices for models in which both the three generations of left-handed (LH) matter fields $\psi$ and right-handed (RH) matter fields $\psi^c$ are assigned to a singlet plus a doublet of $Q_{8}\rtimes S_{3}$. Without loss of generality, we assume that the first generation is a singlet and the other two generations form a doublet for both LH and RH matter fields, i.e.
\begin{equation}
\psi_{1}\sim\bm{1_{a}}, \quad \psi^{c}_{1}\sim\bm{1_{b}}, \quad \psi_{d}=(\psi_{2},\psi_{3})^T\sim\bm{2_{x}}, \quad \psi^{c}_{d}=(\psi^{c}_{2},\psi^{c}_{3})^T\sim\bm{2_{y}}, \qquad a,b,x,y=0,1\,.
\end{equation}
The modular weights of $\psi_{1}$, $\psi^{c}_{1}$, $\psi_{d}$ and $\psi^{c}_{d}$  are denoted as $k_{\psi_{1}}$, $k_{\psi^{c}_{1}}$, $k_{\psi_{d}}$ and $k_{\psi^c_{d}}$, respectively. In the approach of EFG, one needs to introduce flavon fields to break the traditional flavor symmetry. In the following, the flavon $\Phi$ with modular weight $k_{\Phi}$ can  comprise all possible irreducible multiplets of EFG $Q_{8}\rtimes S_{3}$, i.e. $\Phi$ can be singlet $\bm{1_{a}}$ (or no flavon), doublet $\bm{2^{0}}$, doublets $\bm{2_{a}}$, triplets $\bm{3_{a}}$ or quartet $\bm{4}$.  The charged lepton/quark Dirac mass terms invariant under $Q_{8}\rtimes S_{3}$ can be generally written as
\begin{eqnarray}
\nonumber \mathcal{W}_{D}&=&\frac{1}{\Lambda}\sum_{\bm{r},s}\alpha_{\bm{r},s}\left(Y^{(k_1)}_{\bm{r}}\Phi \psi^{c}_{1}\psi_{1}\right)_{\bm{1_{0}},s}H_{u/d} +\frac{1}{\Lambda}\sum_{\bm{r},s}\beta_{\bm{r},s}\left(Y^{(k_2)}_{\bm{r}}\Phi \psi^{c}_{1}\psi_{d}\right)_{\bm{1_{0}},s}H_{u/d} \\
\label{eq:Q8S3_gen_superpotential}&&+\frac{1}{\Lambda}\sum_{\bm{r},s}\gamma_{\bm{r},s}\left(Y^{(k_3)}_{\bm{r}}\Phi \psi^{c}_{d}\psi_{d}\right)_{\bm{1_{0}},s}H_{u/d} +\frac{1}{\Lambda}\sum_{\bm{r},s}\varepsilon_{\bm{r},s}\left(Y^{(k_4)}_{\bm{r}}\Phi \psi^{c}_{d}\psi_{1}\right)_{\bm{1_{0}},s}H_{u/d}\,,
\end{eqnarray}
where we have summed over all modular multiplets of modular weights $k_{i}$ ($i=1,2,3,4$) and all independent singlet contractions labeled by the index $s$. In Eq.~\eqref{eq:Q8S3_gen_superpotential}, the modular weight $k_{i}$ should fulfill $k_1=k_{\psi_{1}}+k_{\psi^{c}_{1}}+k_{\Phi}$, $k_2=k_{\psi_{d}}+k_{\psi^{c}_{1}}+k_{\Phi}$, $k_3=k_{\psi_{d}}+k_{\psi^{c}_{d}}+k_{\Phi}$ and $k_4=k_{\psi_{1}}+k_{\psi^{c}_{d}}+k_{\Phi}$. The coupling constants $\alpha_{\bm{r},s}$, $\beta_{\bm{r},s}$, $\gamma_{\bm{r},s}$ and $\varepsilon_{\bm{r},s}$ in Eq.~\eqref{eq:Q8S3_gen_superpotential} are generally complex parameters without gCP symmetry.  If the EFG $Q_{8}\rtimes S_{3}$ is extended to combine with the gCP symmetry, they are further constrained to be real in our working basis. Now we discuss the superpotential for all independent assignments of the matter fields and flavons one by one.

\begin{table}[t!]
\centering
\small
\renewcommand{\arraystretch}{0.8}
\renewcommand{\tabcolsep}{1.0mm}
\begin{tabular}{|c|c|c|c|c|c|c|c|}  \hline \hline

& gCP & $\langle\phi_{\bm{2_{b}}}\rangle/v_{\phi}$  &  $\langle\phi_{\bm{3_{0}}}\rangle/v_{\phi}$  &  $\langle\phi_{\bm{3_{1}}}\rangle/v_{\phi}$  &   $\langle\phi_{\bm{2^{0}}}\rangle/v_{\phi}$ \\ \hline

\multirow{2}{*}{$Z^{A^{2a}T}_{2}$} &  $1$  & $(1,(-1)^{a+b}\sqrt{2}-1)^T$  & \multirow{2}{*}{$(1,1,x)^T$}&  \multirow{2}{*}{$(1,-1,0)^T$}  & \multirow{2}{*}{$(1,0)^T$}  \\ \cline{2-3}
& $A^2$ & $i(1,(-1)^{a+b}\sqrt{2}-1)^T$    &  &    &   \\ \hline

\multirow{2}{*}{$Z^{A^{2a}S}_{2}$} & $B$   & $i^{a+b}e^{7\pi i/8}(1,(-1)^{a+b}\xi^3)^T$    & \multirow{2}{*}{$(i,-x,ix)^T$} &  \multirow{2}{*}{$(0,1,i)^T$}  & \multirow{2}{*}{$(1,-\sqrt{3})^T$}  \\ \cline{2-3}
& $B^3$   & $i^{a+b}e^{3\pi i/8}(1,(-1)^{a+b}\xi^3)^T$    &  &    &   \\ \hline

\multirow{2}{*}{$Z^{A^{2a+1}BT}_{2}$} & $1$   & $(1,(-1)^{a+b}\sqrt{2}+1)^T$    & \multirow{2}{*}{$(1,-1,x)^T$} &  \multirow{2}{*}{$(1,1,0)^T$}  & \multirow{2}{*}{$(1,0)^T$}  \\ \cline{2-3}
& $A^2$   & $i(1,(-1)^{a+b}\sqrt{2}+1)^T$  & &    &   \\ \hline

\multirow{2}{*}{$Z^{A^{2a}TST}_{2}$} & $A$   & $\xi(1,(-1)^{a+b+1}\sqrt{2}i-i)^T$   & \multirow{2}{*}{$(1,-ix,-i)^T$} &  \multirow{2}{*}{$(1,0,i)^T$}  & \multirow{2}{*}{$(1,\sqrt{3})^T$}  \\ \cline{2-3}
& $A^3$   & $\xi^3(1,(-1)^{a+b+1}\sqrt{2}i-i)^T$    &  &   &   \\ \hline
\multirow{2}{*}{$Z^{A^{2a+1}S}_{2}$} & $B$  & $i^{a+b}e^{\pi i/8}(1,(-1)^{a+b}\xi)^T$    & \multirow{2}{*}{$(i,x,ix)^T$} &  \multirow{2}{*}{$(0,1,-i)^T$}  & \multirow{2}{*}{$(1,-\sqrt{3})^T$}  \\ \cline{2-3}
& $BS$ & $i^{a+b}e^{5\pi i/8}(1,(-1)^{a+b}\xi)^T$   &  &    &   \\ \hline

\multirow{2}{*}{$Z^{B^{2a+1}TST}_{2}$} &  $A$   & $\xi(1,(-1)^{a+b+1}\sqrt{2}i+i)^T$   &  \multirow{2}{*}{$(1,ix,i)^T$} &  \multirow{2}{*}{$(1,0,-i)^T$}  & \multirow{2}{*}{$(1,\sqrt{3})^T$}  \\  \cline{2-3}
& $ATST$  & $\xi^3(1,(-1)^{a+b+1}\sqrt{2}i+i)^T$    &  &   &   \\ \hline

\multirow{2}{*}{$Z^{ST}_{3}$,} \multirow{2}{*}{$Z^{A^2ST}_{6}$}  & $ATS$    & \multirow{2}{*}{---}  & \multirow{2}{*}{$(i,i,1)^T$} &  $(i,i,1)^T$  & \multirow{2}{*}{---}  \\ \cline{2-2} \cline{5-5}
& $ATST$      &   &  &  $(1,1,-i)^T$  &   \\ \hline

\multirow{2}{*}{$Z^{ATS}_{3}$,} \multirow{2}{*}{ $Z^{A^3TS}_{6}$}  & $T$  &  \multirow{2}{*}{---}  & \multirow{2}{*}{$(i,-i,1)^T$} &  $(1,-1,-i)^T$  & \multirow{2}{*}{---}  \\ \cline{2-2} \cline{5-5}
& $TS$  &     &  &  $(i,-i,1)^T$  &  \\ \hline

\multirow{2}{*}{$Z^{BST}_{3}$,} \multirow{2}{*}{ $Z^{B^3ST}_{6}$ } & $T$  &  \multirow{2}{*}{---}  & \multirow{2}{*}{$(i,-i,-1)^T$} &  $(1,-1,i)^T$  & \multirow{2}{*}{---}  \\ \cline{2-2} \cline{5-5}
& $ST$  &     &  &  $(i,-i,-1)^T$  &   \\ \hline

\multirow{2}{*}{$Z^{A^3ST}_{3}$,} \multirow{2}{*}{$Z^{AST}_{6}$}  & $AB$  &  \multirow{2}{*}{---}  & \multirow{2}{*}{$(i,i,-1)^T$} &  $(i,i,-1)^T$  & \multirow{2}{*}{---}  \\ \cline{2-2} \cline{5-5}
& $ABS$    &   &  &  $(1,1,i)^T$  &   \\ \hline

\multirow{2}{*}{$Z^{A}_{4}$} & $1$   &  \multirow{2}{*}{---}  & $(1,0,0)^T$ &  $(1,0,0)^T$  & $(1,x)^T$  \\ \cline{2-2} \cline{4-6}
& $S$   &   & $(i,0,0)^T$ &  $(i,0,0)^T$  & $(1+ix,(-3+ix)/\sqrt{3})^T$  \\ \hline

\multirow{2}{*}{$Z^{B}_{4}$} & $1$   & \multirow{2}{*}{---}  & $(0,1,0)^T$ &  $(0,1,0)^T$  & $(1,x)^T$  \\ \cline{2-2} \cline{4-6}
& $TST$     &   & $(0,i,0)^T$ &  $(0,i,0)^T$  & $(1+ix,(3-ix)/\sqrt{3})^T$  \\ \hline

\multirow{2}{*}{$Z^{AB}_{4}$} & $1$     & \multirow{2}{*}{---} & $(0,0,1)^T$ &  $(0,0,1)^T$  & $(1,x)^T$  \\ \cline{2-2} \cline{4-6}
& $T$     &   & $(0,0,i)^T$ &  $(0,0,i)^T$  & $(1,ix)^T$  \\ \hline

\multirow{2}{*}{ $K^{(A^2,T)}_{4}$}   &  $1$  & \multirow{2}{*}{---}  & $(1,1,x)^T$&  $(1,-1,0)^T$  & \multirow{2}{*}{$(1,0)^T$}  \\ \cline{2-2} \cline{4-5}
 & $AB$ &    &    $(i,i,x)^T$ &  $(i,-i,0)^T$    &   \\ \hline

\multirow{2}{*}{ $K^{(A^2,S)}_{4}$}  & $B$     & \multirow{2}{*}{---} & $(i,-x,ix)^T$ & $(0,1,i)^T$  & \multirow{2}{*}{$(1,-\sqrt{3})^T$}  \\ \cline{2-2} \cline{4-5}
& $BS$     &   &  $(i,ix,x)^T$ &  $(0,i,-1)^T$  &   \\ \hline

 \multirow{2}{*}{  $K^{(A^2,AB^3T)}_{4}$} & $1$   & \multirow{2}{*}{---}  & $(1,-1,x)^T$ &  $(1,1,0)^T$  & \multirow{2}{*}{$(1,0)^T$}  \\ \cline{2-2} \cline{4-5}
 & $T$      &   &  $(i,-i,x)^T$ &  $(i,i,0)^T$  &   \\ \hline

\multirow{2}{*}{   $K^{(A^2,TST)}_{4}$}  & $A$   & \multirow{2}{*}{---}  & $(1,-ix,-i)^T$ &  $(1,0,i)^T$  & \multirow{2}{*}{$(1,\sqrt{3})^T$}  \\ \cline{2-2} \cline{4-5}
 & $ATST$      &   &  $(i,ix,1)^T$ &  $(i,0,-1)^T$  &   \\ \hline

\multirow{2}{*}{  $K^{(A^2,AS)}_{4}$}  & $B$    & \multirow{2}{*}{---}  & $(i,x,ix)^T$ &  $(0,1,-i)^T$  & \multirow{2}{*}{$(1,-\sqrt{3})^T$}  \\ \cline{2-2} \cline{4-5}
& $AB$    &   &  $(i,ix,-x)^T$ &  $(0,i,1)^T$  &   \\ \hline

\multirow{2}{*}{ $K^{(A^2,BTST)}_{4}$ }  & $A$&   \multirow{2}{*}{---}  & $(1,ix,i)^T$ &  $(1,0,-i)^T$  & \multirow{2}{*}{$(1,\sqrt{3})^T$}  \\  \cline{2-2} \cline{4-5}
& $AB$   &     &  $(i,ix,-1)^T$ &  $(i,0,1)^T$  &   \\ \hline

\multirow{2}{*}{   $Z^{BS}_{8}$}   & $1$      & \multirow{2}{*}{---}  & \multirow{2}{*}{---} &  $(1,0,0)^T$  &  \multirow{2}{*}{$(1,-\sqrt{3})^T$ } \\ \cline{2-2} \cline{5-5}
& $S$     &  &  &  $(i,0,0)^T$  &   \\ \hline

\multirow{2}{*}{  $Z^{AT}_{8}$}  & $1$     & \multirow{2}{*}{---}  & \multirow{2}{*}{---} &  $(0,0,1)^T$  &  \multirow{2}{*}{$(1,0)^T$ } \\ \cline{2-2} \cline{5-5}
 & $T$    &  &  &  $(0,0,i)^T$  &   \\ \hline

\multirow{2}{*}{  $Z^{ATST}_{8}$ }  & $1$      & \multirow{2}{*}{---}  & \multirow{2}{*}{---} &  $(0,1,0)^T$  &  \multirow{2}{*}{$(1,\sqrt{3})^T$ } \\ \cline{2-2} \cline{5-5}
  & $TST$     &  &  &  $(0,i,0)^T$  &   \\ \hline \hline
\end{tabular}
\caption{\label{tab:VEVs_subs_gCP}
The invariant VEVs of flavons preserving certain Abelian subgroups of $Q_{8}\rtimes S_{3}$, where $\xi=e^{\pi i/4}$ and $x$ is a general real number with $a,b=0,1$. The notation ``---'' denotes the non-existence of  flavon VEVs invariant under the corresponding subgroup.   We give the residual CP transformation $\rho(g)X_{\bm{r}}$ in the second column, and only the element $g$ is shown for simplicity. The overall VEV $v_{\phi}$ is constrained to be real by the residual CP symmetry. We drop the invariant VEVs of flavon transforming as $\bm{4}$, as they are too lengthy to be presented here. Notice the residual symmetry can also constrain the VEV of $\tau$  if it acts nontrivially on $\tau$.}
\end{table}

\begin{description}[labelindent=-0.7em, leftmargin=0.9em]

\item[~~(\lowercase\expandafter{\romannumeral1})] {  $\Phi\sim\bm{1_{c}}$ with $c=0,1$ }

From the Kronecker products of the EFG $Q_{8}\rtimes S_{3}$ in Eq.~\eqref{eq:Q8S3_KP}, we find that only the first and third terms of the superpotential $\mathcal{W}_{D}$ in Eq.~\eqref{eq:Q8S3_gen_superpotential} are not vanishing. The superpotential takes the following form
\begin{equation}\label{eq:WD1}
\hskip-0.08in \mathcal{W}_{D}=\frac{H_{u/d}}{\Lambda} \left[\alpha\left(\Phi \psi^{c}_{1}\psi_{1}\right)_{\bm{1_{[a+b+c]}}}Y^{(k_1)}_{\bm{1_{[a+b+c]}}}+\gamma \Phi \left(\psi^{c}_{d}\psi_{d}\right)_{\bm{1_{[x+y+1]}}}Y^{(k_3)}_{\bm{1_{[x+y+c+1]}}}\right]\,,
\end{equation}
where the explicit expression of $\mathcal{W}_{D}$ only depends on the values of subscripts $[a+b+c]$ and $[x+y+c]$ for certain modular weight assignments. Hence there are four possible $\mathcal{W}_{D}$ corresponding to the four independent values of $[a+b+c]$ and $[x+y+c]$. As an example, we consider the superpotential and mass matrix for the representation assignment with $\psi_{1}, \psi^{c}_{1}\sim \bm{1_{0}}$, $\psi_{d}$, $\psi^{c}_{d}\sim \bm{2_{0}}$, $\Phi=\phi_1\sim \bm{1_{0}}$ which implies $[a+b+c]=0$ and $[x+y+c]=0$, the superpotential and matrix of other three assignments can be obtained analogously. To obtain $Q_{8}\rtimes S_{3}$ invariant mass terms, we find that only the modular form multiplets $Y^{(k_1)}_{\bm{1_{0}}}$ and $Y^{(k_3)}_{\bm{1_{1}}}$ can enter into the first and second terms of Eq.~\eqref{eq:WD1}, respectively. Hence the  $Q_{8}\rtimes S_{3}$ invariant superpotential is
\begin{equation}
\mathcal{W}_{D}=\frac{H_{u/d}}{\Lambda}\left[\alpha Y^{(k_1)}_{\bm{1_{0}}} \phi_{1} \psi^{c}_{1}\psi_{1}+\gamma Y^{(k_3)}_{\bm{1_{1}}} \phi_{1}\left(\psi^c_2 \psi_{3}-\psi^c_3 \psi_{2}\right)\right]\,.
\end{equation}
After the electroweak symmetry breaking by the VEVs of Higgs, we obtain the fermion mass matrix
\begin{equation}\label{eq:Ch_mass1}
M_{\psi}= \frac{v_{u/d}\,\langle\phi_{1}\rangle}{\Lambda}\left(
\begin{array}{ccc}
\alpha Y^{(k_1)}_{\bm{1_{0}}} & 0 & 0 \\
0 & 0 & \gamma Y^{(k_3)}_{\bm{1_{1}}} \\
0 & -\gamma Y^{(k_3)}_{\bm{1_{1}}} & 0 \\
\end{array}
\right)\,,
\end{equation}
where the mass matrix $M_{\psi}$ is defined in the convention $\psi^c M_{\psi}\psi$ in the present work and  $v_{u/d}=\langle H_{u/d}\rangle$ and $\langle\phi_{1}\rangle$ denote the VEVs of Higgs fields and flavon $\phi_{1}$, respectively. Notice that the phases of parameters $\alpha $ and $\gamma$ are unphysical and they can be removed by a field redefinition. Furthermore, the singlet modular multiplets $Y^{(k_1)}_{\bm{1_{0}}}$ and $Y^{(k_3)}_{\bm{1_{1}}}$ in these mass matrices can be absorbed into the couplings $\alpha$ and $\gamma$. If the EFG $Q_{8}\rtimes S_{3}$ and gCP is broken down to an Abelian subgroup and residual gCP compatible with it,  the residual symmetry requires that the VEVs of flavon should take one of the form in table~\ref{tab:VEVs_subs_gCP} \footnote{If the residual symmetry is not a subgroup of $Q_{8}$, the VEV of $\tau$ should take certain value  which is invariant under the action of residual symmetry.}.

For the other three independent assignments with singlet $\Phi$, the corresponding cases are labeled as $\mathcal{W}_{\psi2}$, $\mathcal{W}_{\psi3}$ and $\mathcal{W}_{\psi4}$, respectively. The mass matrices for the three cases can be obtained similarly and the results are listed in table~\ref{tab:Diarc_mms}.  We see that two of the three fermion masses would be degenerate in this case.

\begin{table}[t!]
\centering
\renewcommand{\tabcolsep}{0.5mm}
\renewcommand{\arraystretch}{1.5}
\begin{tabular}{|c|c|c|c|c|c|c|c|c|c|c|c|}  \hline\hline

$\mathcal{W}$ & \multicolumn{3}{c|}{Assignments}  & $Y^{(k_{Y})}_{\bm{r}}$ & $M_{\psi}$ \\  \hline

$\mathcal{W}_{\psi1}$ & \multirow{4}{*}{ $\Phi\sim\bm{1_{c}}$}  & \multirow{2}{*}{$[x+y+c]=0$} & $[a+b+c]=0$  & $Y^{(k_{1})}_{\bm{1_{0}}}$, $Y^{(k_{3})}_{\bm{1_{1}}}$  & $M_{\psi}$ in Eq.~\eqref{eq:Ch_mass1}  \\ \cline{1-1}  \cline{4-6}

$\mathcal{W}_{\psi2}$ &  &   & $[a+b+c]=1$ & $Y^{(k_{1})}_{\bm{1_{1}}}$, $Y^{(k_{3})}_{\bm{1_{1}}}$  & $M_{\psi}\left(Y^{(k_{1})}_{\bm{1_{0}}}\to Y^{(k_{1})}_{\bm{1_{1}}}\right)$  \\ \cline{1-1} \cline{3-6}

$\mathcal{W}_{\psi3}$ & & \multirow{2}{*}{$[x+y+c]=1$}   & $[a+b+c]=0$ &  $Y^{(k_{1})}_{\bm{1_{0}}}$, $Y^{(k_{3})}_{\bm{1_{0}}}$  & $M_{\psi}\left(Y^{(k_{3})}_{\bm{1_{1}}}\to Y^{(k_{3})}_{\bm{1_{0}}}\right)$  \\ \cline{1-1} \cline{4-6}

$\mathcal{W}_{\psi4}$ &  &    &  $[a+b+c]=1$ & $Y^{(k_{1})}_{\bm{1_{1}}}$, $Y^{(k_{3})}_{\bm{1_{0}}}$  & $M_{\psi}\left(Y^{(k_{1})}_{\bm{1_{0}}}\to Y^{(k_{1})}_{\bm{1_{1}}},Y^{(k_{3})}_{\bm{1_{1}}}\to Y^{(k_{3})}_{\bm{1_{0}}}\right)$  \\ \hline

$\mathcal{W}_{\psi5}$ &  \multirow{4}{*}{ $\Phi\sim\bm{2_{c}}$} & \multirow{2}{*}{$[a+y+c]=0$}  & $[b+x+c]=0 $ & $Y^{(k_{2})}_{\bm{1_{1}}}$, $Y^{(k_{4})}_{\bm{1_{1}}}$  & $M^{\prime}_{\psi}$ in Eq.~\eqref{eq:Ch_mass2}  \\ \cline{1-1} \cline{4-6}

$\mathcal{W}_{\psi6}$ &  &  & $[b+x+c]=1$ & $Y^{(k_{2})}_{\bm{1_{0}}}$, $Y^{(k_{4})}_{\bm{1_{1}}}$  & $M^{\prime}_{\psi}\left(Y^{(k_{2})}_{\bm{1_{1}}}\to Y^{(k_{2})}_{\bm{1_{0}}}\right)$  \\ \cline{1-1} \cline{3-6}

$\mathcal{W}_{\psi7}$ &  & \multirow{2}{*}{$[a+y+c]=1$}  & $[b+x+c]=0$ &  $Y^{(k_{2})}_{\bm{1_{1}}}$, $Y^{(k_{4})}_{\bm{1_{0}}}$  & $M^{\prime}_{\psi}\left(Y^{(k_{4})}_{\bm{1_{1}}}\to Y^{(k_{4})}_{\bm{1_{0}}}\right)$  \\ \cline{1-1} \cline{4-6}

$\mathcal{W}_{\psi8}$ &  &  & $[b+x+c]=1$ & $Y^{(k_{2})}_{\bm{1_{0}}}$, $Y^{(k_{4})}_{\bm{1_{0}}}$  & $M^{\prime}_{\psi}\left(Y^{(k_{2})}_{\bm{1_{1}}}\to Y^{(k_{2})}_{\bm{1_{0}}},Y^{(k_{4})}_{\bm{1_{1}}} \to Y^{(k_{4})}_{\bm{1_{0}}}\right)$ \\ \hline

$\mathcal{W}_{\psi9}$ & \multirow{2}{*}{$\Phi\sim\bm{3_{c}}$} &  \multicolumn{2}{c|}{$[x+y+c]=0$} & $Y^{(k_{3})}_{\bm{1_{0}}}$, $Y^{(k_{3})}_{\bm{2^{0}}}$  & $M^{\prime\prime}_{\psi}$ in Eq.~\eqref{eq:Ch_mass3}  \\ \cline{1-1} \cline{3-6}

$\mathcal{W}_{\psi10}$ & &  \multicolumn{2}{c|}{$[x+y+c]=1 $} &   $Y^{(k_{3})}_{\bm{1_{1}}}$, $Y^{(k_{3})}_{\bm{2^{0}}}$  & $M^{\prime\prime}_{\psi}\left(Y^{(k_3)}_{\bm{1_{0}}}\to Y^{(k_3)}_{\bm{1_{1}}},Y^{(k_3)}_{\bm{2^{0}}}\to P_{2}Y^{(k_3)}_{\bm{2^{0}}}\right)$ \\ \hline

$\mathcal{W}_{\psi11}$ &  \multirow{4}{*}{$\Phi\sim\bm{4}$} & \multirow{2}{*}{$[a+y]=0 $} & $[b+x]=0$   & \multirow{4}{*}{$Y^{(k_{2})}_{\bm{2^{0}}}$, $Y^{(k_{4})}_{\bm{2^{0}}}$}  & $M^{\prime\prime\prime}_{\psi}$ in Eq.~\eqref{eq:Ch_mass4}  \\ \cline{1-1} \cline{4-4} \cline{6-6}

$\mathcal{W}_{\psi12}$ & &   &  $[b+x]=1$ &   & $M^{\prime\prime\prime}_{\psi}\left(Y^{(k_{2})}_{\bm{2^{0}}}\to P_{2}Y^{(k_{2})}_{\bm{2^{0}}}\right)$   \\ \cline{1-1} \cline{3-4} \cline{6-6}

$\mathcal{W}_{\psi13}$ &  & \multirow{2}{*}{$[a+y]=1$} & $[b+x]=0$ &    & $M^{\prime\prime\prime}_{\psi}\left(Y^{(k_{4})}_{\bm{2^{0}}}\to P_{2}Y^{(k_{4})}_{\bm{2^{0}}}\right)$   \\ \cline{1-1} \cline{4-4} \cline{6-6}

$\mathcal{W}_{\psi14}$ &  &   &$[b+x]=1 $ &  & $M^{\prime\prime\prime}_{\psi}\left(Y^{(k_{2})}_{\bm{2^{0}}}\to P_{2}Y^{(k_{2})}_{\bm{2^{0}}},Y^{(k_{4})}_{\bm{2^{0}}}\to P_{2}Y^{(k_{4})}_{\bm{2^{0}}}\right)$ \\ \hline \hline
\end{tabular}
\caption{\label{tab:Diarc_mms} The Dirac fermion mass matrices for different representation assignments of matter fields $\psi^{c}_{1}$, $\psi_{1}$, $\psi^{c}_{d}$, $\psi_{d}$ and flavon $\Phi$, where $a,b,c,x,y=0,1$, and the matrix $P_{2}$ is given in Eq.~\eqref{eq:sim_mart}.}
\end{table}

\item[~~(\lowercase\expandafter{\romannumeral2})]  {$\Phi\sim\bm{2_{c}}$ with $c=0,1$}

For this kind of representation assignments of matter fields and flavon $\Phi=(\phi_{1},\phi_{2})^T$, we find that only the second and fourth terms of $\mathcal{W}_{D}$ in  Eq.~\eqref{eq:Q8S3_gen_superpotential} are survived.  Hence the general form of the superpotential $\mathcal{W}_{D}$ can be written as
\begin{equation}\label{eq:WD2}
\mathcal{W}_{D}=\frac{H_{u/d}}{\Lambda}\left[ \beta\psi^{c}_{1}\left(\Phi \psi_{d}\right)_{\bm{1_{[x+c+1]}}}Y^{(k_2)}_{\bm{1_{[b+x+c+1]}}}+\varepsilon \psi_{1} \left(\psi^{c}_{d}\Phi\right)_{\bm{1_{[y+c+1]}}}Y^{(k_4)}_{\bm{1_{[a+y+c+1]}}}\right]\,,
\end{equation}
which depends on the four different values of the representation indices $[b+x+c]$ and $[a+y+c]$. For the case of $[b+x+c]=0$ and $[a+y+c]=0$ which is labeled as $W_{\psi5}$ in table~\ref{tab:Diarc_mms}, only the singlet modular forms $Y^{(k_2)}_{\bm{1_{1}}}$ and $Y^{(k_4)}_{\bm{1_{1}}}$ can enter into the first and the second terms of Eq.~\eqref{eq:WD2} respectively  to form EFG invariant superpotential,
\begin{equation}
\mathcal{W}_{D}=\frac{H_{u/d}}{\Lambda}\left[ \beta Y^{(k_{2})}_{\bm{1_{1}}}\psi^{c}_{1} (\phi_{1}\psi_{3}-\phi_{2} \psi_{2}) + \varepsilon Y^{(k_{4})}_{\bm{1_{1}}} \psi_{1} (\phi_{1}\psi^{c}_{3} -\phi_{2}\psi^{c}_{2})\right]\,,
\end{equation}
which leads to the following mass matrix,
\begin{equation}\label{eq:Ch_mass2}
M^{\prime}_{\psi}= \frac{v_{u/d}}{\Lambda}\left(
\begin{array}{ccc}
0 & -\beta Y^{(k_{2})}_{\bm{1_{1}}}\phi_{2} & \beta Y^{(k_{2})}_{\bm{1_{1}}}\phi_{1} \\
-\varepsilon Y^{(k_{4})}_{\bm{1_{1}}}\phi_{2} & 0 & 0 \\
\varepsilon Y^{(k_{4})}_{\bm{1_{1}}}\phi_{1} & 0 & 0
\end{array}\right)\,,
\end{equation}
where the phases of parameters $\beta $ and $\varepsilon $ can be absorbed by redefinition of the fields $\psi^{c}_{1}$ and $\psi^{c}_{d}$, respectively. The mass matrix $M^{\prime}_{\psi}$ in Eq.~\eqref{eq:Ch_mass2} is of rank two so that one fermion mass is zero.

The other three independent assignments with doublet $\Phi\sim\bm{2_{c}}$ are labeled as $\mathcal{W}_{\psi6}$, $\mathcal{W}_{\psi7}$ and $\mathcal{W}_{\psi8}$, respectively. The corresponding mass matrices are shown in table~\ref{tab:Diarc_mms}. The modular form singlets $Y^{(k_2)}_{\bm{1_{[b+x+c+1]}}}$ and $Y^{(k_4)}_{\bm{1_{[a+y+c+1]}}}$ in the mass matrices can be absorbed into the coupling constants $\beta$ and $\varepsilon$ which can be set real by field redefinition.

\item[~~(\lowercase\expandafter{\romannumeral3})] { $ \Phi\sim\bm{3_{c}}$ with $c=0,1$  }

From the Kronecker product of $Q_{8}\rtimes S_{3}$ in the table~\ref{tab:2O_CG-1st}, we see that only the third term in Eq.~\eqref{eq:Q8S3_gen_superpotential} would give nonzero contribution and the superpotential is given by
\begin{equation}\label{eq:WD9}
\hskip-0.1in\mathcal{W}_{D}=\frac{H_{u/d}}{\Lambda} \left\{\gamma_{1}\left(\Phi \left(\psi^{c}_{d}\psi_{d}\right)_{\bm{3_{[x+y]}}}\right)_{\bm{1_{[x+y+c]}}}Y^{(k_3)}_{\bm{1_{[x+y+c]}}}+\gamma_{2} \left[\left(\Phi \left(\psi^{c}_{d}\psi_{d}\right)_{\bm{3_{[x+y]}}}\right)_{\bm{2^{0}}}Y^{(k_3)}_{\bm{2^{0}}}\right]_{\bm{1_{0}}}\right\}\,.
\end{equation}
The explicit expression of $\mathcal{W}_{D}$ only depends on the value of $[x+y+c]$ which can be equal to $0$ or $1$, and the corresponding two cases are labeled as $\mathcal{W}_{\psi9}$ and $\mathcal{W}_{\psi10}$ in  table~\ref{tab:Diarc_mms}. For the case $[x+y+c]=0$, we find that only the modular form multiplets $Y^{(k_3)}_{\bm{1_{0}}}$ and   $Y^{(k_3)}_{\bm{2^{0}}}$ can enter into superpotential to form  invariant combinations under the EFG $Q_{8}\rtimes S_{3}$. Applying the decomposition rules of $Q_{8}\rtimes S_{3}$, we obtain that the superpotential takes the following form
\begin{equation}
\mathcal{W}_{D}=\frac{H_{u/d}}{\Lambda}\left(\gamma_{1}Y^{(k_3)}_{\bm{1_{0}}}\mathcal{O}_1+\gamma_{2}Y^{(k_3)}_{\bm{2^{0}},1}\mathcal{O}_2 +\gamma_{2}Y^{(k_3)}_{\bm{2^{0}},2}\mathcal{O}_3\right) \,,
\end{equation}
where $Y^{(k_3)}_{\bm{2^{0}},1}$ and $Y^{(k_3)}_{\bm{2^{0}},2}$ represent two components of the modular doublet $Y^{(k_3)}_{\bm{2^{0}}}$, and  the three $Q_{8}$ invariant contractions $\mathcal{O}_{1,2,3}$ are defined as
\begin{eqnarray}
\nonumber \mathcal{O}_{1}&=&\phi_{1}(\psi^{c}_{2} \psi_{3}+\psi^{c}_{3} \psi_{2})+ \psi^{c}_{2} \psi_{2} (\phi_{2}-\phi_{3})-\psi^{c}_{3} \psi_{3} (\phi_{2}+\phi_{3})\,, \\
\nonumber \mathcal{O}_{2}&=& \phi_{1}(\psi^{c}_{2} \psi_{3}+\psi^{c}_{3} \psi_{2})+\psi^{c}_{2} \psi_{2} (\phi_{2}+2 \phi_{3})+\psi^{c}_{3}\psi_{3} (2  \phi_{3}- \phi_{2})\,, \\
\mathcal{O}_{3}&=& -\sqrt{3} [ \phi_{1}(\psi^{c}_{2} \psi_{3}+\psi^{c}_{3} \psi_{2})-\psi^{c}_{2} \psi_{2} \phi_{2}+\psi^{c}_{3} \psi_{3} \phi_{2}]\,.
\end{eqnarray}
The corresponding mass matrix read as
\begin{eqnarray}
\nonumber \hskip-0.95in M^{\prime\prime}_{\psi}&=& \frac{v_{u/d}}{\Lambda}\left[\gamma_{1}Y^{(k_3)}_{\bm{1_{0}}}\left(
\begin{array}{ccc}
0 & 0 & 0 \\
0 & \phi_{2}- \phi_{3} & \phi_{1} \\
0 & \phi_{1} & -\phi_{2}- \phi_{3} \\
\end{array}
\right)\right.\\
\label{eq:Ch_mass3}&&\hskip-0.55in\left.+\gamma_{2}\left(
\begin{array}{ccc}
0 & 0 & 0 \\
0 & 2\phi_{3}Y^{(k_3)}_{\bm{2^{0}},1} +\phi_{2}\left(Y^{(k_3)}_{\bm{2^{0}},1}+\sqrt{3} Y^{(k_3)}_{\bm{2^{0}},2}\right)  & \phi_{1} \left(Y^{(k_3)}_{\bm{2^{0}},1}-\sqrt{3} Y^{(k_3)}_{\bm{2^{0}},2}\right) \\
0 & \phi_{1} \left(Y^{(k_3)}_{\bm{2^{0}},1}-\sqrt{3} Y^{(k_3)}_{\bm{2^{0}},2}\right) & 2\phi_{3}Y^{(k_3)}_{\bm{2^{0}},1} -\phi_{2}\left(Y^{(k_3)}_{\bm{2^{0}},1}+\sqrt{3} Y^{(k_3)}_{\bm{2^{0}},2}\right)  \\
\end{array}
\right)\right]\,.
\end{eqnarray}
Here the phase of $\gamma_{1}$ can be set to be zero,  while the phase  of $\gamma_{2}$ can not be removed by a field redefinition.
Both  $\gamma_{1}$ and $\gamma_{2}$ would be real if gCP symmetry is imposed. Since the first row and first column of $M^{\prime\prime}_{\psi}$ are zero, one fermion mass is vanishing. Furthermore, the fermion mass matrix for $\mathcal{W}_{\psi10}$ can be obtained from that of Eq.~\eqref{eq:Ch_mass3} through the replacements $Y^{(k_3)}_{\bm{1_{0}}}\rightarrow Y^{(k_3)}_{\bm{1_{1}}}$ and $Y^{(k_3)}_{\bm{2^{0}}}\rightarrow P_{2}Y^{(k_3)}_{\bm{2^{0}}}$, where the matrix $P_{2}$ is defined in Eq.~\eqref{eq:sim_mart}. The result is shown in table~\ref{tab:Diarc_mms}.

\item[~~(\lowercase\expandafter{\romannumeral4})]  { $\Phi\sim\bm{4}$   }

In this case, we assume that the Yukawa couplings involve a quartet flavon $\Phi=(\phi_{1},\phi_{2},\phi_{3},\phi_{4})^T$ transforming as $\bm{4}$ under EFG $Q_{8}\rtimes S_{3}$. One can check that the first term and the third term in Eq.~\eqref{eq:Q8S3_gen_superpotential} are vanishing for any modular multiplets $Y^{(k_{1})}_{\bm{r}}$ and $Y^{(k_{3})}_{\bm{r}}$. To obtain nonzero contribution, the modular form multiplets  $Y^{(k_2)}_{\bm{2^0}}$ and $Y^{(k_4)}_{\bm{2^0}}$  should enter into the second and fourth terms in  Eq.~\eqref{eq:Q8S3_gen_superpotential}  and then  the most general invariant superpotential is given by
\begin{equation}\label{eq:WD4}
\mathcal{W}_{D}=\frac{H_{u/d}}{\Lambda} \left\{\beta\psi^{c}_{1}\left[\left(\Phi \psi_{d}\right)_{\bm{2^{0}}}Y^{(k_2)}_{\bm{2^{0}}}\right]_{\bm{1_{b}}}+\varepsilon \psi_{1} \left[\left(\psi^{c}_{d}\Phi\right)_{\bm{2^{0}}}Y^{(k_4)}_{\bm{2^{0}}}\right]_{\bm{1_{a}}}\right\}\,,
\end{equation}
which depends on the four possible values of combinations $[a+y]$ and $[b+x]$. The four possible cases are labeled as $\mathcal{W}_{\psi11}$ to $\mathcal{W}_{\psi14}$ given in table~\ref{tab:Diarc_mms}. We give the results of the case $[a+y]=0$ and $[b+x]=0$, and the results of other three cases can be obtained similarly. For this example case, the eclectic invariant superpotential is given by
\begin{eqnarray}
\nonumber \mathcal{W}_{D}&=&\frac{H_{u/d}}{\Lambda}\left[\beta Y^{(k_{2})}_{\bm{2^{0}},1}\psi^{c}_{1} (\psi_{2}\phi_{4}-\psi_{3}\phi_{2})-\beta Y^{(k_{2})}_{\bm{2^{0}},2}\psi^{c}_{1} (\psi_{2} \phi_{3}-\psi_{3} \phi_{1})+\varepsilon Y^{(k_{4})}_{\bm{2^{0}},1}\psi_{1} (\psi^{c}_{2} \phi_{4}-\psi^{c}_{3} \phi_{2})\right. \\ &&
\left.-\varepsilon Y^{(k_{4})}_{\bm{2^{0}},2}\psi_{1} (\psi^{c}_{2} \phi_{3}-\psi^{c}_{3} \phi_{1})\right]\,,
\end{eqnarray}
which leads to the following mass matrix,
\begin{equation}\label{eq:Ch_mass4}
\hskip-0.13in   M^{\prime\prime\prime}_{\psi}= \frac{v_{u/d}}{\Lambda}\left(
\begin{array}{ccc}
0 & \beta\left(Y^{(k_{2})}_{\bm{2^{0}},1}\phi_{4}-Y^{(k_{2})}_{\bm{2^{0}},2}\phi_{3}\right) & \beta \left(Y^{(k_{2})}_{\bm{2^{0}},2}\phi_{1}-Y^{(k_{2})}_{\bm{2^{0}},1}\phi_{2}\right) \\
\varepsilon\left(Y^{(k_{4})}_{\bm{2^{0}},1}\phi_{4}-Y^{(k_{4})}_{\bm{2^{0}},2}\phi_{3}\right) & 0 & 0 \\
\varepsilon \left(Y^{(k_{4})}_{\bm{2^{0}},2}\phi_{1}-Y^{(k_{4})}_{\bm{2^{0}},1}\phi_{2}\right) & 0 & 0
\end{array}\right)\,,
\end{equation}
where the phases of parameters $\beta $ and $\varepsilon $ are unphysical if the mass terms only involve a single quartet flavon, and they can be removed by a field redefinition. The rank of $M^{\prime\prime\prime}_{\psi}$ is equal to two so that one fermion mass is vanishing. Analogously we list the mass matrices for $\mathcal{W}_{\psi12}$ to $\mathcal{W}_{\psi14}$ in table~\ref{tab:Diarc_mms}.

\item[~~(\lowercase\expandafter{\romannumeral5})]  { $\Phi\sim\bm{2^{0}}$   }

The superpotential can be generally written as
\begin{equation}\label{eq:WD5}
\hskip-0.08in \mathcal{W}_{D}=\frac{H_{u/d}}{\Lambda} \left[\alpha\left( \psi^{c}_{1}\psi_{1}\right)_{\bm{1_{[a+b]}}}\left(\Phi Y^{(k_1)}_{\bm{2^{0}}}\right)_{\bm{1_{[a+b]}}}+\gamma  \left(\psi^{c}_{d}\psi_{d}\right)_{\bm{1_{[x+y+1]}}}\left(\Phi Y^{(k_3)}_{\bm{2^{0}}}\right)_{\bm{1_{[x+y+1]}}}\right]\,.
\end{equation}
The Dirac mass matrix is similar to the mass matrix in Eq.~\eqref{eq:Ch_mass1} and shall give the same predictions for fermion masses and  diagonal matrix. We do not give detailed  results of this case.

\end{description}

In short, two of the three fermion masses are degenerate in cases (i) and (v), and one fermion mass is vanishing in cases (ii), (iii) and (iv) if only one flavon is involved in the Yukawa superpotential. As a result,  at least two flavons are necessary in order to get reasonable fermion mass spectrum in a concrete model.

If a model includes SM singlet $\psi^c$ which can be the right-handed neutrinos or the combination of left-handed leptons and Higgs, the Majorana mass term is allowed.  For the assignment $\psi^c_{1} \sim \bm{1_a}$ and  $\psi^{c}_{d}\sim\bm{2_{b}}$, the Majorana mass term $\mathcal{W}_{M}$ can be obtained from $\mathcal{W}_{D}$ in Eq.~\eqref{eq:Q8S3_gen_superpotential} through the replacements $\psi_{1}\to \psi^{c}_{1}$, $\psi_{d}\to \psi^{c}_{d}$, $\beta_{\bm{r},s}\to \beta_{\bm{r},s}/2$, $\varepsilon_{\bm{r},s}\to \beta_{\bm{r},s}/2$, $Y^{(k_4)}_{\bm{r}}\to Y^{(k_2)}_{\bm{r}}$ and $H_{u/d} /\Lambda\to 1$, i.e.
\begin{equation}
\mathcal{W}_{M}=\sum_{\bm{r},s}\alpha_{\bm{r},s}\left(Y^{(k_1)}_{\bm{r}}\Phi \psi^{c}_{1}\psi^c_{1}\right)_{\bm{1_{0}},s}+\sum_{\bm{r},s}\beta_{\bm{r},s}\left(Y^{(k_2)}_{\bm{r}}\Phi\psi^{c}_{1}\psi^c_{d}\right)_{\bm{1_{0}},s}+\sum_{\bm{r},s}\gamma_{\bm{r},s}\left(Y^{(k_3)}_{\bm{r}}\Phi \psi^{c}_{d}\psi^c_{d}\right)_{\bm{1_{0}},s}\,,
\end{equation}
where  invariance under gCP constrains all coupling constants to be real in our working basis. Then the Majorana mass matrices of $\psi^c$ for the corresponding assignments of flavon $\Phi$  can be obtained from the mass matrices of $\mathcal{W}_{\psi1}\sim \mathcal{W}_{\psi14}$ in table~\ref{tab:Diarc_mms} by dropping the antisymmetric contributions and performing the above replacements.

\section{\label{sec:example-model-EFG}Eclectic lepton model based on $Q_{8}\rtimes S_{3}$}

\begin{table}[h!]
\renewcommand{\tabcolsep}{2.5mm}
\renewcommand{\arraystretch}{1.1}
\centering
\begin{tabular}{|c|c|c|c|c|c|c||c|c|c|c|}\hline \hline
Fields &   $E^{c}_{1}$    & $E^{c}_{d}$    &  $L_{1}$ & $L_{d}$   &  $H_{u}$ & $H_{d}$ & $\zeta$ & $\phi$  &  $\varphi$  &  $\chi$   \\ \hline

$Q_{8}\rtimes S_{3}$ & $\bm{1_{1}}$ & $\bm{2_{0}}$ & $\bm{1_{1}}$ &  $\bm{2_{0}}$ & $\bm{1_{0}}$  & $\bm{1_{0}}$ & $\bm{1_{1}}$ & $\bm{3_{0}}$ & $\bm{2_{0}}$ & $\bm{3_{1}}$   \\ \hline

Modular weight & $0$ & $0$ &$0$ &  $0$ & $0$ & $0$ & $0$ & $0$ & $0$ & $2$   \\  \hline

$Z_5$ & $1$ & $\omega^{2}_{5}$ & $\omega_{5}$ & $\omega_{5}$ & $1$ & $1$ & $\omega^2_{5}$ &  $\omega^2_{5}$   & $\omega^{3}_{5}$  & $\omega^{3}_{5}$  \\ \hline  \hline

\end{tabular}
\caption{\label{tab:model-fields}Assignments of fields under the EFG $Q_{8}\rtimes S_{3}$ and the auxiliary symmetry $Z_{5}$, where $\omega_{5}=e^{\frac{2\pi i}{5}}$.  }
\end{table}

In previous section, we have performed a comprehensive analysis for the general K\"ahler potential and superpotential which are invariant under the EFG $Q_{8}\rtimes S_{3}$ and gCP symmetry. We find that the eclectic models would be severely restricted when the three generations of matter fields are assigned to $\bm{1_{a}}\oplus \bm{2_{b}}$. In this section, we shall construct a concrete lepton model based on the general analysis of section~\ref{sec:mod_general}. In the model, the light neutrino masses arise from the Weinberg operator. The three generations of left-handed lepton doublets and right-handed charged leptons are assigned to the singlet $\bm{1_{1}}$ plus the doublet $\bm{2_{0}}$ of $Q_{8}\rtimes S_{3}$, i.e. the lepton fields $E^{c}_{1}=e^{c}$ and $L_{1}$ transform as singlet $\bm{1_{1}}$, and the other two generations $E^{c}_{d}=(\mu^{c},\tau^{c})^T$ and $L_{d}=(L_{2},L_{3})^{T}$ transform as doublet $\bm{2_{0}}$.  We take all these lepton fields to carry zero modular weight. Furthermore, we also introduce the auxiliary symmetry $Z_5$ which allows us to forbid unwanted operators. The Higgs doublets $H_u$ and $H_d$ are invariant under the actions of the EFG $Q_{8}\rtimes S_{3}$ and the auxiliary symmetry $Z_5$. In the approach of EFG, flavon fields which are standard model singlets are necessary to break the traditional flavor symmetry. In section~\ref{sec:mod_general}, we have pointed out that charged lepton mass terms or neutrino mass terms must involve at least two flavon fields with different number of components to avoid a degenerate lepton mass spectrum or a vanishing lepton mass. In the model, we have introduced four  flavons $\zeta$, $\phi$, $\varphi$ and $\chi$ which transform as $\bm{1_{1}}$, $\bm{3_{0}}$, $\bm{2_{0}}$ and $\bm{3_{1}}$ under $Q_{8}\rtimes S_{3}$, respectively. The former two enter into the charged lepton sector and the latter two enter into the neutrino sector. The field content and their classification under the EFG $Q_{8}\rtimes S_{3}$ and $Z_5$ are listed in table~\ref{tab:model-fields}.

With the fields content in table~\ref{tab:model-fields}, the EFG invariant superpotential for charged lepton and neutrino masses can be written as
\begin{equation}
\mathcal{W}=\mathcal{W}_{e}+\mathcal{W}_{\nu}\,,
\end{equation}
with
\begin{eqnarray}
\nonumber \mathcal{W}_{e}&=&\frac{1}{\Lambda^2}  \left(E^c_{1} L_{1} \right)_{\bm{1_{0}}}\left[\alpha_{1}\left(\zeta\zeta\right)_{\bm{1_{0}}}+\alpha_{2}\left(\phi\phi\right)_{\bm{1_{0}}}\right] H_d+\frac{\beta}{\Lambda}  \left(E^c_{d} L_{d}\zeta \right)_{\bm{1_{0}}} H_d+\frac{\gamma}{\Lambda}  \left(E^c_{d} L_{d}\phi \right)_{\bm{1_{0}}} H_d\,, \\
\label{eq:WO_SP}\mathcal{W}_{\nu}&=&\frac{g_{1}}{\Lambda^2}\left(L_{1}L_{d}\varphi \right)_{\bm{1_{0}}}H_{u}H_{u}+\frac{g_{2}}{\Lambda^2}\left(L_{d}L_{d}\chi Y^{(2)}_{\bm{2^{0}}} \right)_{\bm{1_{0}}}H_{u}H_{u}\,,
\end{eqnarray}
where $\mathcal{W}_{e}$ is Yukawa superpotential of charged leptons, and $\mathcal{W}_{\nu}$ generates the light neutrino masses through the Weinberg operator. We impose gCP symmetry in this model so that all the five couplings $\alpha_{1}$, $\alpha_{2}$, $\beta$, $\gamma$, $g_{1}$ and $g_{2}$ are real. The first two terms of $\mathcal{W}_{e}$ yield the Dirac mass matrix of $\mathcal{W}_{\psi3}$ in table~\ref{tab:Diarc_mms}. The third term of $\mathcal{W}_{e}$ generates the mass matrix $M^{\prime\prime}_{\psi}$ in Eq.~\eqref{eq:Ch_mass3} with modular weight $k_{3}=0$\footnote{Since there is no doublet modular form of level 2 at weight 0,  only the term proportional to $\gamma_1$ in Eq.~\eqref{eq:Ch_mass3} is obtained from the third term of $\mathcal{W}_{e}$.}. The light neutrino mass matrix can be obtained from Eq.~\eqref{eq:Ch_mass2} and $\mathcal{W}_{\psi10}$ in table~\ref{tab:Diarc_mms}.

Both flavons and the modulus $\tau$ are necessary in eclectic flavor models, and their VEVs are the source of symmetry breaking. The VEVs of flavons will spontaneously break the traditional flavor symmetry and the modular symmetry at the same time, and the VEV of $\tau$ only breaks the modular symmetry since $\tau$ is invariant under the traditional flavor symmetry. When a residual symmetry which is a subgroup of EFG is preserved by the mass terms of the charged leptons or neutrinos,  not only VEVs of flavons should be aligned along particular directions in the flavor space, but also the VEV of the modulus $\tau$. It implies that if the residual symmetry is not a subgroup of traditional symmetry, then the modulus $\tau$ must take the value at certain fixed point. In the present model, we assume that the EFG $Q_{8}\rtimes S_{3}$ is broken down to $Z^{B}_{4}\equiv\left\{1, B, B^2,B^3\right\}$ in the charged lepton sector, while the subgroup $Z^{S}_{2}\equiv\left\{1, S\right\}$ is preserved  by vacuum of flavons $\varphi$ and $\chi$ which are associated with neutrino masses.
Hence the flavons $\zeta$, $\phi$, $\varphi$ and $\chi$ will take particular vacuum alignments as follow,
\begin{equation}\label{eq:seesaw_VEVs}
\langle\zeta\rangle=v_{\zeta}, \qquad
\langle\phi\rangle=(0,1,0)^Tv_{\phi}, \qquad    \langle\varphi\rangle=e^{\frac{7\pi i}{8}}(1,\xi^3)^Tv_{\varphi}, \qquad \langle\chi\rangle=(0,1,i)^Tv_{\chi}\,,
\end{equation}
In our model, the residual CP transformation of charged lepton is assumed to be $X_{\bm{r}}$, and the vacuum of flavons $\varphi$ and $\chi$ preserve the CP transformation $\rho_{\bm{r}}(B)X_{\bm{r}}$. Hence all VEV parameters $v_{\zeta}$, $v_{\phi}$, $v_{\varphi}$ and $v_{\chi}$ are real.  As $Z^{B}_{4}$ is a subgroup of the traditional flavor group $Q_{8}$, the charged lepton sector will preserve it for any value of $\tau$. $Z^{S}_{2}$ will be completely broken unless  the modulus at the fixed point $\tau=i$. If the modulus $\tau$ is fixed to be $\tau=i$, one column of the lepton mixing matrix is determined to be $(0,\frac{\sqrt{2-\sqrt{2}}}{2},\frac{\sqrt{2+\sqrt{2}}}{2})^T$ up to reordering and rephasing of the elements. Obviously it is not compatible with experimental data. Hence the modulus $\tau$ can not take value $i$ in this model, and $Q_{8}\rtimes S_{3}$ should be completely broken in neutrino sector. In the following, $\tau$ is treated as a free parameter varying in the fundamental domain $\mathcal{D}=\left\{\tau\in\mathbb{C}\big||\Re(\tau)|\leq1/2, \Im(\tau)>0, |\tau|\geq1\right\}$ in order to adjust the agreement with the data. It is notoriously difficult to dynamically realize vacuum alignment of flavons in traditional discrete flavor symmetry approach. Generally extra new fields and auxiliary symmetry are necessary and the scalar potential has to be cleverly designed. In Appendix~\ref{sec:vev}, we have constructed a flavon potential which leads to the flavon VEV alignments in Eq.~\eqref{eq:seesaw_VEVs}. The standard $F$-term alignment mechanism is employed to arrange the vacuum~\cite{Altarelli:2005yx}, and three additional flavon fields $\psi$, $\rho$, $\kappa$, five driving fields and an auxiliary symmetry $Z_2$ are introduced in the model. We find that the effect of the additional flavon fields on fermion masses and mixing parameters can be absorbed into the parameters of the superpotential, as shown in the Appendix~\ref{sec:vev}. Moreover the mechanism fixing the VEV of the modulus $\tau$ is still an open question, this is the so-called modulus stabilisation problem. There are some evidence that the VEVs of $\tau$ at or in the vicinity of the fixed points are favored~\cite{Cvetic:1991qm,Novichkov:2022wvg,Leedom:2022zdm}. The VEV of $\tau$ is usually taken as a free parameter in the bottom-up approach, likewise we will not attempt to address the modulus stabilisation problem here.

After the electroweak breaking and inserting the vacuum alignments in Eq.~\eqref{eq:seesaw_VEVs}, we can read out the light neutrino mass matrix and the charged lepton mass matrix as follows
{\small\begin{eqnarray}
\nonumber &&\hskip-0.35in M_{\nu}=\frac{v^2_{u}}{\Lambda^2}\left[
g_{1}e^{\frac{7\pi i}{8}}v_{\varphi}\left(
\begin{array}{ccc}
0 & -\xi^3 & 1 \\
-\xi^3 & 0 & 0 \\
1 & 0 & 0 \\
\end{array}
\right)+g_{2}v_{\chi}\left(
\begin{array}{ccc}
0 & 0 & 0 \\
0 & (1+2 i)Y^{(2)}_{\bm{2^{0}},2} -\sqrt{3}  Y^{(2)}_{\bm{2^{0}},1} & 0 \\
0 & 0 & \sqrt{3}  Y^{(2)}_{\bm{2^{0}},1}-(1 -2 i)Y^{(2)}_{\bm{2^{0}},2} \\
\end{array}
\right)\right]\,,  \\
\label{eq:lep_masses} &&\hskip-0.35in M_{e}=\frac{v_d}{\Lambda}\left(
\begin{array}{ccc}
\alpha^{\prime}v^{2}_{\zeta}/\Lambda  & 0 & 0 \\
0 & \gamma v_{\phi}  & -\beta v_{\zeta}  \\
0 & \beta v_{\zeta}  & -\gamma v_{\phi} \\
\end{array}
\right)\,,
\end{eqnarray}}
where $\alpha^{\prime}= \alpha_{1}+\alpha_{2}v^{2}_{\phi}/v^{2}_{\zeta}$, the phase $e^{\frac{7\pi i}{8}}$ is unphysical and it can be removed by redefining the field $L_1$.  Note that the EFG $Q_{8}\rtimes S_{3}$ is broken to the subgroup $Z_4^{B}$ by the VEVs of flavons $\zeta$ and $\phi$ in the charged lepton sector. Then one has  $\rho^\dagger_{\bm{1_{1}}\oplus \bm{2_{0}}}(B)M^\dagger_{e}M_{e}\rho_{\bm{\bm{1_{1}}\oplus \bm{2_{0}}}}(B)=M^\dagger_{e}M_{e}$ which implies that the charged lepton mass matrix is invariant under the residual flavor symmetry $Z_4^{B}$, where $\rho_{\bm{1_{1}}\oplus \bm{2_{0}}}(B)$ is the direct sum of $\rho_{\bm{1_{1}}}(B)$ and $\rho_{\bm{2_{0}}}(B)$,
\begin{equation}
\rho_{\bm{1_{1}}\oplus \bm{2_{0}}}(B)=\left(
\begin{array}{ccc}
1 & 0 & 0 \\
0 & 0 & i \\
0 & i & 0 \\
\end{array}
\right)\,.
\end{equation}
Therefore the unitary transformation $U_{e}$ diagonalizing the hermitian combination $M^\dagger_{e}M_{e}$ via $U_{e}^{\dagger} M^\dagger_{e}M_{e}U_{e}=\text{diag}(m^2_e,m^2_{\mu},m^2_{\tau})$ is of the form
\begin{equation}\label{eq:Ue1}
U_{e}=\frac{1}{\sqrt{2}}\left(
\begin{array}{ccc}
\sqrt{2} & 0 & 0 \\
0 & -1 & 1 \\
0 & 1 & 1 \\
\end{array}
\right)\,.
\end{equation}
The three charged lepton masses are given by
\begin{equation}
m_{e}=\left|\alpha^{\prime} v^{2}_{\zeta}\right|\frac{ v_d}{\Lambda^2} , \qquad
 m_{\mu}=\left|\beta v_{\zeta}+\gamma v_{\phi} \right|\frac{ v_d}{\Lambda}, \qquad
 m_{\tau}=\left|\beta v_{\zeta}-\gamma v_{\phi} \right|\frac{ v_d}{\Lambda}\,.
\end{equation}
Consequently both the unitary diagonalization matrix $U_{e}$ and the charged lepton masses are independent of the modulus $\tau$. The mass hierarchy between muon and tau requires some fine-tuning in the values of $\beta v_{\zeta}$ and $\gamma v_{\phi}$. Since the different VEVs of flavon fields are related via dimensionless couplings, as shown in Appendix~\ref{sec:vev}, we expect a common order of magnitude for the VEVs. To reproduce the observed hierarchy between the charged lepton masses $m_{e}$ and $m_{\tau}$, we choose
\begin{equation}\label{eq:vevs_magn}
\frac{m_{e}}{m_{\tau}}\sim\frac{|v_{\zeta}|}{\Lambda}\sim \frac{|v_{\phi}|}{\Lambda}\sim\frac{|v_{\varphi}|}{\Lambda}\sim\frac{|v_{\chi}|}{\Lambda}\sim\lambda^5_{C}\,,
\end{equation}
where $\lambda_{C}=0.23$ is the Cabibbo angle~\cite{ParticleDataGroup:2024cfk}. From the neutrino mass matrix $M_{\nu}$ in Eq.~\eqref{eq:lep_masses}, we see that the light neutrino mass is of order $\frac{v^2_{u}}{\Lambda}\frac{v_{\varphi}}{\Lambda}$ or similarly $\frac{v^2_{u}}{\Lambda}\frac{v_{\chi}}{\Lambda}$ which should be approximately 0.1 eV. In the present work, the up-type Higgs VEV $v_{u}=v\sin\beta\simeq 170.6\,\text{GeV}$ which is obtained from the assumption $\tan\beta=5$ and Higgs VEV $v=174\,\text{GeV}$. Hence we can estimate the order of magnitude of the cutoff $\Lambda$ and flavon VEVs to be roughly $10^{11}\,\text{GeV}$ and $10^{8}\,\text{GeV}$, respectively. Hence we expect the charged lepton flavor violation processes such as $\mu\rightarrow e\gamma$, $\mu\rightarrow3e$ and $\mu-e$ conversion are negligible due to the large scale of flavon VEVs~\cite{Petcov:1976ff,Bilenky:1977du,Borzumati:1986qx,Hisano:1995cp,Raidal:2008jk,Feruglio:2008ht}. 

\begin{table}[t!]
\centering
\renewcommand{\tabcolsep}{1.6mm}
\renewcommand{\arraystretch}{1.2}
\begin{tabular}{|c|c|c||c|c|c|c|c|}\hline \hline
Mixings  &  $\text{bf}\pm1\sigma$   & $3\sigma$ region  &   Mass ratios &  $\text{bf}\pm1\sigma$  & $3\sigma$ region     \\ \hline

$\sin^2\theta_{13}$ & $0.02224^{+0.00056}_{-0.00057}$ & $[0.02047,0.02397]$ &  $\Delta m^2_{21}/\Delta m^2_{31}$ & $0.0296^{+0.00087}_{-0.00087}$  & $[0.0263,0.0331]$ \\ [0.050in]

$\sin^2\theta_{12}$ & $0.307^{+0.012}_{-0.011}$ & $[0.275,0.344]$ & $m_e/m_{\mu}$ &$0.004737^{+0.00004}_{-0.00004}$ & --- \\ [0.050in]

$\sin^2\theta_{23}$  & $0.454^{+0.019}_{-0.016}$  & $[0.411,0.606]$ &  $m_{\mu}/m_{\tau}$ & $0.05876^{+0.000465}_{-0.000465}$ & ---\\ [0.050in]

$\delta_{CP}/\pi$  & $1.289^{+0.217}_{-0.139}$ & $[0.772,1.944]$& & & \\ [0.050in] \hline \hline

\end{tabular}
\caption{\label{tab:bf_13sigma_data}
The global best fit values and $1\sigma$ ranges and $3\sigma$ ranges for mixing parameters and lepton mass ratios, where the experimental data and errors of the lepton mixing parameters and neutrino masses for NO neutrino mass spectrum are obtained from NuFIT5.3 with Super-Kamiokande atmospheric data~\cite{Esteban:2020cvm}, the charged lepton mass ratios are taken from~\cite{Antusch:2013jca} with $M_{\text{SUSY}}=10\,\text{TeV}$ and $\tan\beta=5$.  }
\end{table}

\begin{figure}[t!]
\centering
\begin{tabular}{c}
\includegraphics[width=0.95\linewidth]{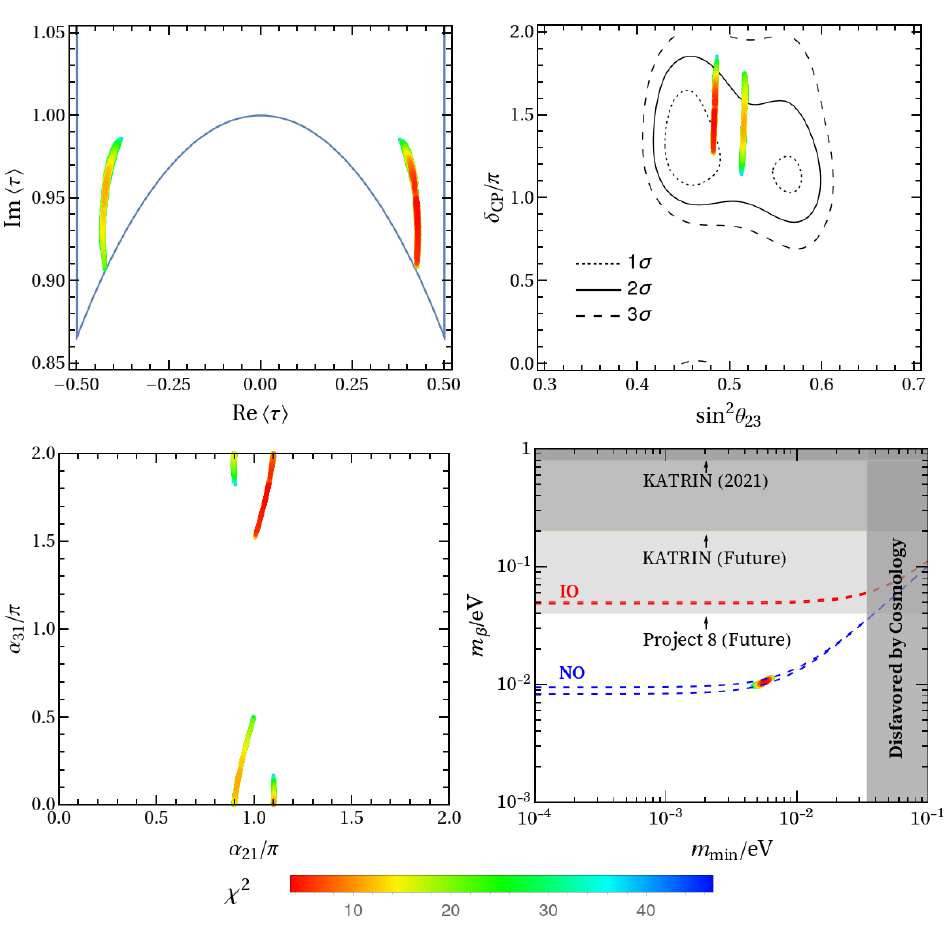}
\end{tabular}
\caption{\label{fig:M_WO} The predicted correlations among the input free parameters, neutrino mixing angles, CP violation phases and the effective mass in beta decay. The gray, reddish and light reddish shaded regions represent the KATRIN upper bound ($m_{\beta}<0.8\,\text{eV}$ at $90\%$ CL)~\cite{KATRIN:2021uub}, KATRIN future bound ($m_{\beta}<0.2\,\text{eV}$ at $90\%$ CL)~\cite{KATRIN:2021dfa} and Project 8 future bound ($m_{\beta}<0.04\,\text{eV}$)~\cite{Project8:2022wqh} respectively. }
\end{figure}

As gCP symmetry is imposed on the model, all the couplings and VEV parameters are enforced to be real. Then the following nine dimensionless observable quantities
\begin{equation}\label{eq:obs_qua}
m_{e}/m_{\mu}, \quad m_{\mu}/m_{\tau},\quad     \sin^2\theta_{12},\quad \sin^2\theta_{13},\quad \sin^2\theta_{23}, \quad \delta_{CP},\quad \Delta m^2_{21}/\Delta m^2_{31}, \quad \alpha_{21}, \quad \alpha_{31}\,,
\end{equation}
depend on  the five dimensionless real parameters $\Re{\tau}$, $\Im{\tau}$, $\beta\Lambda/(\alpha^{\prime} v_{\zeta})$, $\gamma v_{\phi}\Lambda/(\alpha^{\prime} v^2_{\zeta})$ and $g_{2}v_{\chi}/(g_{1}v_{\varphi})$. The remaining two overall input parameters $\big|\alpha^{\prime} v^2_{\zeta}v_{d}/\Lambda^2\big|$ and $\big|g_{1}v^2_uv_{\varphi}/\Lambda^2\big|$ can be determined by the measured electron mass and the solar neutrino mass square difference $\Delta m^2_{21}$. In order to quantitatively measure how well the present model can describe the experimental data, we define a $\chi^2$ function to estimate the goodness-of-fit of a set of chosen values of the input parameters
\begin{equation}\label{eq:chisq}
\chi^2 = \sum_{i=1}^{7} \left( \frac{P_i-O_i}{\sigma_i}\right)^2\,,
\end{equation}
where $O_i$ and $\sigma_{i}$ represent the global best fit values and $1\sigma$ uncertainties of the former seven observable quantities in Eq.~\eqref{eq:obs_qua} respectively, and the corresponding data are summarized in table~\ref{tab:bf_13sigma_data}. $P_{i}$ are the theoretical predictions for the seven observable quantities for certain values of the input parameters. For each set of values of the input parameters, one can obtain the predictions for  $P_{i}$ and $\chi^2$, and  the optimum input parameters yield the lowest $\chi^2$. After performing a comprehensive numerical analysis, we find that this model can describe the experimental data very well, as the global minimum of the function $\chi^2$ is quite small $\chi^2_{\text{min}}=3.504$, and the corresponding values of the input parameters at the $\chi^2_{\text{min}}$ are
\begin{eqnarray}
\nonumber &&    \Re\langle\tau\rangle=0.426, \qquad \Im\langle\tau\rangle=0.944, \qquad \beta\Lambda/(\alpha v_{\zeta})=1901.92, \qquad \gamma v_{\phi}\Lambda/(\alpha^{\prime} v^2_{\zeta})=-1690.82,  \\
&& g_{2}v_{\chi}/(g_{1}v_{\varphi})=17.61, \qquad     \big|\alpha^{\prime} v^2_{\zeta}v_{d}/\Lambda^2\big|=0.511\,\text{MeV}\,, \qquad        \big|g_{1}v^2_uv_{\varphi}/\Lambda^2\big|=7.443\,\text{meV}\,.
\end{eqnarray}
We see that the predictions for the lepton masses and mixing parameters agree rather well with their measured values
\begin{eqnarray}
\nonumber && \sin^2\theta_{13}=0.02231, \qquad \sin^2\theta_{12}=0.3021, \qquad \sin^2\theta_{23}=0.4836, \qquad \delta_{CP}=1.488\pi\,, \\
\nonumber &&  \alpha_{21}=1.068\pi,\qquad \alpha_{31}=1.781\pi, \qquad m_1=5.729\,\text{meV} ,\quad m_2=10.34\,\text{meV} \,,\\
\nonumber &&    m_3=50.47\,\text{meV} , \qquad \sum^3_{i=1}m_{i}=66.54\,\text{meV} , \qquad     m_{\beta\beta}=0\,\text{meV}, \qquad m_{\beta}=10.53\,\text{meV}\,,\\
&& m_{e}=0.511\,\text{MeV}\,, \qquad   m_{\mu}=107.9\,\text{MeV} , \qquad   m_{\tau}=1.836\,\text{GeV} \,.~~~~~~~
\end{eqnarray}
Here $m_{\beta\beta}$ and $m_{\beta}$ refer to the effective mass in neutrinoless double beta decay and the kinematical mass in beta decay respectively, and they are defined as
\begin{equation}
m_{\beta\beta}=\left|\sum^3_{i}m_{i}U^{2}_{1i}\right|\,, \qquad m_{\beta}=\left[\sum^3_{i}m^2_{i}\left|U_{1i}\right|^{2}\right]^{1/2}\,,
\end{equation}
where $U$ represents the lepton mixing matrix. Note that the best fit value of  the three neutrino mass sum is predicted to be $\sum^3_{i=1}m_{i}=66.54\,\text{meV}$ which is much below the current most stringent limit $\sum m_{i}<120\,\text{meV}$ from the Planck collaboration~\cite{Planck:2018vyg}. Moreover, the best fit value of the effective Majorana mass is predicted to be exactly vanishing with $m_{\beta\beta}=0$ which is compatible with the latest result $m_{\beta\beta}<(28-122)\,\text{meV}$ of KamLAND-Zen~\cite{KamLAND-Zen:2024eml}.  In order to explain the result of vanishing $m_{\beta\beta}$, we transfer to the charged lepton diagonal basis by redefining the LH lepton fields $L=(L_{1},L_{2},L_{3})^{T}\to U_{e}L$, then the neutrino mass matrix is given by
{\small\begin{equation}
M^{\prime}_{\nu}=U^{T}_{e}M_{\nu}U_{e}=\frac{v^2_{u}}{\sqrt{2}\Lambda^2}\left(
\begin{array}{ccc}
0 &  g_{1}e^{\frac{7\pi i}{8}} v_{\varphi}\left(\xi ^3+1\right) &  g_{1}e^{\frac{7\pi i}{8}} v_{\varphi}\left(1-\xi ^3\right) \\
g_{1}e^{\frac{7\pi i}{8}}v_{\varphi} \left(\xi ^3+1\right) & 2\sqrt{2} i  g_{2}v_{\chi} Y^{(2)}_{\bm{2^{0}},2} & \sqrt{2} g_{2}v_{\chi} \left(\sqrt{3} Y^{(2)}_{\bm{2^{0}},1}-Y^{(2)}_{\bm{2^{0}},2}\right) \\
g_{1}e^{\frac{7\pi i}{8}}v_{\varphi} \left(1-\xi ^3\right) & \sqrt{2} g_{2}v_{\chi} \left(\sqrt{3} Y^{(2)}_{\bm{2^{0}},1}-Y^{(2)}_{\bm{2^{0}},2}\right) & 2\sqrt{2} i  g_{2}v_{\chi} Y^{(2)}_{\bm{2^{0}},2} \\
\end{array}
\right)\,.
\end{equation}}
In the charged lepton diagonal basis, the effective Majorana mass $m_{\beta\beta}$ is the absolute value of the (11) entry of the neutrino mass matrix, i.e.
\begin{equation}
m_{\beta\beta}=\left|\left(M^{\prime}_{\nu}\right)_{11}\right|=0\,.
\end{equation}
It implies that $m_{\beta\beta}$  is always predicted to be zero and it is independent of the values of input parameters. The best fit value of $m_{\beta}$ is $10.53\,\text{meV}$ which is below the KATRIN upper bound $0.8\,\text{eV}$~\cite{KATRIN:2021uub}, KATRIN future bound $0.2\,\text{eV}$ ~\cite{KATRIN:2021dfa} and Project 8 future bound $0.04\,\text{eV}$~\cite{Project8:2022wqh}. It can be measured through the analysis of the electron energy spectrum near its end point. In order to quantitatively assess how well the model can describe the experimental data on the nine dimensionless observable quantities in Eq.~\eqref{eq:obs_qua}, we comprehensively scan the parameter space and require all the observables lie in their experimentally preferred $3\sigma$ regions, some interesting correlations among the input parameters and observables are obtained, and the predictions for kinematical mass  $m_{\beta}$ versus the lightest neutrino mass are shown in figure~\ref{fig:M_WO}.

Furthermore, we find that the subleading corrections to the superpotential $\mathcal{W}_{e}$ and $\mathcal{W}_{\nu}$ are of the following form:
\begin{eqnarray}
\nonumber&&\hskip-0.3in \Delta \mathcal{W}_{e}=\frac{H_{d}}{\Lambda^3}\left[\left(E^{c}_{1}L_{1}\phi^3_{\nu}Y^{(k)}_{\bm{r}}\right)_{\bm{1_{0}}}+\left(E^{c}_{1}L_{d}\phi^3_{\nu}Y^{(k)}_{\bm{r}}\right)_{\bm{1_{0}}}+\left(E^{c}_{d}L_{1}\phi^2_{l}\phi_{\nu}Y^{(k)}_{\bm{r}}\right)_{\bm{1_{0}}}+\left(E^{c}_{d}L_{d}\phi^2_{l}\phi_{\nu}Y^{(k)}_{\bm{r}}\right)_{\bm{1_{0}}}\right]\,,\\
\label{eq:WO_NLO}&&\hskip-0.3in \Delta \mathcal{W}_{\nu}=\frac{H_{u}H_{u}}{\Lambda^4}\left[\left(L_{1}L_{1}\phi_{l}\phi^{2}_{\nu}Y^{(k)}_{\bm{r}}\right)_{\bm{1_{0}}}+\left(L_{1}L_{d}\phi_{l}\phi^{2}_{\nu}Y^{(k)}_{\bm{r}}\right)_{\bm{1_{0}}}+\left(L_{d}L_{d}\phi_{l}\phi^{2}_{\nu}Y^{(k)}_{\bm{r}}\right)_{\bm{1_{0}}}\right]\,,
\end{eqnarray}
where $\phi_{l}=\{\zeta,\phi\}$ and $\phi_{\nu}=\{\varphi,\chi\}$, the couplings of each independent contraction are dropped. As a result, the subleading corrections to the electron mass are suppressed by $\langle\Phi\rangle/\Lambda$, while other lepton masses and mixing parameters receive corrections suppressed by $\langle\Phi\rangle^2/\Lambda^2$ with respect to LO results, where $\Phi=\{\zeta,\phi,\varphi,\chi \}$ denotes a generic flavon in the model, and all flavon VEVs are expected to be of a common order of magnitude. Hence the subleading corrections are small enough to be neglected.  From the assignments of flavon fields in table~\ref{tab:model-fields},  the corrections to the leading order flavor universal kinetic terms can be estimated by using the general results of section~\ref{sec:mod_general}. We find that the flavor non-universal off-diagonal elements of the K\"ahler metric arise from the NNLO terms $\mathcal{K}_{\rm NNLO}$. The corresponding corrections are found to be suppressed by $\langle\Phi\rangle^2/\Lambda^2$, and they should be negligible. A non-zero $m_{\beta\beta}$ could be generated by the subleading corrections, and it is estimated to be of order $4\times 10^{-8}\,\text{eV}$ which is much below the latest bound $m_{\beta\beta}<(28-122)\,\text{meV}$ of KamLAND-Zen~\cite{KamLAND-Zen:2024eml}, and the next generation experiments sensitivity ranges $m_{\beta\beta}<(9-21)\,\text{meV}$ from LEGEND-1000~\cite{LEGEND:2017cdu} and $m_{\beta\beta}<(4.7-20.3)\,\text{meV}$ from nEXO~\cite{nEXO:2017nam}.

\section{\label{sec:conclusion} Conclusion}

The approach of EFG is a quite predictive scheme for constructing fermion mass models. It is semi-direct product of a traditional flavor group with a corresponding finite modular group, and it combines the advantages of them while largely avoiding their limitations.  If traditional flavor group is fixed to be non-Abelian group, the minimal EFG is $Q_{8}\rtimes S_{3}\cong GL(2,3)$ which is semi-direct product of the traditional flavor group $Q_{8}$ with the finite modular group $S_{3}$~\cite{Nilles:2020nnc}. Notice that if Abelian traditional flavor groups are considered, we find that the minimal EFG is $(Z_2\times Z_2)\rtimes S_3\cong S_4$ which is the extension of the traditional flavor symmetry $Z_{2}\times Z_{2}$  by the finite modular group $S_{3}$. We expect that the $S_4$ eclectic model should be different from the flavor models based on either traditional flavor symmetry $S_{4}$ or modular symmetry $S_{4}$. In the present work, we shall focus on the minimal EFG $Q_{8}\rtimes S_{3}$. All its irreducible representations of $Q_{8}\rtimes S_{3}$ have been induced from the irreducible representations of $Q_{8}$ and $S_{3}$, and the results are summarized in table~\ref{tab:EFG_Q8S3_Reps}. Moreover, we have studied the consistency conditions between EFG and gCP symmetry. The corresponding results for the two symmetry transformations acting on the modulus $\tau$ and the matter fields are derived. The gCP symmetry consistent with the minimal EFG $Q_{8}\rtimes S_{3}$ is considered. We found that the gCP transformation is a unit matrix up to eclectic transformations in our work basis and all coupling constants are real if gCP invariance is imposed.

We have performed a comprehensive analysis for the general K\"ahler potential and superpotential of models based on the minimal EFG $Q_{8}\rtimes S_{3}$. We find that the eclectic models would be severely restricted when the three generations of matter fields are assigned to  a singlet $\bm{1_{a}}$ plus a doublet $\bm{2_{b}}$.  According to this kind of representation assignments of matter fields, we find that the corrections to fermion masses and mixing parameters from non-canonical K\"ahler terms are suppressed by $\langle\Phi\rangle^2/\Lambda^2$ in the case of without multiplet flavon which is chargeless under the auxiliary group in a model. Furthermore, the general Dirac mass matrices are achieved for each possible assignment of matter fields and flavon, and the results are listed in table~\ref{tab:Diarc_mms}. Then one can obtained the Dirac mass matrix and Majorana mass matrix for any assignment of matter fields and any number of flavons from table~\ref{tab:Diarc_mms}.

Guided by most general analysis, a concrete lepton model which gives rise to both EFG $Q_{8}\rtimes S_{3}$ and gCP was constructed. In our model, the EFG $Q_{8}\rtimes S_{3}$ is broken down to $Z^{B}_{4}$ in the charged lepton sector and it is broken completely in the neutrino sector. The resulting contributions to lepton masses and mixing parameters from non-canonical K\"ahler terms are suppressed by $\langle\Phi\rangle^2/\Lambda^2$, and they can be safely neglected. We have studied the flavor phenomenology of the lepton sector emerging from the model. In order to fully explore the phenomenology of the model, we have performed a numerical analysis of the model. We find that the 9 dimensionless observable quantities listed in Eq.~\eqref{eq:obs_qua} can be well fitted by adjusting five free input parameters. Furthermore, we give the predictions for the three neutrino mass sum $\sum_{i}m_{i}$, the kinematical mass $m_{\beta}$ in beta decay and the effective mass $m_{\beta\beta}$ in neutrinoless double beta decay. The former two are predicted to be safely below the present upper limit and the last one is always predicted to be zero. The VEV of flavons is estimated to be of order $\mathcal{O}(10^{8})\,\text{GeV}$, consequently the branch ratios of the charged lepton flavor violation processes should be suppressed, and a detailed study is beyond the scope of the present paper.

\section*{Acknowledgements}

CCL is supported by Natural Science Basic Research Program of Shaanxi (Program No. 2024JC-YBQN-0004), the National Natural Science Foundation of China under Grant Nos. 12005167, 12247103,  and the Young Talent Fund of Association for Science and Technology in Shaanxi, China. JNL is supported by the Grants No. NSFC-12147110 and the China Post-doctoral Science Foundation under Grant No. 2021M70. GJD is supported by the National Natural Science Foundation of China under Grant Nos.~12375104, 11975224.

\newpage

\section*{Appendix}

\setcounter{equation}{0}
\renewcommand{\theequation}{\thesection.\arabic{equation}}

\begin{appendix}

\section{\label{sec:EFGQ8S3_group}Group theory of EFG $Q_{8}\rtimes S_{3}\cong GL(2,3)$}

In the present work, the EFG $Q_{8}\rtimes S_{3}\cong GL(2,3)$ are defined in terms of four generators $A$, $B$, $S$ and $T$ fulfilling the following relations
\begin{eqnarray}
\nonumber && A^4=S^2=T^2=(ST)^3=1,\qquad B^2=A^2,\qquad B^{-1}AB=A^{-1}, \qquad \\
\label{eq:Q8S3_mul_ruls} && SAS = A^{3},\qquad SBS =AB, \qquad TAT = B, \qquad TBT = A\,,
\end{eqnarray}
where $A$ and $B$ generate the traditional flavor group $Q_{8}$, and $S$ and $T$ generate the finite modular group $S_{3}$. Note that $A^2$ is the center of $Q_{8}\rtimes S_{3}$ and it commutes with all group elements. The 48 elements of EFG $Q_{8}\rtimes S_{3}$  belong to the following eight conjugacy classes
\begin{eqnarray}
\nonumber 1C_1&=&\{1\}\,, \\
\nonumber  1C_2&=&\{A^2\}\,,\\
\nonumber 12C_{2}&=& \{S,T,TST,A^2S,A^2T,A^2TST,AS,A^3S,BTST,B^3TST,ABT,AB^3T\}\,,\\
\nonumber   8C_{3}&=& \{ST,TS,ATS,A^3ST,BST,B^3TS,ABTS,AB^3ST\}\,, \\
\nonumber   6C_{4}&=& \{A,A^3,B,B^3,AB,AB^3\}\,, \\
\nonumber  8C_{6}&=& \{A^2ST,A^2TS,AST,A^3TS,BTS,B^3ST,ABST,AB^3TS\}\,, \\
\nonumber  6C_{8}&=& \{AT,A^3TST,BT,B^3S,ABTST,AB^3S\}\,, \\
\label{eq:Q8S3_CC} 6C^{\prime}_{8}&=& \{ATST,A^3T,BS,B^3T,ABS,AB^3TST\}\,.
\end{eqnarray}
where $kC_{n}$ denotes a conjugacy class which contains $k$ elements with order $n$.
\begin{table}[t!]
\begin{center}
\renewcommand{\tabcolsep}{2.8mm}
\renewcommand{\arraystretch}{1.1}
\begin{tabular}{|c|c|c|c|c|c|c|c|c|}\hline\hline
\text{Classes} & $1C_{1}$ & $1C_{2}$& $12C_{2}$ & $8C_3$ & $6C_4$ & $8C_6$ & $6C_8$ & $6C^{\prime}_8$     \\ \hline
\text{Representatives} & $1$ & $A^{2}$& $S$ & $ST$ & $A$ & $A^2ST$ & $AT$ & $BS$     \\ \hline
$\bm{1_{a}} $ & $ 1 $ & $ 1 $ & $ (-1)^{a} $ & $ 1 $ & $ 1 $ & $ 1 $ & $ (-1)^{a} $  & $ (-1)^{a}$ \\ \hline
$\bm{2^0} $ & $ 2 $ & $ 2 $ & $ 0 $ & $ -1 $ & $ 2 $ & $ -1 $& $ 0 $  & $ 0$ \\ \hline
$\bm{2_{a}} $ & $ 2 $ & $ -2 $ & $ 0 $ & $ -1 $ & $ 0 $ & $ 1 $& $ (-1)^{1-a}\sqrt{2} i$  & $ (-1)^{a}\sqrt{2}i$ \\ \hline
$\bm{3_{a}} $ & $ 3 $ & $ 3 $ & $ (-1)^{a} $ & $ 0 $ & $ -1 $ & $ 0 $ & $ (-1)^{1-a} $  & $ (-1)^{1-a}$ \\ \hline
$\bm{4} $ & $ 4 $ & $ -4 $ & $ 0 $ & $ 1 $ & $ 0 $ & $ -1 $ & $ 0 $  & $ 0$ \\ \hline   \hline
\end{tabular}
\caption{\label{tab:character_Q8S3} The character table of the EFG $Q_{8}\rtimes S_{3}\cong GL(2,3)$, where $a=0,1$.}
\end{center}
\end{table}
The  representation matrices of the generators $A$, $B$, $S$ and $T$ for all eight irreducible representations of $Q_{8}\rtimes S_{3}$  are listed in table~\ref{tab:EFG_Q8S3_Reps}. We can derive the $Q_{8}\rtimes S_{3}$ character table by taking traces over the relevant representation matrices. The results are displayed in table~\ref{tab:character_Q8S3}. The Kronecker products between various irreducible representations follow immediately
\begin{eqnarray}
\nonumber&&\bm{1_{0}}\otimes \bm{r}=\bm{r}, \quad \bm{1_{1}}\otimes \bm{1_{1}}=\bm{1_{0}}, \quad \bm{1_{1}}\otimes\bm{2^{0}}=\bm{2^{0}},\quad \bm{1_{1}}\otimes\bm{2_{a}}=\bm{2_{[a+1]}}, \quad \bm{1_{1}}\otimes\bm{3_{a}}=\bm{3_{[a+1]}}, \\
\nonumber && \bm{1_{1}}\otimes\bm{4}=\bm{4}, \quad \bm{2^{0}}\otimes \bm{2^{0}}=\bm{1_{0}}\oplus \bm{1_{1}}\oplus\bm{2^{0}}, \quad \bm{2^{0}}\otimes \bm{2_{a}}= \bm{4}, \quad \bm{2^{0}}\otimes \bm{3_{a}}= \bm{3_{0}}\oplus \bm{3_{1}},  \\
\nonumber && \bm{2^{0}}\otimes \bm{4}= \bm{2_{0}}\oplus \bm{2_{1}}\oplus \bm{4}, \quad  \bm{2_{a}}\otimes \bm{2_{b}}=\bm{1_{[a+b+1]}}\oplus\bm{3_{[a+b]}}, \quad
\bm{2_{a}}\otimes \bm{3_{b}}=\bm{2_{[a+b+1]}}\oplus\bm{4},  \\
\nonumber &&\bm{2_{a}}\otimes \bm{4}=\bm{2^{0}}\oplus\bm{3_{0}}\oplus \bm{3_{1}}, \quad  \bm{3_{a}}\otimes \bm{3_{b}}=\bm{1_{[a+b]}}\oplus \bm{2^{0}}\oplus\bm{3_{0}}\oplus \bm{3_{1}},  \\
\label{eq:Q8S3_KP} && \bm{3_{a}}\otimes \bm{4}=\bm{2_{0}}\oplus \bm{2_{1}}\oplus \bm{4_{i}}\oplus \bm{4_{ii}}, \quad  \bm{4}\otimes \bm{4}=\bm{1_{0}}\oplus \bm{1_{1}}\oplus\bm{2^{0}}\oplus\bm{3_{0i}}\oplus\bm{3_{0ii}}\oplus \bm{3_{1i}}\oplus \bm{3_{1ii}}\,,
\end{eqnarray}
where $\bm{r}$ denotes any irreducible representation of $Q_{8}\rtimes S_{3}$, the irreducible representation $\bm{r_{[i]}}$ refers to $\bm{r_{\,i\, \text{mod}\,2}}$ in the case of the lower index $i>1$, and $\bm{3_{0i}}$, $\bm{3_{0ii}}$, $\bm{3_{1i}}$, $\bm{3_{1ii}}$, $\bm{4_{i}}$ and $\bm{4_{ii}}$ stand for the two $\bm{3_{0}}$, two $\bm{3_{1}}$ and two $\bm{4}$ representations which appear in the Kronecker products, respectively. In the following, we list the CG coefficients in our basis. All CG coefficients would be reported in the form of $\alpha\otimes \beta$. We shall  use $\alpha_{i}$ ($\beta_{i}$) stands for the elements of the first (second) representation of the tensor product. Furthermore,  the following notations are used to facilitate the expressions of CG coefficients
\begin{equation}\label{eq:sim_mart}
P_{2}=\begin{pmatrix}
 0 & -1 \\ 1 &0
\end{pmatrix}, \qquad
P_{4}=\begin{pmatrix}
P_{2} &\mathbb{0}_{2}  \\
\mathbb{0}_{2} & P_{2} \\
\end{pmatrix}\,.
\end{equation}
The results of the CG coefficients are summarized in table~\ref{tab:2O_CG-1st}.

From the irreducible representation matrices of $Q_{8}\rtimes S_{3}$ summarized in table~\ref{tab:EFG_Q8S3_Reps}, we find that the representations $\bm{1_{a}}$ and $\bm{2^{0}}$ are real, the representations $\bm{3_{a}}$ and $\bm{4}$ are  self-conjugate, and the doublet representations $\bm{2_{0}}$  and $\bm{2_{1}}$ are the complex conjugate of each other.  We derive the following  similarly transformations between the representations $\bm{2_{a}}$, $\bm{3_{a}}$ and $\bm{4}$ and their complex conjugation representations
\begin{equation}
\rho^*_{\bm{2_{a}}}(\mathfrak{g})=U^{\dagger}_{2}\rho_{\bm{2_{[a+1]}}}(\mathfrak{g})U_{2}, \quad \rho^*_{\bm{3_{a}}}(\mathfrak{g})=U^{\dagger}_{3}\rho_{\bm{3_{a}}}(\mathfrak{g})U_{3}, \quad \rho^*_{\bm{4}}(\mathfrak{g})=U^{\dagger}_{4}\rho_{\bm{4}}(\mathfrak{g})U_{4}, \qquad \forall \mathfrak{g}\in Q_{8}\rtimes S_{3}\,,
\end{equation}
with
\begin{equation}
U_{2}=P_{2}=\begin{pmatrix}
0 & -1 \\ 1 &0
\end{pmatrix}, \qquad
U_{3}=\begin{pmatrix}
1 & 0 & 0  \\ 0 & 1 &0 \\ 0 &0 & -1
\end{pmatrix}, \qquad
U_{4}=\begin{pmatrix}
0 & 0 & 0 & -1 \\ 0 & 0& 1 &0 \\ 0 & 1 & 0 &0 \\ -1 & 0 &0 & 0
\end{pmatrix}.
\end{equation}
It implies that
\begin{equation}\label{eq:Reps_conju}
U_{2}\alpha^*\sim \bm{2_{[a+1]}} ~~\text{for}~~ \alpha\sim  \bm{2_{a}}, \qquad  U_{3}\alpha^*\sim \bm{3_{a}} ~~\text{for}~~ \alpha\sim  \bm{3_{a}}, \qquad      U_{4}\alpha^*\sim \bm{4} ~~\text{for}~~ \alpha\sim  \bm{4}\,.
\end{equation}
Then those contractions involving fields which transform as $\bm{\bar{2}_{a}}$, $\bm{\bar{3}_{a}}$ and $\bm{\bar{4}}$ can be easily obtained  from the CG coefficients in table~\ref{tab:2O_CG-1st}, where $\bm{\bar{2}_{a}}$, $\bm{\bar{3}_{a}}$ and $\bm{\bar{4}}$ represent the complex conjugation of $\bm{2_{a}}$, $\bm{3_{a}}$ and $\bm{4}$, respectively.

\newpage

\begin{center}
\renewcommand{\arraystretch}{1.15}
\begin{small}
\setlength\LTcapwidth{\textwidth}
\setlength\LTleft{-0.42in}
\setlength\LTright{0pt}
\begin{longtable}{|c|c|c|c|c|c|c|c|c|c|c|c|}
\caption{\label{tab:2O_CG-1st}
Tensor products and the corresponding CG coefficients for the EFG $Q_{8}\rtimes S_{3}$. Here $\alpha_i$ and $\beta_i$ denote the elements of the first and second representations respectively in the tensor product.  Note that $\mathbf{1}\otimes \bm{r}=\bm{r}$ for any irreducible representation $\bm{r}$, and the matrices $P_2$ and $P_4$ are given in Eq.~\eqref{eq:sim_mart}.} \\
\midrule
\specialrule{0em}{1.0pt}{1.0pt}

\endfirsthead

\multicolumn{12}{c}
{{\bfseries \tablename\ \thetable{} -- continued from previous page}} \\
\hline

\endhead

\caption[]{continues on next page}\\
\endfoot

\endlastfoot

\hline
\multicolumn{3}{|c}{~~~~~~$\bm{1_{1}}\otimes\bm{2^{0}}=\bm{2^{0}}$~~~~~~} & \multicolumn{3}{|c}{$\bm{1_{1}}\otimes\bm{2_{a}}=\bm{2_{[a+1]}}$} & \multicolumn{3}{|c}{$\bm{1_{1}}\otimes\bm{3_{a}}=\bm{3_{[a+1]}}$}&\multicolumn{3}{|c|}{$\bm{1_{1}}\otimes\bm{4}=\bm{4}$}  \\ \hline
\multicolumn{3}{|c}{$\bm{2^{0}}:
~\alpha_{1}\begin{pmatrix}\beta _2 \\-\beta _1
\end{pmatrix}$} &
\multicolumn{3}{|c}{$\bm{2_{[a+1]}}:~ \alpha_{1}\begin{pmatrix}\beta _1 \\ \beta _2
\end{pmatrix}$} &
\multicolumn{3}{|c}{$\bm{3_{[a+1]}}:~ \alpha_{1}\begin{pmatrix}\beta _1 \\ \beta _2 \\ \beta _3 \end{pmatrix}$} &
\multicolumn{3}{|c|}{$\bm{4}:~ \alpha_{1}\begin{pmatrix}\beta _2 \\ -\beta _1 \\\beta _4 \\ -\beta _3 \end{pmatrix}$}  \\ \hline

\multicolumn{3}{|c}{$\bm{2^{0}}\otimes \bm{2^{0}}=\bm{1_{0}}\oplus \bm{1_{1}}\oplus\bm{2^{0}}$} & \multicolumn{3}{|c}{$\bm{2^{0}}\otimes\bm{2_{a}}=\bm{4}$} & \multicolumn{3}{|c}{$\bm{2^{0}}\otimes\bm{3_{a}}=\bm{3_{0}}\oplus \bm{3_{1}}$}&\multicolumn{3}{|c|}{$\bm{2^{0}}\otimes\bm{4}=\bm{2_{0}}\oplus \bm{2_{1}}\oplus \bm{4} $}  \\ \hline

\multicolumn{3}{|c}{  $ \begin{array}{l}
\bm{1_{0}}:~\alpha_{1} \beta_{1}+\alpha_{2} \beta_{2} \\
\bm{1_{1}}:~\alpha_{1} \beta_{2}-\alpha_{2} \beta_{1} \\
\bm{2_{0}}:~\begin{pmatrix}\alpha_{2} \beta_{2}-\alpha_{1} \beta_{1} \\
\alpha_{1} \beta_{2}+\alpha_{2} \beta_{1}
\end{pmatrix} \\
\end{array} $ } &
\multicolumn{3}{|c}{  $ \bm{4}:
P^{a}_{4}\begin{pmatrix}\alpha_{1}\beta_{1} \\ \alpha_{2}\beta_{1} \\
        \alpha_{1}\beta_{2} \\ \alpha_{2}\beta_{2}
\end{pmatrix} $ } &
\multicolumn{3}{|c}{  $ \begin{array}{l}
\bm{3_{a}}:
\begin{pmatrix}\beta_{1} \left(\alpha_{1}-\sqrt{3} \alpha_{2}\right)\\
\beta_{2} \left(\alpha_{1}+\sqrt{3} \alpha_{2}\right)\\
-2 \alpha_{1} \beta_{3}
\end{pmatrix} \\
\bm{3_{[a+1]}}:
\begin{pmatrix}\beta_{1} \left(\sqrt{3} \alpha_{1}+\alpha_{2}\right)\\
        \beta_{2} \left(\alpha_{2}-\sqrt{3} \alpha_{1}\right)\\
-2 \alpha_{2} \beta_{3}
\end{pmatrix}  \\ \end{array} $ } &
\multicolumn{3}{|c|}{  $ \begin{array}{l}
\bm{2_{0}}:~ \begin{pmatrix}\alpha_{1} \beta_{1}+\alpha_{2} \beta_{2}\\
\alpha_{1} \beta_{3}+\alpha_{2} \beta_{4}
\end{pmatrix}\\
\bm{2_{1}}:
\begin{pmatrix}
\alpha_{1} \beta_{2}-\alpha_{2} \beta_{1}\\
\alpha_{1} \beta_{4}-\alpha_{2} \beta_{3}
\end{pmatrix}\\
\bm{4}:
\begin{pmatrix}
\alpha_{2} \beta_{2}-\alpha_{1} \beta_{1}\\
\alpha_{1} \beta_{2}+\alpha_{2} \beta_{1}\\
\alpha_{2} \beta_{4}-\alpha_{1} \beta_{3}\\
\alpha_{1} \beta_{4}+\alpha_{2} \beta_{3}
\end{pmatrix} \\ \end{array} $ } \\ \hline

\multicolumn{4}{|c}{$\bm{2_{a}}\otimes\bm{2_{b}}=\bm{1_{[a+b+1]}}\oplus\bm{3_{[a+b]}}$} & \multicolumn{4}{|c}{$\bm{2_{a}}\otimes\bm{3_{b}}=\bm{2_{[a+b+1]}}\oplus\bm{4}$} & \multicolumn{4}{|c|}{$\bm{2_{a}}\otimes\bm{4}=\bm{2^{0}}\oplus\bm{3_{0}}\oplus \bm{3_{1}}$}  \\ \hline

\multicolumn{4}{|c}{  $ \begin{array}{l}
 \bm{1_{[a+b+1]}}:\alpha_{1} \beta_{2}-\alpha_{2} \beta_{1} \\
 \bm{3_{[a+b]}}:\begin{pmatrix}\alpha_{1} \beta_{2}+\alpha_{2} \beta_{1}\\
\alpha_{1} \beta_{1}-\alpha_{2} \beta_{2}\\
\alpha_{1} \beta_{1}+\alpha_{2} \beta_{2}
\end{pmatrix}\\
\end{array} $ } &
\multicolumn{4}{|c}{  $ \begin{array}{l}
 \bm{2_{[a+b+1]}}:  \begin{pmatrix}\alpha_{2} (\beta_{2}+\beta_{3})-\alpha_{1} \beta_{1}\\
\alpha_{1} (\beta_{2}-\beta_{3})+\alpha_{2} \beta_{1}
\end{pmatrix} \\
 \bm{4}:
P^{[a+b]}_{4}\begin{pmatrix}-\sqrt{3} (\alpha_{1} \beta_{1}+\alpha_{2} \beta_{2})\\
\alpha_{2} (\beta_{2}-2 \beta_{3})-\alpha_{1} \beta_{1}\\
\sqrt{3} (\alpha_{2} \beta_{1}-\alpha_{1} \beta_{2})\\
\alpha_{1} (\beta_{2}+2 \beta_{3})+\alpha_{2} \beta_{1}
\end{pmatrix}\\
\end{array} $ } &
\multicolumn{4}{|c|}{  $ \begin{array}{l}
 \bm{2^{0}}: P^{a}_{2}\begin{pmatrix}\alpha_{1} \beta_{4}-\alpha_{2} \beta_{2}\\
\alpha_{2} \beta_{1}-\alpha_{1} \beta_{3}
\end{pmatrix} \\
  \bm{3_{[a]}}:
\begin{pmatrix}\sqrt{3} (\alpha_{1} \beta_{4}+\alpha_{2} \beta_{2})-\alpha_{1} \beta_{3}-\alpha_{2} \beta_{1}\\
\sqrt{3}(\alpha_{2}\beta_{4}-\alpha_{1}\beta_{2})-\alpha_{1}\beta_{1}+\alpha_{2}\beta_{3}\\
2 (\alpha_{1} \beta_{1}+\alpha_{2} \beta_{3})
\end{pmatrix}\\
 \bm{3_{[a+1]}}:
\begin{pmatrix}
\sqrt{3}(\alpha_{1}\beta_{3}+\alpha_{2}\beta_{1})+\alpha_{1}\beta_{4}+\alpha_{2}\beta_{2}\\
\sqrt{3} (\alpha_{2} \beta_{3}-\alpha_{1} \beta_{1})+\alpha_{1} \beta_{2}-\alpha_{2} \beta_{4}\\
-2 (\alpha_{1} \beta_{2}+\alpha_{2} \beta_{4})
\end{pmatrix}\\
\end{array} $ } \\ \hline

\multicolumn{6}{|c}{$\bm{3_{a}}\otimes\bm{3_{b}}=\bm{1_{[a+b]}}\oplus \bm{2^{0}}\oplus\bm{3_{0}}\oplus \bm{3_{1}}$} & \multicolumn{6}{|c|}{$\bm{3_{a}}\otimes\bm{4}=\bm{2_{0}}\oplus \bm{2_{1}}\oplus \bm{4_{i}}\oplus \bm{4_{ii}}$} \\ \hline

\multicolumn{6}{|c}{  $ \begin{array}{l}
 \bm{1_{[a+b]}}:~\alpha_{1} \beta_{1}+\alpha_{2} \beta_{2}-\alpha_{3} \beta_{3}\\
 \bm{2^{0}}:~P^{a+b}_{2}\begin{pmatrix}\alpha_{1} \beta_{1}+\alpha_{2} \beta_{2}+2 \alpha_{3} \beta_{3}\\
\sqrt{3} (\alpha_{2} \beta_{2}-\alpha_{1} \beta_{1})
        \end{pmatrix}\\
 \bm{3_{[a+b]}}:
~\begin{pmatrix}\alpha_{2} \beta_{3}+\alpha_{3} \beta_{2}\\
\alpha_{1} \beta_{3}+\alpha_{3} \beta_{1}\\
-\alpha_{1} \beta_{2}-\alpha_{2} \beta_{1}
\end{pmatrix} \\
 \bm{3_{[a+b+1]}}:
~\begin{pmatrix}\alpha_{2} \beta_{3}-\alpha_{3} \beta_{2}\\
\alpha_{3} \beta_{1}-\alpha_{1} \beta_{3}\\
\alpha_{2} \beta_{1}-\alpha_{1} \beta_{2}
\end{pmatrix} \\
\end{array} $ } &
\multicolumn{6}{|c|}{  $ \begin{array}{l}
 \bm{2_{a}}: ~ \begin{pmatrix}
\sqrt{3}(\alpha_{1}\beta_{1}+\alpha_{2} \beta_{3})+\alpha_{1}\beta_{2}-\alpha_{2} \beta_{4}+2 \alpha_{3} \beta_{4}\\
\sqrt{3} (\alpha_{2} \beta_{1}-\alpha_{1} \beta_{3})-\alpha_{1}\beta_{4}-\alpha_{2} \beta_{2}-2 \alpha_{3} \beta_{2}
\end{pmatrix} \\
 \bm{2_{[a+1]}}:
~\begin{pmatrix}
\sqrt{3} (\alpha_{1} \beta_{2}+\alpha_{2} \beta_{4})-\alpha_{1} \beta_{1}+\alpha_{2} \beta_{3}-2 \alpha_{3} \beta_{3} \\
\sqrt{3}(\alpha_{2}\beta_{2}-\alpha_{1}\beta_{4})+\alpha_{1}\beta_{3}+\alpha_{2}\beta_{1}+2 \alpha_{3} \beta_{1}
\end{pmatrix}\\
 \bm{4_{i}}:
~P^{a}_{4}\begin{pmatrix} -\alpha_{1} \beta_{2}+\alpha_{2}\beta_{4}+\alpha_{3}\beta_{4}\\
\alpha_{1} \beta_{1}-\alpha_{2}\beta_{3} -\alpha_{3}\beta_{3} \\
\alpha_{1} \beta_{4}+\alpha_{2} \beta_{2}-\alpha_{3} \beta_{2}\\
-\alpha_{1} \beta_{3}-\alpha_{2} \beta_{1}+\alpha_{3} \beta_{1}
\end{pmatrix} \\
 \bm{4_{ii}}:
~P^{a}_{4}\begin{pmatrix}
\sqrt{3} (\alpha_{1} \beta_{1}+ \alpha_{2} \beta_{3})-3 \alpha_{3} \beta_{4}\\
-\sqrt{3}(\alpha_{1}\beta_{2}+\alpha_{2} \beta_{4})-2\alpha_{1}\beta_{1}+2 \alpha_{2} \beta_{3}-\alpha_{3} \beta_{3} \\
\sqrt{3} (\alpha_{2} \beta_{1}-\alpha_{1} \beta_{3})+3 \alpha_{3} \beta_{2}\\
\sqrt{3} (\alpha_{1} \beta_{4}-\alpha_{2} \beta_{2})+2 \alpha_{1} \beta_{3}+2 \alpha_{2} \beta_{1}+\alpha_{3} \beta_{1}
\end{pmatrix} \\
\end{array} $ } \\ \hline

\multicolumn{12}{|c|}{$\bm{4}\otimes \bm{4}=\bm{1_{0}}\oplus \bm{1_{1}}\oplus\bm{2^{0}}\oplus\bm{3_{0i}}\oplus\bm{3_{0ii}}\oplus \bm{3_{1i}}\oplus \bm{3_{1ii}}$}  \\ \hline
\multicolumn{12}{|c|}{  $ \begin{array}{l}
\bm{1_{0}}:~\alpha_{1} \beta_{4}-\alpha_{2} \beta_{3}-\alpha_{3} \beta_{2}+\alpha_{4} \beta_{1}\\
\bm{1_{1}}:~\alpha_{1} \beta_{3}+\alpha_{2} \beta_{4}-\alpha_{3} \beta_{1}-\alpha_{4} \beta_{2} \\
\bm{2^{0}}:~\begin{pmatrix}
\alpha_{1} \beta_{4}+\alpha_{2} \beta_{3}-\alpha_{3} \beta_{2}-\alpha_{4} \beta_{1}\\
\alpha_{1} \beta_{3}-\alpha_{2} \beta_{4}-\alpha_{3} \beta_{1}+\alpha_{4} \beta_{2}
\end{pmatrix} \\
\bm{3_{0i}}:~\begin{pmatrix}
\alpha_{1} \beta_{3}+\alpha_{2} \beta_{4}+\alpha_{3} \beta_{1}+\alpha_{4} \beta_{2}\\
\alpha_{1} \beta_{1}+\alpha_{2} \beta_{2}-\alpha_{3} \beta_{3}-\alpha_{4} \beta_{4}\\
\alpha_{1} \beta_{1}+\alpha_{2} \beta_{2}+\alpha_{3} \beta_{3}+\alpha_{4} \beta_{4}
\end{pmatrix} \\
\bm{3_{0ii}}:~\begin{pmatrix}
\sqrt{3} (\alpha_{1} \beta_{4}+ \alpha_{2} \beta_{3}+\alpha_{3} \beta_{2}+ \alpha_{4} \beta_{1})-2 \alpha_{2} \beta_{4} -2 \alpha_{4} \beta_{2}\\
-\sqrt{3}(\alpha_{1} \beta_{2}+\alpha_{2}\beta_{1}-\alpha_{3} \beta_{4}-\alpha_{4} \beta_{3})-2\alpha_{2} \beta_{2} +2 \alpha_{4} \beta_{4}\\
-3 \alpha_{1} \beta_{1}+\alpha_{2} \beta_{2}-3 \alpha_{3} \beta_{3}+\alpha_{4} \beta_{4}
\end{pmatrix} \\
\bm{3_{1i}}:~\begin{pmatrix}
\alpha_{1} \beta_{4}-\alpha_{2} \beta_{3}+\alpha_{3} \beta_{2}-\alpha_{4} \beta_{1}\\
\alpha_{1} \beta_{2}-\alpha_{2} \beta_{1}-\alpha_{3} \beta_{4}+\alpha_{4} \beta_{3}\\
\alpha_{1} \beta_{2}-\alpha_{2} \beta_{1}+\alpha_{3} \beta_{4}-\alpha_{4} \beta_{3}
\end{pmatrix} \\
\bm{3_{1ii}}:~\begin{pmatrix}
\sqrt{3}(\alpha_{1}\beta_{3}-\alpha_{2}\beta_{4}+\alpha_{3}\beta_{1}-\alpha_{4}\beta_{2})-\alpha_{1}\beta_{4}-\alpha_{2}\beta_{3}-\alpha_{3}\beta_{2}-\alpha_{4}\beta_{1} \\
-\sqrt{3}(\alpha_{1}\beta_{1}+\alpha_{2} \beta_{2}+\alpha_{3} \beta_{3}-\alpha_{4} \beta_{4})-\alpha_{1}\beta_{2}-\alpha_{2} \beta_{1}+\alpha_{3} \beta_{4}+\alpha_{4} \beta_{3} \\
2 (\alpha_{1} \beta_{2}+\alpha_{2} \beta_{1}+\alpha_{3} \beta_{4}+\alpha_{4} \beta_{3})
\end{pmatrix}
\\
\end{array} $ } \\
\hline

\specialrule{0em}{1.0pt}{1.0pt}
\midrule
\end{longtable}
\end{small}
\end{center}

\section{\label{sec:vev} Vacuum alignment of flavons }

\begin{table}[h!]
\renewcommand{\tabcolsep}{2.5mm}
\renewcommand{\arraystretch}{1.1}
\centering
\begin{tabular}{|c|c|c|c|c|c|c|c||c|c|c|c|c|c|c|c|}\hline \hline
Fields &   $\zeta$ & $\phi$  &  $\varphi$  &  $\chi$ & $\psi$  & $\rho$ & $\kappa$  & $\zeta^0$ & $\phi^0$ & $\psi^0$ &   $\varphi^0$ & $\chi^0$   \\ \hline

$Q_{8}\rtimes S_{3}$ & $\bm{1_{1}}$ & $\bm{3_{0}}$ & $\bm{2_{0}}$ & $\bm{3_{1}}$  & $\bm{3_{1}}$ &$\bm{1_{0}}$ & $\bm{1_{0}}$ & $\bm{1_{0}}$ &   $\bm{3_{0}}$ &  $\bm{3_{1}}$ &    $\bm{2_{1}}$ &  $\bm{2^{0}}$ \\ \hline

Modular weight &  $0$ & $0$ & $0$ & $2$  & $2$  & $2$ & $0$ & $0$ & $0$ &  $-2$  &  $-4$  & $-4$\\  \hline

$Z_5$ &  $\omega^2_{5}$ &  $\omega^2_{5}$   & $\omega^{3}_{5}$  & $\omega^{3}_{5}$ & $\omega^{3}_{5}$  & $\omega^{3}_{5}$ &  $\omega^2_{5}$ &  $\omega_{5}$ &  $\omega_{5}$ &  $1$ &   $\omega^4_{5}$ &  $\omega^4_{5}$ \\ \hline

$Z_2$ &  $+$ &  $+$   & $+$  & $+$ & $-$ & $+$ & $-$ &  $+$ &  $+$   & $+$  & $+$  & $+$ \\ \hline

$U(1)_{R}$ &  $0$ &  $0$   & $0$  & $0$ & $0$ & $0$ & $0$ &  $2$ &  $2$   & $2$  & $2$  & $2$ \\ \hline \hline

\end{tabular}
\caption{\label{tab:flavon_driving_fields}Assignments of flavon fields and driving fields under the EFG $Q_{8}\rtimes S_{3}$ and the auxiliary symmetry $Z_{5}\times Z_{2}$, where $\omega_{5}=e^{\frac{2\pi i}{5}}$.  }
\end{table}

In this Appendix, we employ the now-standard $F$-term alignment mechanism~\cite{Altarelli:2005yx} to generate the appropriate vacuum alignments of the EFG symmetry breaking flavons fields in Eq.~\eqref{eq:seesaw_VEVs}. A global $U(1)_{R}$ continuous symmetry is required in this approach, and all terms in the superpotential must carry two units of $R$ charge. The matter fields, flavon fields and Higgs, and driving fields are assumed to carry one, zero and two units of $R-$charge, respectively. Three additional flavon fields $\psi$, $\rho$, $\kappa$ together with five driving fields are introduced in the model. Another auxiliary $Z_2$ symmetry is introduced to forbid the dangerous operators in the driving superpotential. The necessary flavon fields and driving fields which are indicated with the superscript ``0'',  and their transformation properties under the flavor symmetry $(Q_{8}\rtimes S_{3})\times Z_5\times Z_2$ are given in table~\ref{tab:flavon_driving_fields}. Notice that all the flavon and driving fields are SM singlets. The minimum of the scalar potential is determined by vanishing $F$-terms of the driving fields. We can read out leading order (LO) driving superpotential $w_{d}$ invariant under $(Q_{8}\rtimes S_{3})\times Z_5\times Z_2$ as follows
\begin{eqnarray}
\nonumber w_{d}&=&f_1\zeta^0\zeta\zeta+f_2\zeta^0\left(\phi\phi\right)_{\bm{1_{0}}}+f_3\zeta^0\kappa\kappa+f_{4}\left(\phi^0\left(\phi\phi\right)_{\bm{3_{0}}}\right)_{\bm{1_{0}}}+f_5\kappa \left(\psi^{0}\psi\right)_{\bm{1_{0}}}+f_6\left(\psi^0\left(\phi\chi\right)_{\bm{3_{1}}}\right)_{\bm{1_{0}}} \\
\label{eq:wd} &&+f_{7}\left(\chi^0\left(\chi\chi\right)_{\bm{2^{0}}}\right)_{\bm{1_{0}}}+f_{8}\left(\chi^0\left(\psi\psi\right)_{\bm{2^{0}}}\right)_{\bm{1_{0}}}+f_{9}\rho \left(\varphi^{0}\varphi\right)_{\bm{1_{0}}} +f_{10}\left(\varphi^0\left(\varphi\chi\right)_{\bm{2_{0}}}\right)_{\bm{1_{0}}}  \,,
\end{eqnarray}
where $(\ldots)_{\bm{r}}$ stands for a contraction into the $Q_{8}\rtimes S_{3}$ irreducible representation $\bm{r}$, and all the coupling constants  $f_i$ ($i=1,\cdots,10$) are real since the theory is required to be invariant under the gCP transformations. In the exact supersymmetric limit, the $F-$terms of the driving fields have to vanish at the minimum of the scalar potential such that the vacuum of the flavon fields is aligned. In the charged lepton sector, the equations for the vanishing of the derivatives of $w_d$ with respect to each component of the driving fields $\zeta^0$ and $\phi^0$ are:
\begin{eqnarray}
\label{eq:align_ch}
\nonumber&&\frac{\partial w_{d}}{\partial\zeta^0}=f_{1} \zeta^2+f_{2} \left(\phi_{1}^2+\phi_{2}^2-\phi_{3}^2\right)+f_{3} \kappa ^2=0\,,\\
\nonumber&&\frac{\partial w_{d}}{\partial\phi^{0}_{1}}=2 f_{4} \phi_{2} \phi_{3}=0\,,\\
\nonumber&&\frac{\partial w_{d}}{\partial\phi^{0}_{2}}=2 f_{4} \phi_{1} \phi_{3}=0\,,\\
&&\frac{\partial w_{d}}{\partial\phi^{0}_{3}}=2 f_{4} \phi_{1} \phi_{2}=0\,.
\end{eqnarray}
One solution to these equations is
\begin{equation}\label{eq:vev_phi}
\langle\phi\rangle=v_{\phi}(0,1,0)^T\,, \qquad  \langle\zeta\rangle=v_{\zeta}\,,\qquad \langle\kappa\rangle=v_{\kappa}\qquad  \text{with} \qquad v_{\zeta}=\sqrt{\frac{f_{2}v_{\phi}^2+f_{3}v^2_{\kappa}}{f_{1}}}\,,
\end{equation}
where $v_{\phi}$ is undetermined. In the neutrino sector, the $F$-term conditions of the driving fields $\psi^0$ and $\chi^0$ give the vacuum alignments of $\chi$ and $\psi$,
\begin{eqnarray}
\nonumber&&\frac{\partial w_{d}}{\partial\psi^{0}_{1}}=f_{5} \kappa  \psi_{1}+f_{6} (\chi_{3} \phi_{2}+\chi_{2} \phi_{3})=0\,,\\
\nonumber&&\frac{\partial w_{d}}{\partial\psi^{0}_{2}}=f_{5} \kappa  \psi_{2}+f_{6} (\chi_{3} \phi_{1}+\chi_{1} \phi_{3})=0\,,\\
\nonumber&&\frac{\partial w_{d}}{\partial\psi^{0}_{3}}=f_{6} (\chi_{2} \phi_{1}+\chi_{1} \phi_{2})-f_{5} \kappa  \psi_{3}=0\,,\\
\nonumber &&\frac{\partial w_{d}}{\partial\chi^{0}_{1}}=f_{7} \left(\chi_{1}^2+\chi_{2}^2+2 \chi_{3}^2\right)+f_{8} \left(\psi_{1}^2+\psi_{2}^2+2 \psi_{3}^2\right)=0\,, \\
&&\frac{\partial w_{d}}{\partial\chi^{0}_{1}}=\sqrt{3} f_{7} \left(\chi_{2}^2-\chi_{1}^2\right)+\sqrt{3} f_{8} \left(\psi_{2}^2-\psi_{1}^2\right)=0\,.
\end{eqnarray}
Given the vacuum of $\phi$ in Eq.~\eqref{eq:vev_phi}, we find the alignments of $\chi$ and $\psi$ are
\begin{equation}\label{eq:vev_chi}
\langle\chi\rangle=v_{\chi}\left(0, 1,i\right)^T\,, \qquad  \langle\psi\rangle=v_{\psi}\left(1, 0,0\right)^T\,,
\end{equation}
with
\begin{equation}
v_{\psi}=\sqrt{\frac{f_{7}}{f_{8}}}v_{\chi}\,,  \qquad v_{\kappa}=\sqrt{-\frac{f_{5}^2f_{7}}{f_{6}^2f_{8}}}v_{\phi}
\end{equation}
The $F$-flatness condition of the driving field $\varphi^{0}$ leads to
\begin{eqnarray}
\nonumber&&\frac{\partial w_{d}}{\partial\varphi^{0}_{1}}=f_{10} \left[\varphi_{1} (\chi_{2}-\chi_{3})+\varphi_{2} \chi_{1}\right]+f_{9} \rho \varphi_{2}=0\,,\\
&& \frac{\partial w_{d}}{\partial\varphi^{0}_{2}}=f_{10} \left[\varphi_{1} \chi_{1}-\varphi_{2} (\chi_{2}+\chi_{3})\right]-f_{9} \rho \varphi_{1}=0\,,
\end{eqnarray}
from which we can extract the vacuum expectation values of $\varphi$ and $\rho$
\begin{equation}\label{eq:vev_varphi}
 \langle\varphi\rangle=v_{\varphi}\left(1, \xi^3\right)^T, \qquad \langle\rho\rangle=v_{\rho}\,,\qquad\text{with} \qquad v_{\rho} =\frac{\sqrt{2}f_{10}}{f_{9}}v_{\chi}\,.
\end{equation}
We see that the vacuum alignments in Eq.~\eqref{eq:seesaw_VEVs} are reproduced.
When considering all seven flavon fields listed in Table~\ref{tab:flavon_driving_fields}, an additional term, $\frac{1}{\Lambda^{2}}(E^{c}_{1}L_{1})_{\bm{1_{0}}}(\kappa\kappa)_{\bm{1_{0}}}H_{d}$, arises in $\mathcal{W}_{e}$ in Eq.~\eqref{eq:WO_SP}. Its contribution can be absorbed by redefining the parameter $\alpha'$ in Eq.~\eqref{eq:lep_masses}. Moreover, the additional subleading corrections to the superpotential terms $\mathcal{W}_{e}$ and $\mathcal{W}_{\nu}$ take the following forms:
\begin{eqnarray}
\nonumber&&\hskip-0.3in \Delta \mathcal{W}^{\prime}_{e}=\frac{H_{d}}{\Lambda^3}\left[\left(E^{c}_{1}L_{1}\phi^2_{\nu}\rho Y^{(k)}_{\bm{r}}\right)_{\bm{1_{0}}}+\left(E^{c}_{1}L_{1}\psi^2\phi_{\nu} Y^{(k)}_{\bm{r}}\right)_{\bm{1_{0}}}+\left(E^{c}_{1}L_{1}\psi^2\rho Y^{(k)}_{\bm{r}}\right)_{\bm{1_{0}}}+\left(E^{c}_{1}L_{1}\rho^3 Y^{(k)}_{\bm{r}}\right)_{\bm{1_{0}}}\right.\\
\nonumber &&+\left(E^{c}_{1}L_{d}\phi^2_{\nu}\rho Y^{(k)}_{\bm{r}}\right)_{\bm{1_{0}}}+\left(E^{c}_{1}L_{d}\psi^2\phi_{\nu} Y^{(k)}_{\bm{r}}\right)_{\bm{1_{0}}}+\left(E^{c}_{1}L_{d}\psi^2\rho Y^{(k)}_{\bm{r}}\right)_{\bm{1_{0}}}+\left(E^{c}_{1}L_{d}\phi_{\nu}\rho^2 Y^{(k)}_{\bm{r}}\right)_{\bm{1_{0}}}\\
\nonumber &&+\left(E^{c}_{d}L_{1}\phi^2_{l}\rho Y^{(k)}_{\bm{r}}\right)_{\bm{1_{0}}}+\left(E^{c}_{d}L_{1}\phi_{l}\kappa \psi Y^{(k)}_{\bm{r}}\right)_{\bm{1_{0}}}+\left(E^{c}_{d}L_{1}\kappa^2\phi_{\nu} Y^{(k)}_{\bm{r}}\right)_{\bm{1_{0}}} \\
\nonumber &&\left.+\left(E^{c}_{d}L_{d}\phi^2_{l}\rho Y^{(k)}_{\bm{r}}\right)_{\bm{1_{0}}}+\left(E^{c}_{d}L_{d}\phi_{l}\kappa \psi Y^{(k)}_{\bm{r}}\right)_{\bm{1_{0}}}+\left(E^{c}_{d}L_{d}\kappa^2\phi_{\nu} Y^{(k)}_{\bm{r}}\right)_{\bm{1_{0}}}+\left(E^{c}_{d}L_{d}\kappa^2\rho Y^{(k)}_{\bm{r}}\right)_{\bm{1_{0}}}\right]\,,\\
\nonumber &&\hskip-0.3in \Delta \mathcal{W}^{\prime}_{\nu}=\frac{H_{u}H_{u}}{\Lambda^4}\left[\left(L^2_{1}\phi_{l}\phi_{\nu}\rho Y^{(k)}_{\bm{r}}\right)_{\bm{1_{0}}}+\left(L^2_{1}\kappa\phi_{\nu}\psi Y^{(k)}_{\bm{r}}\right)_{\bm{1_{0}}}+\left(L^2_{1}\phi_{l}\psi^2 Y^{(k)}_{\bm{r}}\right)_{\bm{1_{0}}}+\left(L^2_{1}\phi_{l}\rho^2 Y^{(k)}_{\bm{r}}\right)_{\bm{1_{0}}}\right. \\
\nonumber &&+ \left(L_{1}L_{d}\phi_{l}\phi_{\nu}\rho Y^{(k)}_{\bm{r}}\right)_{\bm{1_{0}}}+\left(L_{1}L_{d}\kappa\phi_{\nu}\psi Y^{(k)}_{\bm{r}}\right)_{\bm{1_{0}}}+\left(L_{1}L_{d}\phi_{l}\psi^2 Y^{(k)}_{\bm{r}}\right)_{\bm{1_{0}}} +\left(L_{1}L_{d}\phi_{l}\rho^2 Y^{(k)}_{\bm{r}}\right)_{\bm{1_{0}}}\\
\nonumber && + \left(L^2_{d}\phi_{l}\phi_{\nu}\rho Y^{(k)}_{\bm{r}}\right)_{\bm{1_{0}}}+\left(L^2_{d}\kappa\phi_{\nu}\psi Y^{(k)}_{\bm{r}}\right)_{\bm{1_{0}}}+\left(L^2_{d}\kappa\rho\psi Y^{(k)}_{\bm{r}}\right)_{\bm{1_{0}}}\\
&& \left. +\left(L^2_{d}\phi_{l}\psi^2 Y^{(k)}_{\bm{r}}\right)_{\bm{1_{0}}} +\left(L^2_{d}\phi_{l}\rho^2 Y^{(k)}_{\bm{r}}\right)_{\bm{1_{0}}}\right]\,,
\end{eqnarray}
whose contributions can be absorbed by $\Delta \mathcal{W}_{e}$ and $\Delta \mathcal{W}_{\nu}$ in Eq.~\eqref{eq:WO_NLO}, respectively. Here the coefficients of each operator are omitted. In short, the EFG can severely restricts K\"ahler potential as well as the superpotential and it leads to a highly constrained framework, the prize is the reintroduction of traditional favour symmetry and flavons which make the model more complicated.

\end{appendix}


\providecommand{\href}[2]{#2}\begingroup\raggedright\endgroup

\end{document}